\documentclass[12pt]{article}
\pdfoutput=1
\usepackage{jheppub}
\usepackage{amsmath}
\usepackage{amsfonts}
\usepackage{amssymb}
\usepackage{graphicx}

\usepackage[export]{adjustbox}
\newcommand{\bmat}{\left(\begin{array}}
\newcommand{\emat}{\end{array}\right)}

\def\yzero{\smash{\hbox{$y\kern-4pt\raise1pt\hbox{${}^\circ$}$}}}

\def\beq{\begin{equation}}
\def\eeq{\end{equation}}
\def\beqa{\begin{eqnarray}}
\def\eeqa{\end{eqnarray}}

\def\-{\hphantom{-}}
\def\ov{\overline}
\def\s2{\frac{1}{\sqrt2}}

\def\beq{\begin{equation}}
\def\eeq{\end{equation}}
\def\beqa{\begin{eqnarray}}
\def\eeqa{\end{eqnarray}}
\def\tr{{\rm tr \,}}

\def\diag{{\rm diag \,}}
\def\II{\relax{\rm I\kern-.18em I}}

\def\Dsl{\,\raise.15ex\hbox{/}\mkern-13.5mu D} 
\def\id{{\rm {I}}}
\def\aD9{{\ov{\rm D9}}}

\def\IC{{\bf{C}}}

\def\IS{{\bf {S}}}
\def\IR{{\bf {R}}}
\def\IZ{{\bf {Z}}}
\def\IX{{\bf {X}}}

\def\CN{{\cal {N}}}

\def\NN{{\cal {N}}}

\newcommand{\drawsquare}[2]{\hbox{%
\rule{#2pt}{#1pt}\hskip-#2pt
\rule{#1pt}{#2pt}\hskip-#1pt
\rule[#1pt]{#1pt}{#2pt}}\rule[#1pt]{#2pt}{#2pt}\hskip-#2pt
\rule{#2pt}{#1pt}}

\newcommand{\fund}{\,\raisebox{-.5pt}{\drawsquare{6.5}{0.4}}\,}
\newcommand{\antifund}{\overline{\fund}}

\catcode`\@=11   
\newdimen\@rotdimen
\newbox\@rotbox  

\def\@vspec#1{\special{ps:#1}}
\def\@rotstart#1{\@vspec{gsave currentpoint currentpoint translate
   #1 neg exch neg exch translate}}
\def\@rotfinish{\@vspec{currentpoint grestore moveto}}
\def\@rotr#1{\@rotdimen=\ht#1\advance\@rotdimen by\dp#1%
   \hbox to\@rotdimen{\hskip\ht#1\vbox to\wd#1{\@rotstart{90 rotate}%
   \box#1\vss}\hss}\@rotfinish}
\def\@rotl#1{\@rotdimen=\ht#1\advance\@rotdimen by\dp#1%
   \hbox to\@rotdimen{\vbox to\wd#1{\vskip\wd#1\@rotstart{270 rotate}%
   \box#1\vss}\hss}\@rotfinish}%
\def\@rotu#1{\@rotdimen=\ht#1\advance\@rotdimen by\dp#1%
   \hbox to\wd#1{\hskip\wd#1\vbox to\@rotdimen{\vskip\@rotdimen
   \@rotstart{-1 dup scale}\box#1\vss}\hss}\@rotfinish}%
\def\@rotf#1{\hbox to\wd#1{\hskip\wd#1\@rotstart{-1 1 scale}%
   \box#1\hss}\@rotfinish}%
\def\rotate{\@ifnextchar[{\@rotate}{\@rotate[l]}}
\def\@rotate[#1]#2{\setbox\@rotbox=\hbox{#2}\@nameuse{@rot#1}\@rotbox}

\catcode`\@=12

\begin{document}

\makeatletter
\@addtoreset{equation}{section}
\makeatother
\renewcommand{\theequation}{\thesection.\arabic{equation}}
\pagestyle{empty}
\rightline{IFT-UAM/CSIC-25-001}
\vspace{.5cm}
\begin{center}
\Large{\bf Relative Quantum Gravity:\\
Localized Gravity and the Swampland 
}
\\

\large{Edoardo Anastasi, Roberta Angius, Jes\'us Huertas, \\Angel M. Uranga, Chuying Wang \\[4mm]}
\footnotesize{Instituto de F\'{\i}sica Te\'orica IFT-UAM/CSIC,\\[-0.3em] 
C/ Nicol\'as Cabrera 13-15, 
Campus de Cantoblanco, 28049 Madrid, Spain}\\ 
\footnotesize{\href{edo.anastasi@virgilio.it}{edo.anastasi@virgilio.it}, \href{roberta.angius@csic.es}{roberta.angius@csic.es}, \href{j.huertas@csic.es}{j.huertas@csic.es}, \href{mailto:angel.uranga@csic.es}{angel.uranga@csic.es}, \href{mailto:chuying.wang@ift.csic.es}{chuying.wang@ift.csic.es}}

\vspace*{8mm}

\small{\bf Abstract} \\
\end{center}
\begin{center}
\begin{minipage}[h]{\textwidth}
\small{We perform a systematic study of the applicability of swampland constraints to theories of localized gravity. We find that these gravity theories can violate swampland constraints, but can be reconciled with them when coupled to a higher-dimensional gravity theory. They realize what we call {\em relative quantum gravity}: to become consistent at the quantum level, these gravity theories must be defined as {\em relative} to a host higher-dimensional gravity theory. 

We show that these theories can admit global symmetries, even anomalous ones; they can violate the cobordism, completeness, weak gravity, and distance conjectures; they may admit stable non-supersymmetric AdS vacua, or dS vacua. All swampland constraints are however satisfied when these gravity theories are regarded as relative and completed by coupling them to a higher-dimensional one.

We discuss these properties in $d$-dimensional gravity theories localized on Karch-Randall End of the World (ETW) boundaries of AdS$_{d+1}$ spacetime. For AdS$_d$ ETW branes we use the formalism of double holography to describe the appearance of the species scale and the emergence of gauge dynamics from the quantum backreaction of CFT$_d$ modes. We also study microscopically the swampland constraints in localized gravity in explicit string theory models. Concretely, we exploit the 10d supergravity solutions describing AdS$_4$ ETW branes for AdS$_5\times\IS^5$, holographically dual to semi-infinite D3-branes ending on NS5- and D5-brane configurations, realizing 4d $\NN=4$ $SU(N)$ on half-space coupled to a 3d Gaiotto-Witten  superconformal boundary CFT$_3$.}

\newpage

\end{minipage}
\end{center}
\newpage
\setcounter{page}{1}
\pagestyle{plain}
\renewcommand{\thefootnote}{\arabic{footnote}}
\setcounter{footnote}{0}

\vspace*{-1cm}

\tableofcontents

\vspace*{1cm}


\section{Introduction}
\label{sec:intro}

Some of the main advances in Theoretical Physics in the recent decades stem from the realization that it is possible to localize degrees of freedom on subspaces of spacetime. This encompasses from the advent of D-branes and other non-perturbative extended branes in string theory \cite{Polchinski:1995mt,Polchinski:1998rr} to the phenomenological building of BSM models with (potentially) large extra dimensions \cite{Arkani-Hamed:1998jmv,Antoniadis:1998ig,Randall:1999ee,Randall:1999vf}. Localization of scalars, fermions and gauge bosons arises fairly naturally, in particular in UV complete setups like string theory, where it has provided a formidable tool in the study of quantum field theories localized on the brane worldvolumes \cite{Giveon:1998sr}. 

This naturally opens up the exploration of mechanisms to localize gravity on lower dimensional subspaces of spacetime, successfully initiated from a bottom-up perspective in several setups \cite{Randall:1999vf,Gregory:2000jc,Dvali:2000hr,Karch:2000ct,Karch:2001cw}. Some of these mechanisms admit well-motivated top-down embeddings in string theory \cite{Verlinde:1999fy,Karch:2001cw} (see also \cite{DHoker:2007zhm,DHoker:2007hhe,Aharony:2011yc,Assel:2011xz,Bachas:2017rch,Bachas:2018zmb}, also \cite{Raamsdonk:2020tin,VanRaamsdonk:2021duo,Demulder:2022aij,Karch:2022rvr,DeLuca:2023kjj,Huertas:2023syg,Chaney:2024bgx} for recent applications). However, general criteria for the consistency or not of the possible setups of localized gravity in UV complete theories of quantum gravity are lacking. 

A natural candidate to provide such criteria is the swampland program \cite{Vafa:2005ui}. This attempts to provide the constraints that an effective theory of gravity (coupled to other fields) must satisfy in order to admit a UV completion in quantum gravity. This has produced a remarkable web of interconnected conjectured constraints, extensively explored in the recent literature (see \cite{Palti:2019pca,vanBeest:2021lhn,Grana:2021zvf,Agmon:2022thq} for reviews), proposed to hold for any theory of gravity admitting such a completion. It is therefore an important question to explore the applicability of these swampland constraints to localized gravity theories, with a two-fold motivation. On one hand, test the conjectures in this unexplored arena of UV completing gravity theories, not by just additional fields, but by coupling them to a new higher-dimensional gravity sector. On the other hand, to provide model independent criteria for the consistency of localized gravity setups, and of their embedding in string theory or quantum gravity in general.

This work is a systematic exploration of the application of swampland constraints to theories of gravity localized on lower-dimensional subspaces of spacetime (for a different approach, see \cite{Geng:2023iqd}). Our analysis reveals a captivating picture: these gravity theories typically violate the usual swampland constraints, but can be reconciled with them if they are coupled to a higher-dimensional gravity theory. Therefore, to become consistent theories at the quantum level these gravity theories must be defined as {\em relative} to a host higher-dimensional gravity theory\footnote{One could envision scenarios of localized gravity theories on subspaces of higher-dimensional non-gravitational theories, sometimes referred to as  emergent (or induced) gravity, see e.g. \cite{Gherghetta:2005se,Kiritsis:2006ua} also \cite{Sindoni:2011ej} for a review. We will not consider these setups (except for the emergence of gravity in the sense of holographic dualities).}. We dub them {\em relative quantum gravity} theories.

For concreteness we focus on gravity theories localized on codimension 1 defects, a setup which in fact includes the main scenarios of localized gravity considered in the literature. Many such setups also localize gravity on the boundary of a gravity theory in a spacetime in one dimension higher, cutoff at the location of the brane, or can be obtained by back-to-back gluing of two copies of such setup \cite{Randall:1999vf}. Hence, we focus on this basic building block of localizing gravity on the boundary of spacetime, which we will refer to as End of the World (ETW) brane. 

This is a further motivation for our use of the term {\em relative}. In formal QFT it is becoming increasingly manifest that certain $d$-dimensional theories, dubbed {\em relative quantum field theories} \cite{Freed:2012bs} require, for their complete definition, being regarded as living on the boundary of a $(d+1)$-dimensional theory to which they couple. One familiar example arises in the study of anomalies (see \cite{Alvarez-Gaume:1985zzv} for a review of early methods, \cite{Dai:1994kq} for modern ones, and \cite{Garcia-Etxebarria:2018ajm} for physical applications of the latter). A more recent example arises in the study of generalized symmetries and their associated topological operators (see \cite{McGreevy:2022oyu,Brennan:2023mmt,Gomes:2023ahz,Shao:2023gho,Schafer-Nameki:2023jdn,Bhardwaj:2023kri,Iqbal:2024pee} for reviews), where physical QFTs require, for their complete definition, being regarded as living on the boundary of a Symmetry Topological Field Theory (SymTFTs) extending in one extra dimension (actually, of finite extent, and supplemented with topological boundary conditions on the additional boundary). In this vein, our notion of relative quantum gravity theories is a further incarnation of this concept: gravity theories which, for their compatibility with swampland constraints (hence with a consistent embedding in quantum gravity), require being coupled to a one dimension higher gravity theory.

We find that relative gravity theories can violate essentially all known swampland conjectures, in their standard formulations. We uncover that these $d$-dimensional gravity theories can admit global symmetries, which can even be anomalous; they can fall in totally disconnected cobordism classes with no domain wall connecting them; they can violate the completeness conjecture, the weak gravity conjecture, and diverse variants of distance conjectures; they may admit stable non-supersymmetric AdS vacua, or dS vacua. All the swampland constraints are however satisfied, essentially in their standard formulations, when these $d$-dimensional gravity theories are regarded as relative, i.e. completed by coupling them to a higher-dimensional gravity theory.

\medskip

We focus our exploration in $d$-dimensional gravity theories localized on the ETW boundary of AdS$_{d+1}$ spacetime. In particular we focus on the Karch-Randall (KR) setup of an AdS$_d$ ETW boundary of AdS$_{d+1}$ \cite{Karch:2000ct} (many conclusions extend similarly to the Randall-Sundrum (RS) scenario of $d$-dimensional Minkowski ETW boundary of AdS$_{d+1}$ \cite{Randall:1999vf}, and to KR dS$_d$ ETW boundaries of AdS$_{d+1}$). Although the localized graviton is massive in the AdS ETW setup, it can be made parametrically lighter than any other scale by tunning $L_d\gg L_{d+1}$ for the curvature length scales of the AdS spaces. Hence the setup describes localized AdS$_d$ gravity in a parametrically vast regime of energies.

The motivation to focus on AdS geometries is that they allow for dual descriptions using the  holographic dictionary, and in particular AdS$_d$ ETW boundaries of AdS$_{d+1}$ realize the scenario known as double holography \cite{Karch:2000gx,Karch:2000ct,Takayanagi:2011zk}, in which there are two holographic dual descriptions. There is a holographic picture in terms of a CFT$_d$, dual to the AdS$_{d+1}$ gravity theory, defined on a half $d$-dimensional flat space with boundary conditions given by a boundary CFT (BCFT$_{d-1}$), dual to the AdS$_d$ gravity theory on the ETW brane. The localized AdS$_d$ graviton is dual to the energy-momentum tensor of the BCFT$_{d-1}$, and its non-conservation due to its coupling to the CFT$_d$ implies the non-zero graviton mass. The parametrically light graviton mass regime corresponds to the BCFT$_{d-1}$ having parametrically larger number of degrees of freedom than the CFT$_d$. This boundary picture will be useful in reformulating the idea of relative quantum gravity in terms of swampland conjectures for CFT/BCFT theories. 

A second alternative holographic description, known as intermediate picture, is obtained by replacing only the BCFT$_{d-1}$ by its gravity dual. It is given by an AdS$_d$ gravity theory coupled to the CFT$_d$ with a UV cutoff, with its conformal boundary coupled with transparent boundary conditions to the CFT$_d$ in flat spacetime with no dynamical gravity. 
In terms of this intermediate picture, one can reformulate the statements about relative quantum gravity as follows. We have that a theory of AdS$_d$ gravity (plus other fields) may violate some swampland constraints, but it can be made consistent it is coupled to a {\em suitable} CFT$_d$ (with, in practice, ``suitable'' meaning ``admitting a gravity dual''). Such theories are thus relative quantum gravity theories in the sense that they are defined as complete UV theories only as relative to a larger theory including this CFT$_d$ sector. As we will discuss, the quantum effects of the CFT$_d$ play, via emergence, a crucial role in reconciling the initial $d$-dimensional gravity theory with the swampland constraints. 

In addition to these bottom-up description of Karch-Randall setups, we also discuss explicit top-down string theory solutions describing gravity localization on supersymmetric AdS$_4$ ETW branes of AdS$_5\times\IS^5$ \cite{DHoker:2007zhm,DHoker:2007hhe,Aharony:2011yc,Assel:2011xz,Bachas:2017rch,Bachas:2018zmb} (see also \cite{Raamsdonk:2020tin,VanRaamsdonk:2021duo,Demulder:2022aij,Karch:2022rvr,Huertas:2023syg,Chaney:2024bgx} for recent applications). They are given by 10d supergravity solutions obtained as near-horizon limits of a stack of $N$ semi-infinite D3-branes ending on configurations of NS5- and D5-branes, which define  3d $\NN=4$ superconformal BCFT$_3$ boundary conditions for 4d $\NN=4$ $SU(N)$ SYM in 4d half-space \cite{Gaiotto:2008sa,Gaiotto:2008ak}. These models play a crucial role to provide an explicit microscopic realization of some of the bottom-up mechanisms we explore, and also allow for an explicit realization of double holography.

An important observation regarding these string theory configurations is that they have no scale separation between the curvature scales of AdS$_5$ and of $\IS^5$ (and similarly of the AdS$_4$ and of the corresponding internal space $\IX_6$). This implies that for certain purposes they may not be considered to produce models of genuinely 4d localized gravity in a genuine 5d bulk. In this respect, we actually regard our string configurations as toy versions of possible genuinely scale separated setups (possibly based on the scale separated AdS vacua of the kind in \cite{DeWolfe:2005uu,Camara:2005dc}) realizing the same mechanisms. In addition, some of our arguments are topological in nature, and are hence insensitive to the size of the compact spaces, and hence, of scale separation issues.

\medskip

{\bf Summary or results and organization of the paper}

The discussion and results of the different swampland constraints for relative gravity theories are organized in the paper as follows: In section \ref{sec:ads} we review the construction of gravity localized models on $d$-dimensional ETW boundaries of AdS$_{d+1}$. In section \ref{sec:double-holography} we review the bottom-up Karch-Randall setup and its double holography interpretation, while in section \ref{sec:d3-ns5-d5} we review the string construction of AdS$_4$ ETW configurations for AdS$_5\times\IS^5$, and its double holographic descriptions in terms of D3-branes ending on 5-brane configurations.
The following sections contain the discussions of the different swampland constraints. Although the discussion is organized following a definite logical flow, the different sections can also be considered in a fairly independent way, hopefully allowing the reader to move freely among them.

In section \ref{sec:global-swampland} we discuss swampland constraints related to global symmetries in relative gravity theories. In section \ref{sec:no-global-symmetries} we discuss the no global symmetry conjecture. We show that localized gravity theories admit exact global symmetries, as long as they are gauged in their higher-dimensional hosts. We show this both in the bottom-up Karch-Randall setup and in top-down explicit string theory constructions, and for both usual and generalized symmetries, focusing on the 1-form symmetries of 4d $\NN=4$ $SU(N)$ SYM. In section \ref{sec:anomaly} we show that the global symmetries in localized gravity theories can be anomalous, if the higher-dimensional host provides an anomaly inflow mechanism to make the gauge version of the symmetry consistent. We present a bottom-up Karch-Randall discussion, and an explicit string theory construction of AdS$_4$ ETW configurations for AdS$_5\times\IS^5$ with extra flavour branes. In section \ref{sec:emergence} we describe the global symmetries in the intermediate picture of double holography, using the formalism of holographic renormalization, and show that the quantum backreaction of the fast CFT$_d$ modes above the UV cutoff leads to the gauging of the symmetry, by generating via emergence the required gauge kinetic terms. The discussion allows for an explicit connection with the species scale, and its various definitions: as a measure of the number of degrees of freedom of the theory, as the scale controlling higher curvature corrections, and as setting the scale of the smaller black holes. In section \ref{sec:cobordism} we discuss the cobordism conjecture. We provide explicit (bottom-up and top-down) realizations of gravity theories which are not cobordant (due to a mismatch of 't Hooft anomalies for some global symmetries present in the theories), but can be connected once embedded in  higher-dimensional host theories. This motivates the concept of {\em relative defect}, i.e. one that exists in a $d$-dimensional relative theory only as the boundary of a higher-dimensional defect in the $(d+1)$-dimensional host theory.

In section \ref{sec:gauge-swampland} we discuss swampland constraints on gauge symmetries of the $d$-dimensional theory, which must necessarily be localized on the ETW boundary. In section \ref{sec:completeness} we explore the completeness conjecture, and argue that it can be violated in the purely $d$-dimensional gravity theory. Completeness is however satisfied by the coupling to the $(d+1)$-dimensional theory, with the missing charges being provided by relative defects, namely defects of the $d$-dimensional theory which are defined as boundaries of defects extending in the extra dimension of the $(d+1)$-dimensional theory. In section \ref{sec:weak-gravity} we study the weak gravity conjecture. We explain that, although it can be violated in the $d$-dimensional theory, e.g. due to the very absence of charged states, the new relative defects after coupling to the $(d+1)$-dimensional theory can allow to satisfy it in the full theory. We also make general remarks about the description of this behavior in the boundary picture, in terms of the CFT charge convexity conjecture. 

In section \ref{distance-swampland} we discuss distance conjectures in relative gravity theories. In section \ref{sec:ads-distance} we focus on the AdS distance conjecture. We use the bottom-up construction to show that the flat space limit of the AdS$_d$ theory corresponds to having the Karch-Randall ETW brane approach the holographic boundary of the host AdS$_{d+1}$ spacetime. We then use a top-down string theory construction of ETW branes in AdS$_5\times\IS^5$ to reproduce this limit and characterize the infinite tower of light modes, which corresponds to KK modes arising due to the the absence of scale separation in the AdS$_4$ ETW configuration. In section \ref{sec:distance-conjecture-cft} we center on the distance conjecture and make some general remarks about its formulation in the boundary description in terms of the CFT distance conjecture, which may be violated in the BCFT$_{d-1}$ even if satisfied in the full CFT/BCFT system.

In section \ref{sec:vacua} we discuss swampland constraints on vacua of the $d$-dimensional gravity theory when regarded as a relative gravity theory. In section \ref{sec:no-nonsusy-ads} we explore the no stable non-supersymmetric AdS conjecture, and focus on the specific question of the appearance of new decay channels in non-supersymmetric AdS$_d$ vacua when embedded as ETW boundaries in AdS$_{d+1}$ host spacetimes, even if the latter are supersymmetric. In particular we study this phenomenon in explicit top-down string theory models of ETW configurations with parametrically controlled supersymmetry breaking of AdS$_5\times \IS^5/\IZ_k$ orbifolds, uncovering a novel tachyon condensation decay channel. In section \ref{sec:desitter} we study the de Sitter conjecture, which as we explain can in principle be violated in the $d$-dimensional gravity theory by using Karch-Randall dS$_d$ ETW branes in AdS$_{d+1}$. As opposed to its AdS or Minkowski ETW counterparts, there is no known top-down string theory embedding of this setup. We argue, based on the addition of flavour branes and the Festina Lente bound, that such hypothetical embeddings face additional challenges, whose solution may guide the search for explicit realizations.

In section \ref{sec:conclusions} we offer some final remarks and open directions. Some more technical computations and constructions are collected in appendices. In appendix \ref{sec:intermediate-action} we describe the computation of the AdS$_d$ brane gravity effective action, including gravity and gauge fields, using holographic renormalization. In appendix \ref{sec:bhs} we collect some results on black hole solutions in the bulk and quantum black holes in intermediate picture, used at some points in the main text. In appendix \ref{sec:string-embedding} we review the details of the 10d supergravity solution describing 5-brane ETW configurations for AdS$_5\times \IS^5$. In section \ref{sec:anomaly-d7s} we provide a fully supersymmetric string theory realization of the ideas in section \ref{sec:anomaly}, about anomalous global symmetries in relative gravity theories, adding D7-brane flavour branes to 5-brane ETW configurations for AdS$_5\times\IS^5/\IZ_k$ orbifolds.

\section{AdS realizations of Localized Quantum Gravity}
\label{sec:ads}

As already mentioned, there are various bottom-up approaches to realize localized gravity (see e.g. \cite{Randall:1999vf,Gregory:2000jc,Dvali:2000hr,Karch:2000ct,Karch:2001cw}). In this section we review the realization of localized gravity in ETW boundaries of AdS$_{d+1}$ spacetime, with emphasis on the case of AdS$_d$ ETW branes. We carry out the discussion both from the perspective of bottom-up Karch-Randall constructions, and from the top-down string theory perspective via 10d solutions for ETW branes for AdS$_5\times\IS^5$. These setups will provide the basic arena for our realization of relative gravity theories, to explore the interplay between swampland constraints and completion via coupling to higher-dimensional gravity, in the rest of this work. 

\subsection{Bottom-up Karch-Randall ETW Branes and Double Holography}
\label{sec:double-holography}

In this section we review the localization of gravity in Karch-Randall ETW branes \cite{Karch:2000ct,Karch:2001cw}, in particular for the AdS ETW boundaries, which allows an optimal use of holography. The embedding in string theory will be discussed in next section. We start the discussion by following \cite{Karch:2000ct} in conventions adapted to the review \cite{Panella:2024sor} for later use. 

\subsubsection{Karch-Randall Braneworlds}
\label{sec:kr}

Consider a bulk $(d+1)$-dimensional Einstein gravity with cosmological constant $\Lambda_{d+1}$, with action
\beq
S_{\text{bulk}}=\frac{1}{16\pi G_{d+1}}\int_{\mathcal{M}} d^{d+1}x\sqrt{-\hat{g}}\left(\hat{R}-2\Lambda_{d+1}\right)\,.
\label{eq:BulkTheory}
\eeq
Here $G_{d+1}$ is the $(d+1)$-dimensional Newton's constant, $\hat{g}_{ab}$ is the $(d+1)$-dimensional metric, and $\Lambda_{d+1}$ is the cosmological constant. In this expression we have ignored a boundary term, see appendix \ref{sec:intermediate-action} for details. We consider an AdS$_{d+1}$ spacetime of curvature length scale $L_{d+1}$, with $\Lambda_{d+1}=-d(d-1)/2L_{d+1}^{2}$. We write its metric as a foliation into maximally symmetric $d$-dimensional slices  (in the mostly plus signature)
\beqa
ds_{d+1}^2=e^{2A(r)}\, g_{ij} dx^i dx^j + dr^2\, ,
\label{ads-foliation}
\eeqa 
where the $d$-dimensional metric $g_{ij}$ can be AdS$_d$, $d$-dimensional Minkowski, or dS$_d$. 
This is a solution for the following warp factors\footnote{The numerical factors differ from \cite{Karch:2000ct} due to a different convention in the relation between $\Lambda_{d+1}$ and $L_{d+1}$.}
\beqa
\label{warps}
&{\rm AdS}_d: & \quad e^{2A}=-\frac{2\Lambda_d L_{d+1}^2}{(d-1)(d-2)} \cosh^2\frac{c-r}{L_{d+1}}\nonumber\\
&M_d: & \quad e^{2A}=\exp\Big(2\,\frac{c-r}{L_{d+1}}\Big)\\
& {\rm dS}_d:& \quad e^{2A} =\frac{2\Lambda_d L_{d+1}^2}{(d-1)(d-2)} \sinh^2\frac{c-r}{L_{d+1}}\, ,\nonumber
\eeqa
where $c$ is an integration constant to be fixed soon. Most of our discussion will be carried out for AdS Karch-Randall branes, but we will also mention the Randall-Sundrum setup (i.e. Minkowski slices) \cite{Randall:1999vf}; the dS Karch-Randall setup is also discussed in section \ref{sec:desitter}.

\begin{figure}[htb]
\begin{center}
\includegraphics[scale=.35]{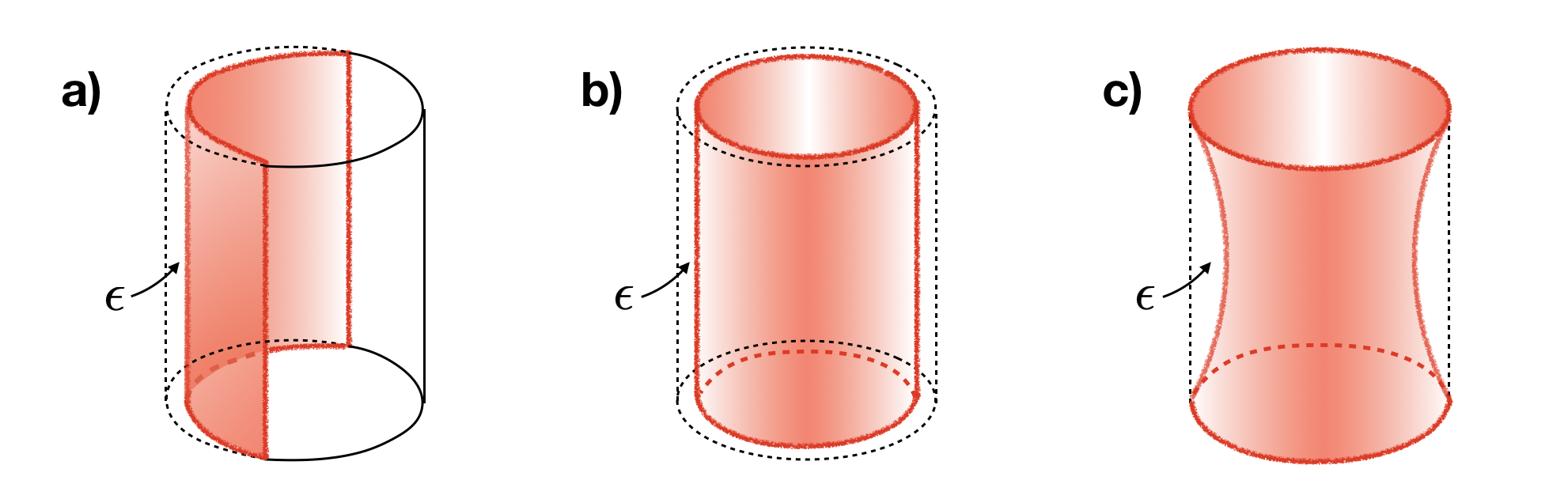}
\caption{\small The bulk AdS$_{d+1}$ spacetime (black lines) and three different kinds of ETW branes (red surfaces) in the AdS Karch-Randall (a), Randall-Sundrum (b), and dS Karch-Randall (c) scenarios. In all cases $\epsilon$ indicates the region (dashed black lines) excised from spacetime.}
\label{fig:ETWs}
\end{center}
\end{figure}

We now locate a brane $\mathcal{B}$ at $r=0$, and take it to be purely tensional, with action
\beq 
S_{\tau}=-\tau \int_{\mathcal{B}}d^{d}x\sqrt{-h}\,.
\label{eq:braneaction}
\eeq
We consider the region `behind the brane' to be excised from spacetime, so that the brane becomes an ETW boundary\footnote{In the literature, a second AdS region is often glued to the first along the brane, which corresponds to replacing $r\to|r|$ in (\ref{warps}) and changing some factors or 2 in some of the following expressions.}, see figure \ref{fig:ETWs}. The nature and location of the ETW brane are determined by the junction conditions, hence by the brane tension. This implies a relation between the tension and the constant $c$ in (\ref{warps}), given by
\beqa
&{\rm AdS}_d: & \quad \tau=\frac {d-1}{8\pi G_{d+1} L_{d+1}}\tanh\frac c{L_{d+1}}\nonumber\\
&M_d: & \quad \tau=\frac {d-1}{8\pi G_{d+1} L_{d+1}}\\
& {\rm dS}_d:& \quad \tau=\frac {d-1}{8\pi G_{d+1} L_{d+1}}\coth\frac c{L_{d+1}}\, ,\nonumber
\eeqa
We note that the Randall-Sundrum scenario is obtained for a critical value of the tension, while the AdS or dS Karch-Randall cases correspond to tension lower or bigger than this critical value, respectively.

As shown in \cite{Karch:2000ct}, in the case of dS and Minkowski slices, the above setup leads to an exactly massless localized graviton mode. In an analogue quantum mechanics formulation for the graviton propagation in the extra dimension, it corresponds to a wavefunction bound to a delta function `crater' in a `volcano' potential decreasing monotonically with the distance to the brane. In the case of AdS slices, the localized graviton arises in the particular regime in which the ETW brane is close to the holographic boundary of the host AdS. It corresponds to a wavefunction bound to a delta function crater in a volcano potential whose slopes have a turning point at $|r|=c$, beyond which they increase indefinitely. This behaviour makes the graviton massive, although the mass can be made parametrically small as the ETW brane gets closer to the holographic boundary. In fact, this scaling was characterized in the $d=4$ case in \cite{Miemiec:2000eq}, as follows. Rewrite (\ref{ads-foliation}) in so-called conformal coordinates 
\beqa
ds_5^2=e^{2A(z)}[g_{ij}dx^i dx^j+dz^2]\quad , \quad e^{2A}=\frac{L_5^2|\Lambda_4|}{\sin^2[\sqrt{|\Lambda_4|}(z+z_0)]}\, ,
\label{conformal}
\eeqa
with $z_0$ determined by $c$. Then the mass follows the relation 
\beqa
m_{g}^2\simeq\frac 32 |\Lambda_4|^2 z_0^2\, .
\eeqa
So the graviton can be made parametrically light for small $z_0$, i.e. when the brane is close to the holographic boundary. In the bottom-up setup this can tuned in a continuous way, by dialing the brane tension, whereas in the top-down models discussed in the next section the brane location depends on discrete microscopic parameters of the system, but it is still tunable.

The main features of these setups hold for general $d$. The brane theory is effectively a $d$-dimensional gravity theory (for the AdS KR case, if one is well above the graviton mass scale) up to the scale $1/L_{d+1}$, at which higher-dimensional gravity kicks in. Hence this setup provides a useful arena for the realization of relative gravity theories and the exploration of their interplay with swampland constraints. In the following we will focus in this case of AdS Karch-Randall brane in this regime, which allows for interesting holographic dual pictures, as we study next.

\subsubsection{Double Holography}
\label{sec:actual-double}

Our discussion follows the excellent review \cite{Panella:2024sor}, to which we refer the reader for details, see also appendix \ref{sec:intermediate-action} for some related computations.

The description of localized gravity in the previous section is referred to as the {\em bulk description}. One main advantage of the implementation of localized gravity in terms of AdS geometries is that it allows for interesting complementary realizations via holography. Specifically, the setup of AdS$_d$ ETW branes in AdS$_{d+1}$ allows for two different holographic dual descriptions, the so-called double holography \cite{Karch:2000gx,Karch:2000ct,Takayanagi:2011zk}. The first is called the {\em boundary picture}, and replaces the bulk spacetime AdS$_{d+1}$ by its holographically dual CFT$_d$ living on a half space, with a $(d-1)$-dimensional boundary at which it couples to a boundary CFT$_{d-1}$ (BCFT$_{d-1}$), which provides the holographic dual of the ETW AdS$_d$ brane at the boundary of AdS$_{d+1}$. The localized graviton mode on the ETW AdS$_d$ brane is dual to the stress-energy tensor of the BCFT$_{d-1}$, and its non-conservation due to its coupling to the CFT$_d$ is dual to the fact that the graviton is massive. The regime in which the graviton is parametrically light corresponds to the situation in which the central charge of the CFT$_d$ is small compared with that of the BCFT$_{d-1}$, so that the leakage of stress-energy is small due to the scarcity of degrees of freedom of the CFT$_d$.

\begin{figure}[htb]
\begin{center}
\includegraphics[scale=.35]{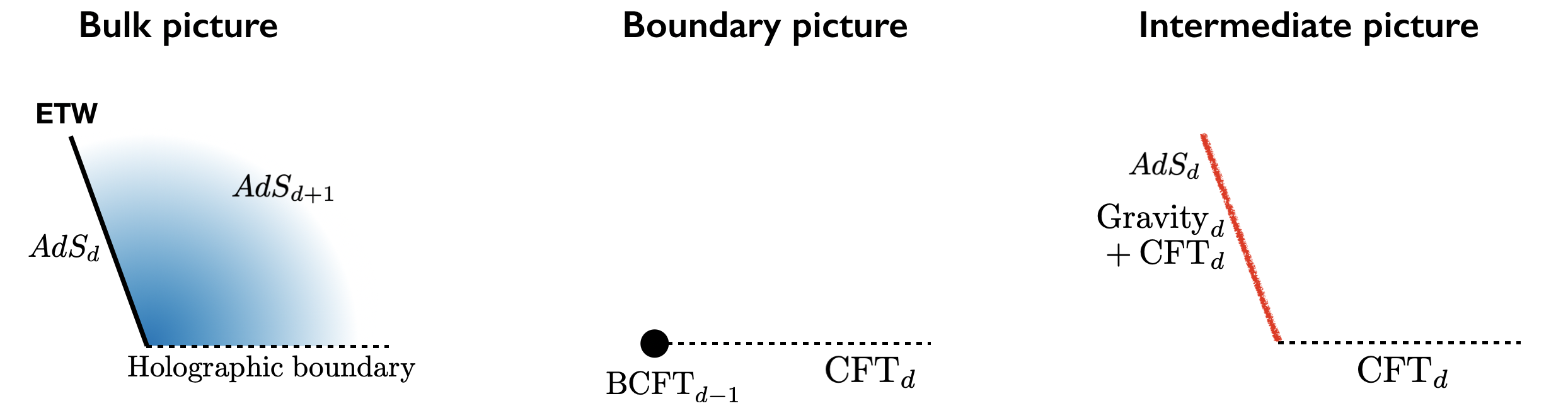}
\caption{\small The three pictures in double holography: (a) The bulk picture, (b) the boundary picture, and (c) the intermediate picture.}
\label{fig:double-holography}
\end{center}
\end{figure}

In double holography there is yet another holographic dual picture, called the {\em intermediate picture}, which is obtained from the boundary picture by replacing the BCFT$_{d-1}$ by its gravitational AdS$_d$ dual. The structure of the gravitational theory living on this AdS$_d$ follows from the observation \cite{Gubser:1999vj} that AdS$_{d+1}$ with an IR cutoff is equivalent to the holographic dual CFT$_d$ coupled to $d$-dimensional dynamical gravity with a UV cutoff related to the dual IR one. This generalizes to the setup of  AdS$_{d+1}$ with a boundary provided by an ETW AdS$_d$ brane as follows. The theory that lives on the  AdS$_d$ ETW brane is the holographic CFT$_d$ coupled to dynamical $d$-dimensional gravity on AdS$_d$ and with a UV cutoff (related to the brane location). On the other hand, we also have the holographic dual CFT$_d$ (without dynamical gravity and with no UV cutoff) on the half-space at the holographic boundary of the AdS$_{d+1}$. Both theories are coupled with transparent boundary conditions at their common boundary, $(d-1)$-dimensional Minkowski space. We illustrate the different pictures in Figure \ref{fig:double-holography}.

This picture, in the $d=2$ case, is the main arena for the quantum island program (see \cite{Almheiri:2020cfm} for a review), to recover the Page curve for black hole evaporation (see e.g. \cite{Almheiri:2019psy,Uhlemann:2021nhu,Demulder:2022aij} for discussions in higher dimensions). There the ETW brane plays the role of a gravitational solution (an evaporating black holes), while the holographic boundary describes a non-gravitating thermal bath to which the Hawking radiation escapes. 

Coming back to our setup, let us focus on the theory on the worldvolume of the AdS$_d$ ETW brane. The presence of a strongly coupled CFT$_d$ with a large number of degrees of freedom implies strong quantum effects in the corresponding effective action, due to fast modes above the UV cutoff, which are to be integrated out. These would be hard to compute directly, but can be easily obtained by exploiting the holographic duality with the bulk picture. Indeed, following the technique of holographic renormalization (see e.g. \cite{Skenderis:2002wp,Papadimitriou:2004ap} for reviews, and appendix \ref{sec:intermediate-action} for computations in our case), the effective action for gravity on the AdS$_d$ KR ETW brane, including the quantum backreaction of the CFT$_d$ fast modes, can be obtained by evaluating the $(d+1)$-dimensional gravitational action in the wedge region between the holographic boundary and the KR brane (i.e. removed from AdS$_{d+1}$) with a general $d$-dimensional metric as boundary condition. The result is given by (see (\ref{master-eq}) in the case with no gauge fields)
\beqa
S_{d}=\frac{1}{16\pi G_{d}}\int_{\mathcal{B}} \hspace{-1mm}d^{d}x\sqrt{-h}\biggr[R-2\Lambda_{d}+\frac{L_{d+1}^{2}}{(d-4)(d-2)}\left(R_{ij}^{2}-\frac{d\,R^{2}}{4(d-1)}\right)+\cdots\biggr]\, .\quad 
\label{brane-action-general}
\eeqa
The ellipsis corresponds to higher curvature terms, entering with higher powers of $L^{2}_{d+1}$, which in principle can be computed systematically. Here $G_{d}$ represents the effective brane Newton's constant inherited from the bulk 
\beq 
G_{d}=\frac{d-2}{L_{d+1}}G_{d+1} \,,
\label{eq:effGd}
\eeq
and the cosmological constant $\Lambda_{d}$ and curvature length scale $L_{d}$ are given by
\beq
\Lambda_{d}=-\frac{(d-1)(d-2)}{2L_{d}^{2}}\quad ,\quad
\frac{1}{L_{d}^2}=\frac{2}{L_{d+1}^2}\left(1-\frac{8\pi G_{d+1}L_{d+1}}{d-1}\tau\right)\,.
\label{eq:curvscale}
\eeq
Here \eqref{eq:effGd} and \eqref{eq:curvscale} differ in factors of 2 from \cite{Panella:2024sor}, because there one integrates out the bulk on both sides of the brane, whereas in our ETW case we have only one.
In addition, the brane supports the CFT$_d$ with UV cutoff, as already mentioned.
The double holographic setup provides a very efficient computational tool for the quantum backreaction of fast modes of a strongly coupled CFT$_d$ on gravitational solutions, in particular black holes. Such so-called quantum black holes (see \cite{Panella:2024sor} for a review) are solutions of the above $d$-dimensional gravity theories in the intermediate picture. The $d$-dimensional solutions are directly obtained from solutions of the $(d+1)$-dimensional bulk theory obeying the corresponding junction conditions at the location of the KR brane. The most extensive exploration of quantum black holes has been carried out for $d=3$, and include neutral, rotating and charged quantum black holes and their thermodynamical properties. In Appendix \ref{sec:bhs} we review some of these solutions and the properties most relevant for our purposes. Some partial results also exist for higher-dimensional setups. The $d=2$ case has been most extensively studied from the perspective of the quantum island program (see \cite{Almheiri:2020cfm} for a review).

\subsection{Top-down String Theory realization of ETW Branes in AdS$_5\times\IS^5$}
\label{sec:d3-ns5-d5}

In this section we study the string theory realization of the ideas in the previous section. The key idea, using the double holography picture reviewed above, is to consider the bulk dual picture of a CFT/BCFT boundary theory. Namely, we consider a CFT$_d$ with known gravity dual, of the form AdS$_{d+1}$ (times some internal space), and to study it on half-space with boundary conditions given by a BCFT$_{d-1}$, with known gravity dual, of the form AdS$_d$ (times some internal space). The AdS$_d$ theory thus corresponds to an ETW brane of the AdS$_{d+1}$ bulk.

This is in general difficult to carry out. However, very remarkably, for 4d $\NN=4$ $SU(N)$ SYM there is class of half-supersymmetric boundary conditions, defined in terms of D3-branes ending on NS5- and D5-brane configurations \cite{Gaiotto:2008sa,Gaiotto:2008ak}, for which the gravity dual has been explicitly described \cite{DHoker:2007zhm,DHoker:2007hhe,Aharony:2011yc,Assel:2011xz,Bachas:2017rch,Bachas:2018zmb} (see also \cite{Raamsdonk:2020tin,VanRaamsdonk:2021duo,Demulder:2022aij,Karch:2022rvr,DeLuca:2023kjj,Huertas:2023syg,Chaney:2024bgx} for recent applications). In this section we describe the brane configurations and describe the 10d supergravity dual background providing AdS$_4$ ETW branes for AdS$_5(\times\IS^5)$. They constitute our prototype of top-down string theory embeddings of the bottom-up Karch-Randall scenario.

\subsubsection{D3-branes ending on NS5- and D5-branes}
\label{sec:d3-ns5-d5-hw}

A useful way to describe boundary conditions for 4d $\NN=4$ $SU(N)$ SYM is to consider stacks of D3-branes ending on 5-branes, which define a 3d $\NN=4$ BCFT$_3$. As pioneered in \cite{Gaiotto:2008sa,Gaiotto:2008ak} these systems are efficiently studied by realizing them as a Hanany-Witten brane construction\cite{Hanany:1996ie} of $N$ semi-infinite D3-branes ending on a configuration of NS5- and D5-branes, as we quickly review. 

Consider a stack of $N$ D3-branes along the directions 0123 realizing 4d $\NN=4$ $SU(N)$ SYM, and consider them to have semi-infinite extent in $x^3\geq 0$. We can define this kind of boundary for the theory by letting the D3-branes end on a configuration of NS5- and D5-branes localized in the direction 3, and spanning 012, and some additional directions. In order to preserve the maximum supersymmetry, the NS5-branes span the directions 012$\,$456, while the D5-branes span the directions 012$\,$789, and there can be additional D3-branes suspended among them. The invariant information of the configuration is the total number of asymptotic D3-branes $N$, and the linking numbers of the 5-branes. We define the linking number\footnote{We purposefully use two different notions of linking number for the two kinds of branes, so as to simplify expressions such as (\ref{sum-linkings}), see \cite{Aharony:2011yc} for discussion.} for a NS5-brane (resp. D5-brane) as the net number of D3-branes ending on it from the right plus the number of D5-branes to its left (resp. minus the number of NS5-branes to its right). 

As uncovered in \cite{Gaiotto:2008sa,Gaiotto:2008ak}, when the $x^3$ positions of the NS5- and D5-branes follow a specific ordering, determined by their linking numbers, the limit of sending their distances to zero makes the theory flow to a 3d $\NN=4$ SCFT which provides maximally supersymmetric boundary conditions for the 4d $\NN=4$ $SU(N)$ theory. The ordering rules, in short, require that any D5-brane on which a net number of D3-branes ends is to the right of all NS5-branes, and that 5-branes of each kind are ordered with non-decreasing linking numbers from left to right. These constraints ensure the configuration to admit a description in terms of a gauge theory of the kind in \cite{Hanany:1996ie}. We advice the reader to check the references \cite{Gaiotto:2008sa,Gaiotto:2008ak} for further details.

The 5-branes with the same linking numbers form stacks leading to enhanced non-abelian flavour symmetries.
Labeling NS5- and D5-brane stacks with indices $a$, $b$, respectively, we denote by $n_a$ the multiplicity of NS5-branes with linking number $K_a$ and by $m_b$ that of D5-branes with linking number ${\tilde L}_b$. 
A simple counting from the definition of linking numbers shows that they satisfy
\beqa
N=\sum_a n_aK_a + \sum_b m_b{\tilde L}_b\, ,
\label{sum-linkings}
\eeqa
We refrain from further discussion of the gauge theory perspective, and instead turn to the description of these ingredients in the gravity dual.

\subsubsection{Gravity Duals: End of the World branes for AdS$_5\times\IS^5$}
\label{sec:d3-ns5-d5-dual}

In this section, we briefly review the gravitational duals of the brane configurations described in the last section. They are given by a particular class of explicit 10d supergravity backgrounds studied in \cite{DHoker:2007zhm,DHoker:2007hhe,Aharony:2011yc,Assel:2011xz,Bachas:2017rch,Bachas:2018zmb} (see also \cite{Raamsdonk:2020tin,VanRaamsdonk:2021duo,Demulder:2022aij,Karch:2022rvr,DeLuca:2023kjj,Huertas:2023syg,Chaney:2024bgx} for recent applications).  We here simply describe the key ideas about the resulting configuration, and refer the reader to appendix \ref{sec:string-embedding} and the references for details. 

The most general supergravity solutions with 16 supersymmetries and $SO(2,3)\times SO(3)\times SO(3)$ symmetry were explicitly
constructed in \cite{DHoker:2007zhm,DHoker:2007hhe}. They are gravitational duals of Hanany-Witten brane configurations of NS5- and
D5-branes with stacks of D3-branes as in the previous section \cite{Aharony:2011yc}. The geometries in general have a ``bagpipe''
structure \cite{Bachas:2018zmb}, with an AdS$_4\times \IX_6$ ``bag'', with $\IX_6$ a 6d manifold which is compact  save for a number of AdS$_5\times \IS^5$ ``pipes'' sticking out of it. We will eventually focus on the specific case with only one AdS$_5\times \IS^5$ ending on the bag, which provides the gravity dual of 4d $\NN=4$ $SU(N)$ SYM on a 4d spacetime with boundary.

\begin{figure}[htb]
\begin{center}
\includegraphics[scale=.1]{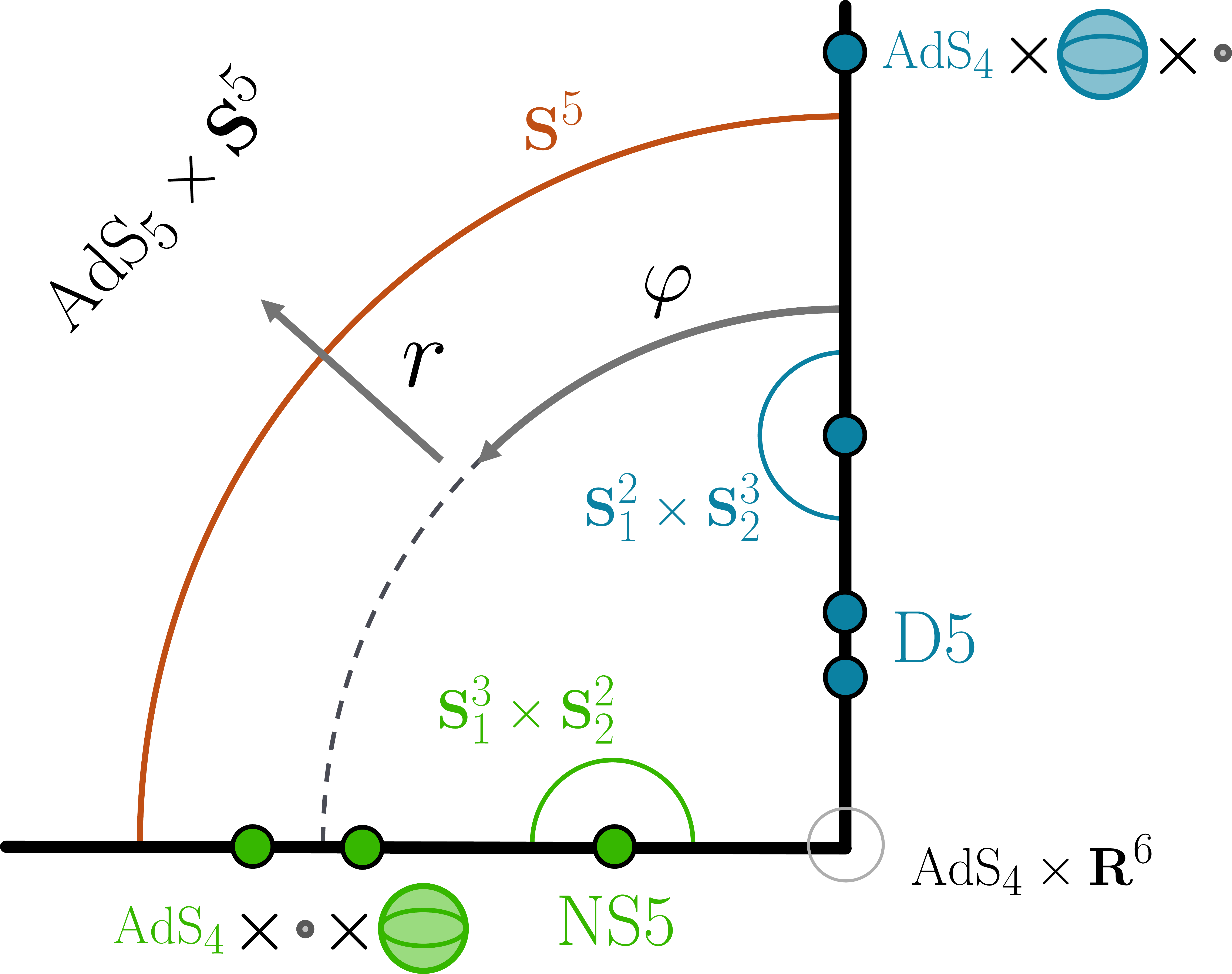}
\caption{\small Riemann surface over which the AdS$_4\times\IS^2_1\times \IS^2_2$ is fibered. The horizontal (resp. vertical) axis is the locus at which the $\IS_1^2$ (resp. $\IS_2^2$) shrinks. The green (resp. blue) dots in the horizontal (resp. vertical) axis correspond to the location of NS5-branes (resp. D5-branes). We have indicated the segments describing $\IS^5$ (purple arc) and $\IS^2\times \IS^3$'s (blue and dark green arcs) in the full fibration.}
\label{fig:quadrant}
\end{center}
\end{figure}

The general supergravity solutions have the structure of a fibration of AdS$_4\times \IS_1^2\times \IS_2^2$ over an oriented Riemann surface $\Sigma$. For the solutions describing one asymptotic AdS$_5\times\IS^5$ region ending on an ETW configuration, the  Riemann surface is the quadrant in the complex plane $w=r e^{i\varphi}$, with $r\in (0,\infty)$ and $\varphi\in \left[\frac{\pi}{2},\pi\right]$. The $\IS_1^2\times \IS_2^2$ is fibered such that $\IS^2_1$ shrinks to zero size over $\varphi=\pi$ (negative real axis) and $\IS^2_2$ shrinks to zero size over $\varphi=\pi/2$ (positive imaginary axis), closing off the geometry over those edges. In particular, the $\IS_1^2\times \IS_2^2$ fibration over an arc parametrized by $\varphi$ at large constant $r$ (away from the 5-brane punctures discussed below) has the structure 
\beqa
(x^4)^2+(x^5)^2+(x^6)^2=\cos^2\varphi\quad , \quad (x^7)^2+(x^8)^2+(x^9)^2=\sin^2\varphi\, ,
\label{s5-split}
\eeqa
(modulo an overall $r$-dependent radius factor). Combining both equations we recover the equation of the round $\IS^5$ in the asymptotic AdS$_5\times\IS^5$. The fibration of $\IS^2\times\IS^2$ over a $\varphi$-arc at any other radius (away from the 5-brane sources discussed below) leads to a similar topological $\IS^5$, see Figure \ref{fig:quadrant}, although in general a squashed one. 

The asymptotic $r\to\infty$ region corresponds to the asymptotic regions AdS$_5\times\IS^5$, which describes the 4d $\CN=4$ $SU(N)$ SYM  on the semi-infinite D3-branes. The solution includes 5-brane sources, corresponding to punctures on $\Sigma$, describing the NS5- and D5-branes. The NS5-branes are along the $\varphi=\pi$ axis, with stacks of multiplicity $n_a$ at positions $w=-k_a$, and the D5-branes are along the $\varphi=\pi/2$ axis, with stacks of multiplicity $m_b$ at positions $w=il_b$. 

The supergravity solution near one of the 5-brane punctures is locally of the form of a NS5-brane spanning AdS$_4\times \IS^2_2$ (respectively a D5-brane spanning AdS$_4\times\IS^2_1$), with $n_a$ units of NSNS 3-form flux (respectively $m_b$ units of RR 3-form flux) on the $\IS^3$ surrounding the 5-brane source. The latter is easily visualised, by simply taking a segment in $\Sigma$ forming a half-circle around the 5-brane puncture, and fibering over it the $\IS^2$ fibre shrinking at the endpoints, so that we get a topological $\IS^3$, see Figure \ref{fig:quadrant}. Hence, near each 5-brane puncture, the metric factorises as AdS$_4\times \IS^2\times \IS^3$ fibered along a local radial coordinate $\tilde{r}$ parametrising the distance to the puncture. The $U(n_a)$, $U(m_b)$ gauge symmetry on the 5-branes is the holographic dual of the enhanced non-abelian flavour symmetries (or their duals) for the BCFT$_3$ mentioned in the previous section.

As explained, in the asymptotic region $r\to\infty$ far away from the 5-branes, the two $\IS^2$'s combine with the coordinate $\varphi$ to form a $\IS^5$, c.f. (\ref{s5-split}), so in this limit we have AdS$_5\times\IS^5$, with $N$ units of RR 5-form flux. The parameter $N$ will shortly be related to other quantities in the solution. 

The asymptotic 5-form flux changes discontinuously across the 5-brane locations, decreasing as one approaches $r=0$, as follows.
The NS5- and D5-branes carry $K_a$ and ${\tilde L}_b$ units of induced D3-brane charge, respectively, which are the gravity description of the linking numbers in the flat space D3-brane construction. In the supergravity background they can be described to arise as bulk fluxes on the local AdS$_4\times\IS^2\times\IS^3$ (times a radial coordinate) near the 5-branes, c.f. (\ref{the-fluxes}). They also admit a description by replacing these local geometries by explicit 5-branes, so the induced D3-brane charge arises from 5-brane worldvolume fluxes, concretely 
\beqa
\int_{\IS^2} F_2^a=K_a\quad \int_{\IS^2}F_2^b={\tilde L}_b\, ,
\label{worldvolume-fluxes}
\eeqa
where the super-index denotes the 5-brane whose worldvolume gauge field we are considering. Morally, these fluxes are related to the number of D3-branes ending on each 5-brane. The induced D3-brane charge on e.g. the D5-branes in the $b^{th}$ stack follows from the worldvolume couplings, 
 \beqa
m_b \int_{6d} C_4 F_2^b \quad \Rightarrow \quad m_b {\tilde L}_b \int_{4d} C_4\, ,
 \eeqa
and analogously for the NS5-branes. Because of these induced D3-brane charges, the RR 5-form flux over $\IS^5$ jumps across the 5-branes (i.e. as the arc in $\varphi$ is moved across some 5-brane puncture in $\Sigma$). Specifically, upon crossing an NS5-brane (resp. D5-brane) the 5-form flux decreases in $n_aK_a$ units (resp. $m_b{\tilde L}_b$), with no sum over indices. The 5-form flux decreases in such crossings until we reach the region $r<|k_a|,|l_b|$, in which there is no leftover flux, so that the $\IS^5$ shrinks and spacetime ends in an smooth way at $r=0$, see Figure \ref{fig:quadrant}. Hence, the asymptotic 5-form flux $N$ and the 5-brane parameters are related as in (\ref{sum-linkings})
\begin{equation}
N=\sum_a n_a K_a+\sum_b m_b{\tilde L}_b\, ,
\label{the-sum}
\end{equation}

Finally, the positions $k_a$, $l_b$ are determined by the linking numbers of the 5-branes $K_a$, ${\tilde L}_b$ in the corresponding stacks. The precise relation is given by (\ref{positions-linking}), but we will not need it for our general discussion.

\begin{figure}[htb]
\begin{center}
\includegraphics[scale=.08]{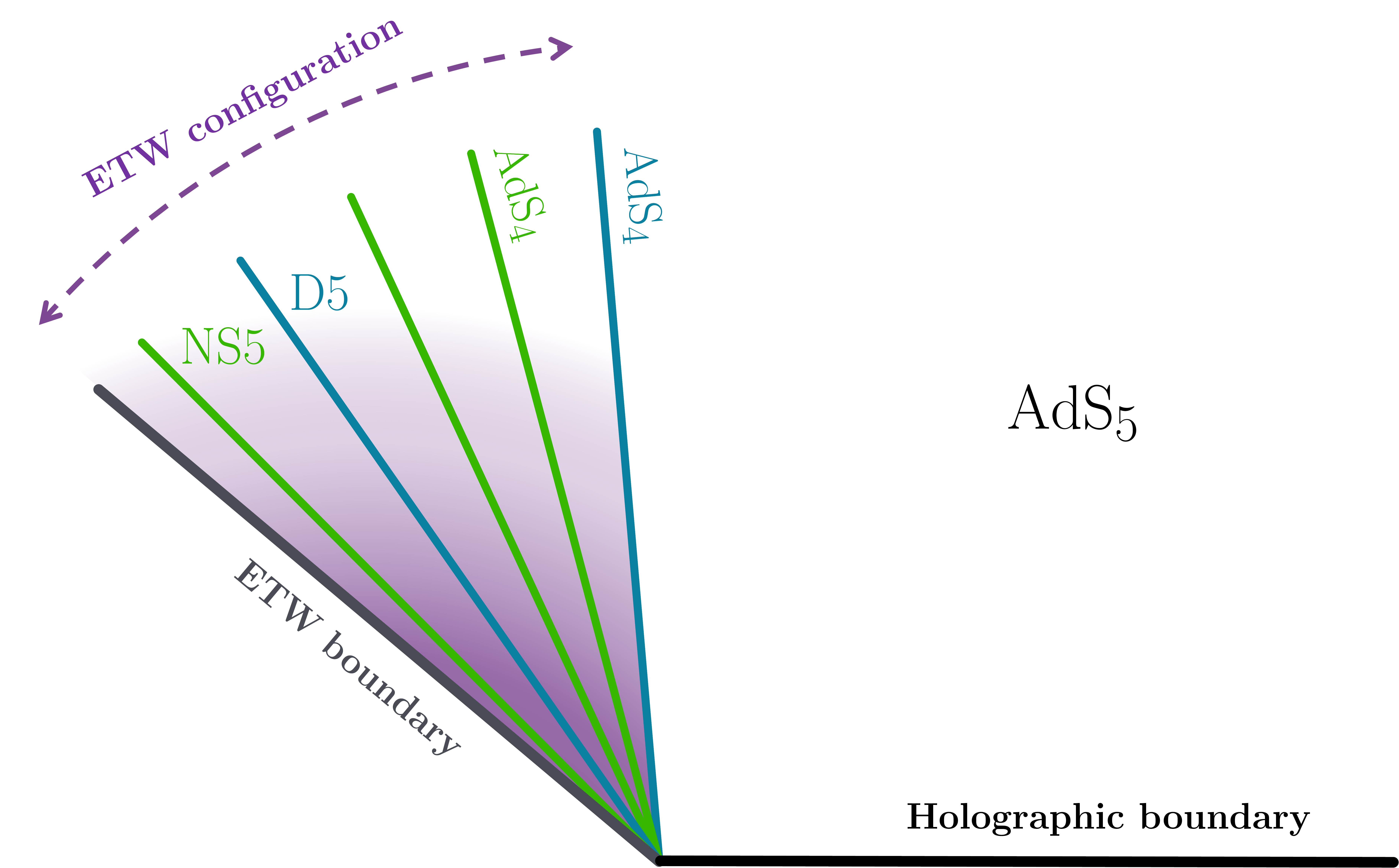}
\caption{\small The solution in Poincar\'e coordinates shows its asymptotic AdS$_5 (\times \IS^5)$ region and the ETW configuration ending spacetime.}
\label{fig:poincare}
\end{center}
\end{figure}

As emphasized in \cite{Huertas:2023syg}, the above supergravity background describes a cobordism to nothing of the asymptotic AdS$_5\times\IS^5$ via an ETW configuration \cite{VanRaamsdonk:2021duo}. This is shown in Figure \ref{fig:poincare} in Poincar\'e coordinates. The holographic boundary is the horizontal line representing 4d half-space. The geometry is a foliation into AdS$_4$ slices parametrized by $r=|w|$ on the Riemann surface of the solution. The asymptotic region $r\to\infty$ corresponds to the radial lines closer to the 4d holographic boundary, and is an AdS$_5$ ($\times\IS^5$) with $N$ units of 5-form flux. As one moves inward in the Riemann surface to values of $r$ corresponding to the 5-brane puncture locations, the 5d geometry hits 5-brane sources spanning AdS$_4$ slices, lines sticking out radially from the 3d boundary of the 4d holographic boundary in a fan-like structure \cite{GarciaEtxebarria:2024jfv}. The 5-form flux over the $\IS^5$ decreases upon crossing the 5-branes until it is completely peeled off and spacetime ends. Hence, the solution describes an AdS$_4$ ETW configuration. 

We conclude with a word of caution about the above 5d picture. As is familiar, these solutions have no scale separation \cite{Lust:2019zwm} (see  \cite{Coudarchet:2023mfs} for a review and further references), namely the scale of the internal geometries are comparable to those of the corresponding AdS geometries, both for the asymptotic AdS$_5\times\IS^5$ and the AdS$_4\times\IX_6$ ``bag'' dual of the BCFT$_3$. This implies that there is no regime of validity of the EFT admitting an actual description in terms of 5d gravity and genuinely localized 4d gravity on the ETW boundary. Hence one might just conclude that the construction in this section is intrinsically 10d, and hence it is not a good UV completion of the 5d/4d bottom-up model in section \ref{sec:no-global-symmetries-bu}.

As explained in the introduction, we actually regard these string configurations as toy versions of possible genuinely scale separated setups. Let us note that there are several solid proposals for AdS vacua with scale separation \cite{DeWolfe:2005uu,Camara:2005dc} (see also \cite{Buratti:2020kda,Farakos:2020phe,Cribiori:2021djm,Apers:2022zjx,Shiu:2022oti,Carrasco:2023hta} for discussion of scale separation in these and other models, and \cite{Coudarchet:2023mfs} for a review). According to the swampland cobordism conjecture \cite{McNamara:2019rup}, these AdS$_{d+1}$ vacua should admit bordisms to nothing,  which may well be ETW configurations of  AdS$_d$ KR kind, possibly sharing the scale separation of their bulk hosts. We thus regard our construction in this section as toy models of such possibly scale separated setups. 

For these purposes, or even more generally, as usually done in holography, we may still use the lower dimensional perspective even in the absence of scale separation by removing the internal space via dimensional reduction, see \cite{Huertas:2023syg}. Also, some of our arguments are topological in nature, and are hence insensitive to the size of the compact spaces, and hence, of scale separation issues. These remarks will be implicit every time we discuss ETW configurations in AdS$_5\times\IS^5$ (or quotients thereof), in particular sections \ref{sec:no-global-symmetries-td}, \ref{sec:generalized}, \ref{sec:anomaly-td}, \ref{sec:anomaly-jumps-td}, \ref{sec:gauge-swampland}, \ref{sec:ads-distance-td}, \ref{sec:no-nonsusy-ads}, and appendices \ref{sec:string-embedding}, \ref{sec:anomaly-d7s}, and will not be explicitly repeated there.

\subsubsection{An illustrative example}
\label{sec:KSU}

In order to illustrate the above ideas, in this section we consider a simple class of examples. As explicitly discussed in \cite{Huertas:2023syg}, the minimal setup to achieve localization of gravity requires at least one stack of NS5-branes and one of D5-branes. We consider a simple configuration of this kind, appeared in \cite{Karch:2022rvr}. 

\medskip

{\bf The brane configuration}

Let us start with the description of the brane configuration in flat space. We consider a configuration of $N$ D3-branes ending on a system of NS5- and D5-branes. For simplicity we consider the same number $N_5$ of 5-branes of each kind, and we also take all 5-branes of each kind to have the same linking number, so that they end up in single stack (of each kind) in the gravity dual. 

\begin{figure}[htb]
\begin{center}
\includegraphics[scale=.38]{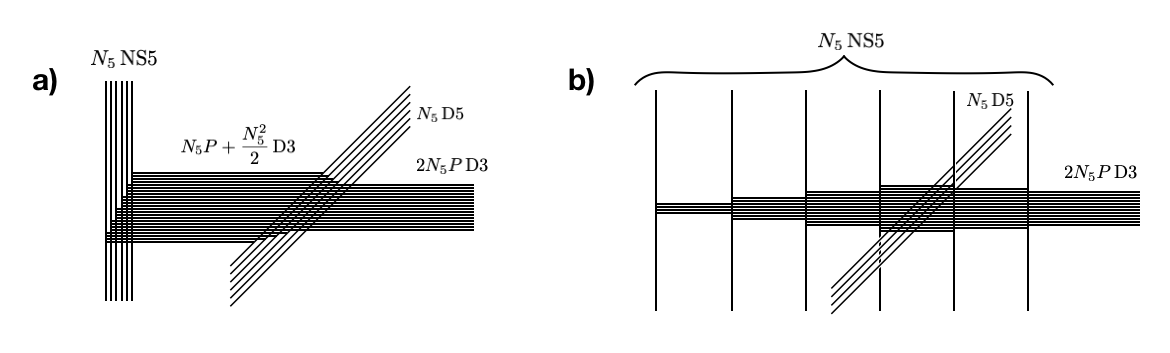}
\caption{\small Example of a system of D3-branes ending on NS5- and D5-branes (for $N_5=6$, $P=1$), in two different 5-brane orderings. a) With all NS5- to the left of all D5-branes. b) With the ordering in \cite{Gaiotto:2008sa,Gaiotto:2008ak}, so as to define a boundary SCFT in the infrared limit of coincident 5-branes.}
\label{fig:example-gw}
\end{center}
\end{figure}

The brane configuration is depicted  in Figure \ref{fig:example-gw}a, represented with all NS5-branes to the left of all D5-branes. Other representations can be obtained by moving the 5-branes, taking into account the brane creation effects \cite{Hanany:1996ie}. Notice that we have expressed the number of D3-branes $N=2N_5P$ in terms of an integer $P$, which is related to the 5-brane linking numbers by
\beqa
K_{\rm NS5}=P+N_5/2\quad,\quad {\tilde L}_{\rm D5}=P-N_5/2\, ,
\label{linking-example}
\eeqa
(we take take $N_5$ even). Note that these satisfy the relation (\ref{sum-linkings}).

As discussed in \cite{Gaiotto:2008sa,Gaiotto:2008ak} and mentioned in section \ref{sec:d3-ns5-d5-hw}, for the configuration to define a SCFT in the infrared limit of coincident 5-branes, we need to choose a specific ordering. For our example, there are two possible structures of the resulting 3d $\NN=4$ theory, depending on whether $N_5\lessgtr 2P$ \cite{Karch:2022rvr}. We are interested in the case $N_5>2P$ (since it allows to access the gravity localization limit), for which the gauge theory has the structure (see figure \ref{fig:example-gw}b, for $N_5=6$, $P=1$). 
\begin{align}\label{quiver3d}
    U(R)-U(2R)-\ldots - &U(R^2) - U(R^2-S)-\ldots - U(2N_5P+S) - \widehat{U(2N_5P)}\, .
	\nonumber\\
	&\ \ \ \vert\\
	\nonumber & \!\! [U(N_5)]	
\end{align}
Here we have introduced $R=N_5/2+P$, $S=N_5/2-P$ (which correspond to the linking numbers (\ref{linking-example}) $K_{\rm NS5}$, and (minus) ${\tilde L}_{\rm D5}$, respectively). The long horizontal sequence of groups describes the gauge symmetry from D3-branes suspended between NS5-branes, the hatted $U(2N_5P)$ factor corresponds to the group on the semi-infinite D3-branes, and the $U(N_5)$ in square brackets in the lower line corresponds to the global symmetry for the flavours introduced by the D5-branes for the $U(R^2)$ gauge factor. The $U(N_5)$ associated to the NS5-branes arises only in the infrared SCFT limit.

The gauge factors increase their rank (in $R$ units per step) as we move to the right, until a maximum is reached at the $U(R^2)$, where one encounters the D5-branes. After that, the ranks start decreasing (in $S$ units per step) until we reach the $2N_5P$ corresponding to the asymptotic D3-branes. The ratio of the number of degrees of freedom in the 3d gauge theory to that of the 4d gauge theory is $N_5/P$. As already explained, we will be interested in the regime of large $N_5/P$ to obtain localization of a light graviton in the gravitational dual, to which we turn next. 

\medskip

{\bf The gravity dual}

Let us now quickly describe the gravity dual, also described in \cite{Karch:2022rvr}. It is given by a 5-brane ETW configuration for an AdS$_5\times\IS^5$ with $N=2N_5P$ units of RR 5-form flux. The geometry is given by a fibration of AdS$_4\times\IS^2\times\IS^2$ over a quadrant of the kind in figure \ref{fig:quadrant} with two 5-brane punctures, each of them with multiplicity $N_5$. As explained above, the fact that all N5- and all D5-branes each cluster in a single stack is because they have the same linking number. 

\begin{figure}
    \centering
    \includegraphics[width=0.9\linewidth]{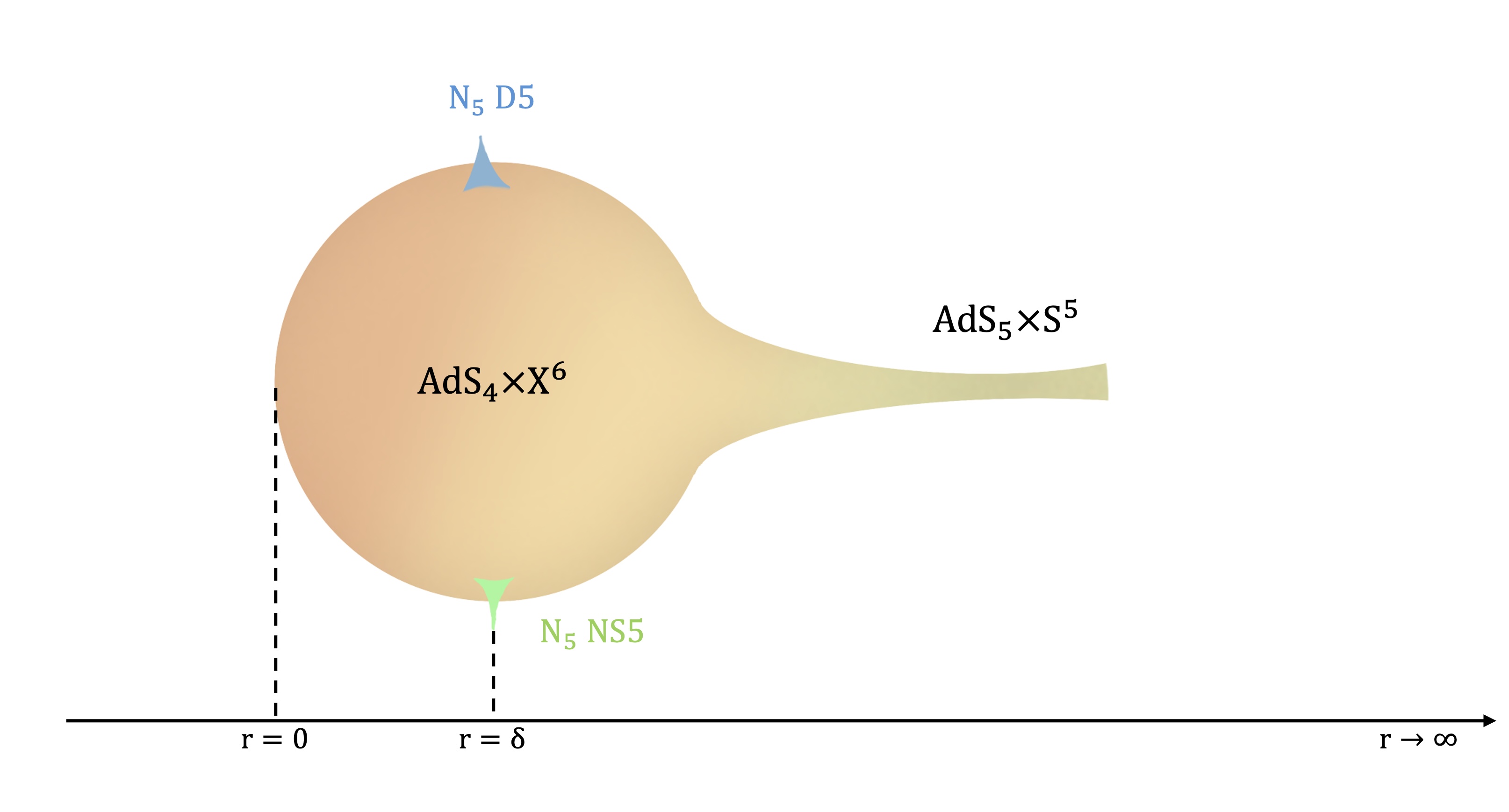}
    \caption{Sketch of the configuration describing an ETW configuration with one 5-brane stack of each kind for AdS$_5\times\IS^5$.}
    \label{fig:draw_bag}
\end{figure}

As explained in appendix \ref{sec:string-embedding} the full gravitational solution is determined in terms of the two harmonic functions $h_1$ and $h_2$ \eqref{the-hs} defined on the quadrant, determined by the positions $k_a$, $l_b$ of the NS5- and D5-brane punctures on the axis of the quadrant, and the coefficients $d_a$, ${\tilde d}_b$ of the corresponding sources, which are related to the 5-brane multiplicities in the stacks. In our present example, we have only one stack of $N_5$ 5-branes of each kind, hence \eqref{quant1} gives $d_a={\tilde d}_b=N_5/32\pi^2\equiv {\rm d}$. Also the positions of the 5-branes are determined by the linking numbers (\ref{linking-example}) via \eqref{positions-linking}, which give $k_a=l_b=P/32\pi\equiv \delta$. The harmonic functions \eqref{the-hs} hence read:
\beqa
h_1&=&4r\sin \varphi +2{\rm d}\log\left(\frac{r^2+\delta^2+2r\delta\sin\varphi}{r^2+\delta^2-2r\delta\sin\varphi}\right)\, ,\\
h_2&=&-4r\cos \varphi -2{\rm d}\log\left(\frac{r^2+\delta^2+2r\delta\cos\varphi}{r^2+\delta^2-2r\delta\cos\varphi}\right)\, .
\label{harm.func}
\eeqa
We see that the position of the punctures in the radial direction of the quadrant is $r=\delta=P/32\pi$.
The geometry of the bag ${\IX_6}$ is thus fully determined by the 5-brane stack position $\delta$ (which encodes their linking numbers, and hence their induced D3-brane charges) and their multiplicity $N_5$. It is schematically shown in figure \ref{fig:draw_bag}.

Let us now recover the regime which relates to the Karch-Randall setup in section \ref{sec:actual-double}, where the ETW brane is close to the holographic boundary, and the localized graviton can be made parametrically light. In the string theory ETW brane it corresponds to the limit of large $N_5/P$, in which the number of degrees of freedom of the 4d CFT is much smaller than that of the 3d BCFT. In fact, as discussed in \cite{Karch:2022rvr} in this class of examples, the angle between the holographic boundary and the ETW brane in Poincar\'e coordinates is effectively given by $P/N_5$ in the 10d picture. We should also notice that in this limit, the volume of (the compact version of) $\IX_6$ is much larger than that of the $\IS^5$, and is related to the AdS$_4$ length scale. This will play an important role in section \ref{sec:ads-distance}. 

The appearance of the parametrically light localized graviton and the computation of its mass from the 10d solution was carried out in \cite{Bachas:2018zmb}, in a fairly model independent way. The localized graviton internal profile is constant in most of the internal space $\IX_6$ and decays rapidly in the neck region gluing it to the AdS$_5\times\IS^5$ throat, where it is zero. The mass was evaluated in full generality for any 5-brane configuration of the kind in section \ref{sec:d3-ns5-d5-dual}, with the sole assumption that $L_4\gg L_5$. In terms of these parameters, its scaling is given by
\beqa
m_gL_4^2\sim \left(\frac{L_5}{L_4}\right)^8\, .
\eeqa 
Hence the graviton still behaves as massless at scales parametrically lower than the 4d cosmological constant. This nicely agrees with the picture from the bottom-up setup in section \ref{sec:kr}. 

One important aspect in the relation of the 10d setup and the bottom-up Karch-Randall approach is the claim (crucial for the use of holographic renormalization) that the AdS$_4$ gravity theory is coupled to the CFT$_4$ (with a cutoff). This is a priori not obvious in the 10d gravity picture. One expects that this should follow from the holographic dual description of the AdS$_4$ brane theory, i.e. in terms of the BCFT$_3$. Although to our knowledge this has not been explicitly shown in the literature, a reasonable proposal was suggested in \cite{VanRaamsdonk:2021duo}, for general Gaiotto-Witten 3d BCFT$_3$'s corresponding to AdS$_4$ ETW branes close to the hologrphic boundary, which we now adapt to our present example. In the limit of large $N_5/P$, we have a very long 3d quiver gauge theory \eqref{quiver3d}, with a number of gauge factors parametrically of order $N_5$. On the other hand, quiver gauge theories in the limit of large number of nodes are able to encode the dynamics of theories in one more dimension, via deconstruction \cite{Arkani-Hamed:2001nha}. Heuristically, gauge invariant operators of the 3d theory given by a long sequence of bifundamentals with contracted gauge indices map to states of the 4d theory carrying momentum in the extra dimension, up to some cutoff energy scale. Hence, the large $N_5$ limit of the above 3d quiver gauge theory can potentially lead to a description of the 4d CFT with a cutoff. It would be interesting to make this map more explicit, and we expect that future progress in this question will serve to continue the quantitative exploration our results on relative quantum gravity in this work.

\subsubsection{Detour: Top-down string embedding of the Randall-Sundrum setup}
\label{sec:rs}

For completeness and future reference, in this section we make a short detour to review the string theory embedding in \cite{Verlinde:1999fy} of the Randall-Sundrum setup \cite{Randall:1999vf} of section \ref{sec:kr}. 

The basic setup is a compactification of type IIB string theory on an orientifold of a CY threefold $\IX_6$ (or F-theory compactified on a CY fourfold $\IX_8$), including ingredients generating a large negative tadpole for the RR 4-form (i.e. negative D3-brane charge). Since we are interested in having 4d Minkowski space, we focus on the constructions in \cite{Giddings:2001yu} (see \cite{Dasgupta:1999ss} for similar earlier work). In the type IIB picture the negative D3-brane charge can arise from several sources, compatible with a (not necessarily supersymmetric) 4d Minkowski vacuum. The first is the introduction of O3-planes (by taking the orientifold action $\Omega R (-1)^{F_L}$, with $\Omega$ being worldsheet parity, $F_L$ being left-moving fermion number, and $R$ being a $\IZ_2$ involution with fixed points, locally around which $R: x^i\to -x^i$, for $i=4,\ldots, 9$). Second, negative induced D3-brane charge can arise on the worldvolume of 7-branes from the Chern-Simons coupling $\int C_4\tr R^2$, such as the $-1$ unit of D3-brane charge on D7-branes wrapped on K3. Finally, let us mention that, although the models can generically include NSNS and RR 3-form fluxes,  they are forced to satisfy an imaginary self-duality condition in order to obtain 4d Minkowski vacua \cite{Dasgupta:1999ss,Giddings:2001yu}, and this forces their contribution to the RR 4-form tadpole to be positive. Hence, these fluxes are not necessary, and they are allowed as long as they do not overshoot the negative D3-brane charge from the previous sources (a similar comment applies to 7-brane worldvolume fluxes). From the F-theory perspective, all the negative D3-brane charge contributions arises from the Euler characteristic of $\IX_8$, from the dual to the 3-form coupling $C_3 \chi(\IX_8)$ in 11d M-theory \cite{Sethi:1996es}. The inclusion of F-theory 4-form fluxes reproduces type IIB 3-form fluxes as well as 7-brane worldvolume fluxes and contribute positively to the 4-form tadpole in 4d Minkowski compactifications.

To make the theory consistent, one then introduces a number of D3-branes, so as to compensate the negative net charge from the earlier sources. This number is relatively small in simple toroidal orbifold/orientifold models, but can be fairly high in more general compactifications, in particular going up to tens or hundred thousand in F-theory on certain CY fourfolds \cite{Candelas:1997eh}. These D3-branes can in principle be located at arbitrary points in the internal space (in the absence of mechanisms stabilizing brane moduli). So, if a large number $N$ of them are located at the same position, they backreact on the geometry, and they are more accurately described as an AdS$_5\times \IS^5$ throat, with $N$ units of RR 5-form flux on the $\IS^5$.

This is precisely the string theory embedding of the Randall-Sundrum scenario. The throat is an AdS$_5 (\times\IS^5)$ bulk, which we can regard as foliated in 4d flat space slices. It is cutoff on its UV end, near the holographic boundary, by the gluing of the compact $\IX_6$ (or $\IX_8$) geometry, which plays the role of the UV or Planck brane, or the ETW brane in our description. The localized graviton is arising simply because of the compactification. There is no extra brane in the IR end of the foliation, so it effectively describes a semi-infinite direction \footnote{It is also possible to realize the RS1 scenario \cite{Randall:1999ee} with an extra IR brane as in \cite{Giddings:2001yu}, by using a Klebanov-Strassler throat \cite{Klebanov:2000hb} as a second ETW brane \cite{Buratti:2021yia}.}, albeit with a finite 4d Planck mass.

This realization of localized gravity shares many feature of the Karch-Randall scenario, but with an exactly massless graviton. One key difference is that in the Randall-Sundrum case the ETW brane does not intersect the holographic boundary of the host AdS$_5$, hence there is no double holography picture. The holographic picture \cite{Gubser:1999vj} is simply 4d gravity in flat space coupled to the CFT$_4$ with a cutoff. The computation of the corresponding effective action is similar to that of the intermediate picture of the Karch-Randall setup. We refer the reader to the literature e.g. \cite{Panella:2024sor} for more detail.

\section{Global Symmetries and Swampland Constraints}
\label{sec:global-swampland}

In this section we discuss the possibility of global symmetries in relative gravity theories, namely in $d$-dimensional gravity theories with global symmetries, which are made consistent (and compatible with the no global symmetry swampland constraint) by their coupling to a higher-dimensional gravity theory, where the symmetries become gauged. We also discuss the interesting novel feature that such global symmetries may be anomalous, with the anomaly canceled thanks to an inflow from the bulk in the higher-dimensional completion.

Although much of the discussion is valid for more general setups, we focus on the case of coupling to a $(d+1)$-dimensional gravity theory. As explained in the introduction, we also focus on AdS$_d$ gravity coupled to Ad$_{d+1}$ bulk gravity. Similar results hold in Randall-Sundrum setups where the ETW brane is $d$-dimensional Minkowski space (for the de Sitter case, see section \ref{sec:desitter}). The realization with AdS$_d$ ETW branes allows for extra insights using the intermediate picture, which the species scale and emergence of gauge symmetries play a key role. We also describe the cobordism conjecture and its interplay with global symmetries.

\subsection{Global Symmetries in Localized Gravity}
\label{sec:no-global-symmetries}

We start by considering the no global symmetries conjecture, which states that global symmetries are either broken or gauged in theories of gravity admitting a quantum UV completion. This conjecture predates the notion of swampland, and is supported by arguments of black hole evaporation and entropy bounds (see e.g. \cite{Giddings:1988cx,Kallosh:1995hi,Arkani-Hamed:2006emk,Banks:2010zn,Harlow:2020bee}) and has been derived also in perturbative string theory \cite{Banks:1988yz,Burgess:2008ri} and from holography \cite{Harlow:2018jwu,Harlow:2018tng}.  

In this section we explore the realization of this conjecture in relative gravity theories. We explore the realization of exact global symmetries in gravity theories, which are nevertheless consistent at the quantum level via coupling to a higher-dimensional gravity theory. This was explored early on in \cite{Kaloper:2000xa} for the Randall-Sundrum scenario. We present a similar bottom-up discussion in the more general context of AdS Karch-Randall branes, and provide a top-down string theory realization in ETW branes in AdS$_5\times\IS^5$. We also consider its extension to generalized symmetries.

\subsubsection{Bottom-up perspective on global symmetries}
\label{sec:no-global-symmetries-bu}

The punchline in this section is that gauge symmetries in bulk gravity theories are effectively observed as global symmetries on the gravity theories localized on a lower-dimensional defect.
We carry out the discussion in the bottom-up setup of gravity localized on an AdS$_d$ Karch-Randall ETW boundary of an AdS$_{d+1}$ gravity theory, section \ref{sec:kr}. 

Let us endow the AdS$_{d+1}$ theory with a gauge symmetry $G$ with $(d+1)$-dimensional gauge coupling $g_{d+1}$. Contrary to the graviton, the $(d+1)$-dimensional gauge bosons  do not have a localized mode on the KR ETW brane \cite{Oda:2000dd,Tachibana:2001te,Tachibana:2001xq} (see e.g. \cite{Pomarol:1999ad} for the analogous result in the Randall-Sundrum setup). To sketch the computation, we start with the 5d gauge action
\beqa
S_{5, gauge}=-\frac 14 \int d^5 x \sqrt{-g} F_{MN} F^{MN}\, .
\label{5d-gauge}
\eeqa 
Using the conformal coordinates in (\ref{conformal}), fixing the gauge $A_z=0$ and focusing on the component $A_\mu(z)=\rho(z)a_\mu(x^\lambda)$, the equation and solution for the lightest mode are
\beqa
\partial_z(e^A\partial_z \rho)=0\quad \Rightarrow \quad \rho(z)=-\frac{\alpha}{L_5(-\Lambda_4)}\cos(\sqrt{-\Lambda_4}(z+z_0))+\beta \, ,
\eeqa
where $\alpha$, $\beta$ are integration constants. The zero mode has an oscillatory form, and is not localized on the ETW brane (or any particular location, for that matter). 

Replacing into (\ref{5d-gauge}) to compute the effective 4d coupling of the gauge fields, one gets a divergent result, due to the absence of a localized mode in the non-compact extra dimension. This implies that, from the perspective of the $d$-dimensional theory the symmetry $G$ is a global symmetry. This is a familiar effect in other localized systems, such as worldvolume theories on branes in string theory, but becomes more striking in this setup of localized gravity due to its clash with the familiar no global symmetry conjecture.

The above can be recast by saying that a theory of gravity can admit global symmetries if we regard it as relative, i.e. if it is coupled to a higher-dimensional gravity theory where the symmetry is gauged. This fits the familiar statement of the conjecture, which requires that symmetries in quantum gravity are either broken (so they are not exact) or gauged. In our present setup, the symmetry is exact and global in the lower-dimensional gravity theory, but it is ultimately gauged in its completion by coupling to the higher-dimensional gravity theory.

This interplay was already discussed in \cite{Kaloper:2000xa} in the analogous Randall-Sundrum setup, where the symmetry was interpreted as effectively global in the $d$-dimensional theory due to screening of charges by the CFT$_d$ degrees of freedom. Bulk charged objects moved on top of the brane still have fluxlines of the gauge field, but they decay too fast to account for a $d$-dimensional charge. The flux through a Gauss sphere in $d$-dimensions decreases with the distance, as if effectively screened. The screening length scale is precisely controlled by the length scale of the bulk AdS$_{d+1}$ to which the theory is coupling. This is consistent with our discussion above: the symmetry is effectively global up to an energy scale in which the coupling to the higher-dimensional theory opens up, and the symmetry is revealed as a gauge one. Note that the gauging of the symmetry in the higher dimension suffices to avoid usual arguments regarding the problems of black hole evaporation and violation of global symmetries.

The screening interpretation due to CFT$_d$ degrees of freedom also holds for our Karch-Randall setup, as we discuss in a 
detailed quantitative manner using the intermediate picture in section \ref{sec:emergence}. We will also use this picture to quantitative characterize the screening of charges via the study of charged black holes.

\subsubsection{The top-down perspective on global symmetries}
\label{sec:no-global-symmetries-td}

In this section we consider a top-down embedding of the above ideas, in the context of the configurations in section \ref{sec:d3-ns5-d5}. Since we need to consider gauge symmetries in the bulk AdS$_5$ theory, one possibility would be to study the dynamics of bulk gauge bosons from isometries of the internal space. Specifically, the 10d supergravity solutions describing 5-brane ETW configurations preserve a full $SO(3)\times SO(3)$ subgroups of the $SO(6)$ isometry of the asymptotic $\IS^5$. However, the same conclusions hold for any mechanism producing 5d bulk gauge symmetries. Therefore, for future use, we prefer to focus on examples where the bulk gauge symmetries arise from extra branes. Specifically, we introduce an additional set of D7-branes, which produce extra flavour symmetries for (4d $\NN=2$ hypermultiplet) quarks coupled to the 4d $\NN=4$ SYM theory \cite{Karch:2002sh}.

Let us start describing the brane configuration in flat space. We consider a stack of $N$ D3-branes along 012 and of semi-infinite extent in the coordinate 3, because they end on a configuration of NS5-branes (along 012456) and D5-branes (along 012789) defining the 3d $\NN=4$ Gaiotto-Witten BCFT$_3$. We now include an additional (small) number $n$ of D7-branes along 0123 4578 and sitting at the D3-brane location in 69. The complete system of branes preserves 4 supercharges, as one can easily check by complexifying the coordinates 47, 58 and 69 as $z^i$, $i=1,2,3$. Indeed, the D3-branes sit at a point in this geometry while, in the language of CY threefolds, the D7-branes wrap the holomorphic 4-cycle $z^3=0$. In the 4d directions 0123, the 5-brane configuration defines a BPS domain wall, since the NS5- and D5-branes wrap special lagrangian 3-cycles on those complex coordinates.

As discussed, the theory on the D3-branes is 4d $\NN=4$ $SU(N)$ SYM coupled to a number $n$ of flavour hypermultiplets, in the fundamental representation of an $SU(n)$ global symmetry. We will work in the quenched approximation $n\ll N$, so that the additional flavours do not spoil conformality of the 4d theory at leading order in $n/N$, so we continue to refer to it as CFT$_4$. The theory is defined on a 4d half-space, with boundary conditions determined by the BCFT$_3$ associated to the 5-brane configuration. 

In the absence of 5-branes, the gravity dual was described in \cite{Karch:2002sh} and corresponds to AdS$_5\times\IS^5$ with $N$ units of RR 5-form flux on the $\IS^5$, and with $n$ D7-branes on AdS$_5\times \IS^3$, introduced as probes (the gravity dual of the quenched approximation). Although the D7-branes wrap a maximal $\IS^3\subset \IS^5$ which is homologically trivial, and the mode describing the sliding off of the $\IS^3$ is tachyonic, it is above the Breitenlohner-Freedman bound, and the system is stable. As discussed, it is actually supersymmetric, as it actually corresponds to a generalized calibration \cite{Gutowski:1999iu,Gutowski:1999tu,Gauntlett:2001ur,Gauntlett:2002sc,Gauntlett:2003cy,Martelli:2003ki,DallAgata:2003txk,Cascales:2004qp}.

The inclusion of the 5-brane configuration is straightforward, since there is no incompatibility with the D7-branes. Hence we obtain a configuration of the kind considered in section \ref{sec:d3-ns5-d5}, with the addition of the $n$ D7-brane probes wrapped on $\IS^3\subset\IS^5$, where the $\IS^5$ is obtained by fibering $\IS^2\times\IS^2$ over the $\varphi$ arc at the corresponding location in the radial direction in the base Riemann surface, recall figure \ref{fig:quadrant}. Recalling (\ref{s5-split}), the D7-branes along 0123 4578 wrap an $\IS^3$ obtained by fibering $\IS^1\times \IS^1$ over the $\varphi$ arc, where the $\IS^1$ is the equator of each $\IS^2$ in the fibration. 

Let us now consider the fate of the $SU(n)$ symmetry from the perspective of the 4d theory at the ETW brane, and connect with the bottom-up discussion in the previous section. As explained in section \ref{sec:d3-ns5-d5-dual}, we can regard the configuration as type IIB on AdS$_4\times \IX_6$ where $\IX_6$ is actually non-compact due to the asymptotic spike producing the AdS$_5\times\IS^5$ throat. The KK reduction of higher-dimensional gauge fields to produce lower-dimensional ones is purely topological, in terms of the homology classes. In our particular case, the would-be lower-dimensional gauge bosons have a non-normalizable internal wavefunction (the harmonic 0-form) due to the non-compactness of $\IX_6$. This reproduces the formal divergence in the 4d gauge coupling discussed in the previous section. Hence, our top-down discussion matches the bottom-up argument there.

As already discussed in section \ref{sec:d3-ns5-d5-dual}, these models have no scale separation, and hence do not provide genuine compactifications to 4d/5d. We would however like to point out that the basic properties underlying the discussion of symmetries are fairly insensitive to the ratio of internal and external length scales, so we expect our conclusions to hold in general, including possible models with scale separation. 

\subsubsection{Generalized Symmetries}
\label{sec:generalized}

In this section we would like to emphasize that a similar picture holds for generalized $p$-form symmetries \cite{Gaiotto:2014kfa} (see \cite{McGreevy:2022oyu,Brennan:2023mmt,Gomes:2023ahz,Shao:2023gho,Schafer-Nameki:2023jdn,Bhardwaj:2023kri,Iqbal:2024pee} for reviews). In particular, the discussion of continuous generalized symmetries from the bottom-up perspective is identical to that carried out for 0-form gauge symmetries in the previous sections (see \cite{Kaloper:2000xa} for an early discussion in the Randall-Sundrum setup), thus resulting in effectively global symmetries in the $d$-dimensional gravity theory up to the energy scale at which the theory is completed by its coupling to $(d+1)$-dimensional gravity, where the symmetry is gauged.

We would like to focus the discussion on a celebrated example of 1-form symmetries of AdS$_5\times\IS^5$. Four-dimensional $\NN=4$ $SU(N)$ SYM theory potentially has electric and magnetic 1-form $\IZ_N$ global symmetries, whose charged operators are Wilson line and 't Hooft line operators. The actual symmetry which is realized is determined by topological choices specifying the allowed line operators \cite{Aharony:2013hda}, such as the global structure of the gauge group. Building on the pioneering work in \cite{Witten:1998wy}, this is most efficiently encoded in the topological sector of the holographic dual (in a realization of the general idea of Symmetry Topological Field Theory or SymTFT \cite{Gaiotto:2014kfa,Gaiotto:2020iye,Ji:2019jhk,Apruzzi:2021nmk,Freed:2022qnc}). In short, the 5d gravitational bulk theory contains a topological sector described by the NSNS and RR 2-forms $B_2$, $C_2$, with 5d topological action
\beqa 
S_5=N\int_{5d} B_2 dC_2\, .
\label{bf-symtft}
\eeqa 
This simply arises from the 10d Chern-Simons coupling $F_5B_2dC_2$, by integrating the RR 5-form over the $\IS^5$. This action makes the 2-form fields $\IZ_N$-valued, and their values near the holographic boundary correspond to the background fields for the actual 1-form global symmetry of the CFT$_4$.

We thus need to consider the implementation of the above picture in setups with ETW boundary configurations on the gravitational AdS$_5 \times \IS^5$ side. This was studied in \cite{GarciaEtxebarria:2024jfv} by extracting the topological sector of the 10d supergravity solutions of 5-brane ETW configurations for AdS$_5 \times \IS^5$ in section \ref{sec:d3-ns5-d5-dual}, to reproduce the behaviour of the 1-form global symmetries of 4d $\NN=4$ $SU(N)$ SYM upon their coupling to a Gaiotto-Witten BCFT$_3$ (see \cite{Huertas:2024mvy} for extension to 3d $\NN=2$ BCFT$_3$). 

The discussion is model dependent, because in general the $\IZ_N$ symmetries of the bulk theory are broken by the 5-brane ETW configuration, morally due to the partitioning of the $N$ semi-infinite D3-branes among the 5-branes on which they are ending. In the gravitational picture, the bulk $\IZ_N$ theory (\ref{bf-symtft}) connects to other $\IZ_{N_i}$ theories via junctions in a structure dubbed SymTFT Fan in \cite{GarciaEtxebarria:2024jfv}. The unbroken symmetry is given by $n={\rm gcd}(N_i)$, which can still be a large subgroup. For illustration, let us focus on a particular case, given by the class of examples in section \ref{sec:KSU}, for which the bulk $\IZ_N$ symmetry, with $N=2N_5P$, is broken to $\IZ_{N_5}$  (if ${\rm gcd}(N_5,P)=1$, for simplicity). This follows because the two stacks of 5-branes reduce the asymptotic 5-form flux $N=2N_5P$ in amounts given by their respective induced D3-brane charges i.e. ($N_5$ times) the linking numbers (\ref{linking-example}), namely $N_5(P+N_5/2)$, $N_5(P-N_5/2)$. 

The bulk generalized symmetries unbroken by the 5-brane ETW configuration correspond to symmetries of the AdS$_4$ gravity theory. As discussed in the bottom-up picture, in analogy with ordinary symmetries, these generalized symmetries are effectively global but microscopically gauged by the coupling to the 5d bulk. We can easily illustrate this using the theories in section \ref{sec:KSU}, as follows. As discussed in \cite{GarciaEtxebarria:2024jfv}, in the worldvolume theory on the D5-brane stack, there is a worldvolume coupling 
\beqa
N_5 \int_{{\rm AdS}_4\times\IS^2} C_2 {\cal F}^2= N_5(P-N_5/2)\int_{{\rm AdS}_4}C_2F_2-N_5(P-N_5/2)\int_{{\rm AdS}_4}C_2B_2\, ,
\label{bf}
\eeqa 
where ${\cal F}=F-B_2$, $F$ is the D5-brane worldvolume $U(1)$ gauge field strength, and we have used that the total monopole charge produces the linking number $\int_{\IS^2}{\cal F}=P-N_5/2$. The second term simply arises from the coupling (\ref{bf-symtft}) taking into account that the effective values of $N$ jump in $N_5(P-N_5/2)$ units in crossing the D5-brane stack. The first term is a coupling of the value of the bulk 2-form at the brane as background field of the global 1-form symmetry. Since the coupling is effectively a localized St\"uckelberg coupling for the D5-brane $U(1)$, it makes the corresponding localized symmetry discrete. Including the effect of similar couplings from the NS5-branes one recovers that generically the bulk 1-form symmetries are broken to a $\IZ_{N_5}$ by the ETW boundary. From the holographic perspective, this corresponds to the unbroken part of the 1-form global symmetry of the CFT$_4$ living on the AdS$_4$ ETW brane.

\subsection{Anomalous Symmetries and Anomaly Inflow}
\label{sec:anomaly}

The fact that global symmetries are admissible in relative gravity theories opens up the possibility that such symmetries may be anomalous, either due to pure gauge anomalies or to mixed gravitational anomalies. This would seem to clash with the fact that the symmetries are ultimately gauged in the higher-dimensional ambient gravity theory, c.f. section \ref{sec:no-global-symmetries}, and the anomaly would render the theory inconsistent. In this section we provide a simple solution of these issues, and argue that this potential problem is actually solved by anomaly inflow mechanisms in the higher-dimensional theory. We start with a bottom-up discussion in the Karch-Randall setup of section \ref{sec:double-holography}, and later provide top-down string theory examples based on those in section \ref{sec:d3-ns5-d5}.

\subsubsection{The bottom-up perspective}
\label{sec:anomaly-bu}

The construction of relative gravity theories with anomalous global symmetries is simple and can be illustrated in the bottom-up setup of AdS KR ETW branes already used in section \ref{sec:no-global-symmetries}. For concreteness we focus on an AdS$_4$ ETW brane on AdS$_5$, which leads to localized 4d gravity, and introduce a 5d $U(1)$ gauge symmetry, and a localized 4d chiral fermion field of charge $+1$. From the perspective of the 4d gravity theory, we have a $U(1)$ global symmetry, and the 4d dynamical fermion leads to a cubic gauge anomaly and a mixed gravitational anomaly. As explained, these anomalies seem not to be dangerous from the perspective of the 4d gravity theory, because they are anomalies for a global symmetry. On the other hand, the 4d gravity theory requires a completion by coupling to the 5d theory, where the $U(1)$ symmetry is gauged. In this 5d theory there is a localized 4d anomaly arising from the localized 4d fermions. This anomaly would make the theory inconsistent, but it is easy to write down a 5d topological coupling which cancels it. It is simply given by a topological 5d Chern-Simons term with the structure (we are cavalier regarding the relative coefficients)
\beqa
S_{{\rm top},5d}=\int_{5d} A_1(F^2-\tr R^2)\, .
\eeqa
The variation of this term under a gauge transformation $\delta_\lambda A_1=d\lambda$ is, integrating by parts and using Stokes' theorem for the 4d boundary of the 5d bulk,
\beqa 
\delta_\lambda S_{{\rm top},5d}=\int_{4d} \lambda (F^2-\tr R^2)\, ,
\eeqa
which precisely cancels the localized 4d anomaly of the localized 4d fermion.

The generalization to general gauge and gravitational anomalies is straightforward, by introducing the anomaly polynomial $Y(F,R)$ and the anomaly descent relations $Y=dY^{(0)}$, $\delta_\lambda Y^{(0)}=\lambda dY^{(1)}$ (see \cite{Alvarez-Gaume:1985zzv} for a review), as usual in anomaly inflow. For the case of 4d localized anomalies, the 5d action and its gauge variation
\beqa
S_{{\rm top},5d}=\int_{5d} Y^{(0)} \quad ,\quad \delta_\lambda S_{{\rm top},5d}=\int_{4d} Y^{(1)}\, ,
\eeqa
which cancels the localized anomaly. Clearly, there is an obvious generalization to discrete symmetries by promoting the setup using the Dai-Freed theorem \cite{Dai:1994kq} (see \cite{Garcia-Etxebarria:2018ajm} for a review for physicists, with applications). In this vein, a formal way to recast the above explanation is that the topological sector of the 5d bulk theory is the anomaly theory of the 4d boundary theory. As mentioned in the introduction, anomalous QFTs regarded as living on the boundary of their anomaly theories provide a classic example of relative QFTs \cite{Freed:2012bs}. The fact that we can recover this picture also in our setup of localized gravity is at the root of our coining of the term relative quantum gravity. 

\subsubsection{An explicit top-down string theory model}
\label{sec:anomaly-td}

In this section we present an explicit top-down realization of the setup in the previous section. The underlying gravitational background is AdS$_5\times \IS^5$ and its ETW configurations studied in section \ref{sec:d3-ns5-d5}, with D9- anti D9-brane pairs as extra ingredients to have a non-trivial anomaly structure. The extra brane pairs are non-supersymmetric, and in fact unstable, but we will not mind because we are mainly looking at topological aspects, for which the construction suffices as toy model. In fact, there are fully supersymmetric models, which work in a similar way, but whose construction is technically more involved and is left for appendix \ref{sec:anomaly-d7s}.

\smallskip

{\bf The flat space D3-brane configuration}

In order to identify the holographic dual 4d theory and its AdS$_5$ bulk dual, we first describe the corresponding flat space D3-brane configuration. We consider type IIB in flat 10d spacetime, and a stack of $N$ D3-branes along 0123. We now add a small number $n\ll N$ of D9- anti D9-brane pairs \cite{Srednicki:1998mq,Witten:1998cd} (denoted $\aD9$-branes in what follows), so there is a 10d gauge group $U(n)\times U(n)'$. There is a also a tachyon whose condensation annihilates the brane-antibrane pairs, but which we artificially maintain uncondensed throughout the whole discussion. The 4d gauge theory is $\NN=4$ $SU(N)$ SYM coupled to a small set of extra 4d chiral fermions in the bifundamental representation $(N;{\ov n},1)+({\ov N};1,n)$ of $SU(N)\times U(n)\times U(n)'$, where the last two factors are the chiral global symmetries from the 4d perspective. There are in addition massive scalars, but no extra tachyons in the D3-D9 or D3-$\aD9$ open string sectors. 

The $SU(N)$ gauge anomalies cancel, but there are localized contributions to the 4d anomalies for the cubic $SU(n)$ and $SU(n)'$, which are canceled by the inflow mechanism in \cite{Green:1996dd}, as we now discuss (there are also anomalies for the $U(1){\rm 's}\subset U(n),U(n)'$), for which the inflow could be discussed analogously). We focus on the $SU(n)$ factor, and the $SU(n)'$ works analogously. On the worldvolume of the D9-branes there is a Chern-Simons coupling
\beqa
S_{D9}=\int_{10d} C_4 \tr F^3\, ,
\label{d9-action}
\eeqa
where $F$ is the $SU(n)$ field strength on the D9-branes. The expression $Y\equiv\tr F^3$ should actually be regarded as the anomaly polynomial of a 4d chiral fermion in the fundamental, which satisfies the descent equations
\beqa
Y=dY^{(0)}\quad ,\quad \delta_\lambda Y^{(0)}=\lambda dY^{(1)}\, .
\label{descent}
\eeqa
Integrating (\ref{d9-action}) by parts and using (\ref{descent}), its gauge variation is
\beqa
\delta_\lambda S_{D9} = -\int_{10d} F_5 \delta_\lambda Y^{(0)}=\lambda\int_{10d} dF_5 Y^{(1)}\, .
\label{variation-d9}
\eeqa
We now use that there are $N$ D3-branes sitting at a point in the transverse $\IR^6$
\beqa
dF_5=N\delta_6({\rm D3})\, ,
\eeqa
where $\delta_6({\rm D3})$ is a bump 6-form Poincar\'e dual to the 4d D3-brane worldvolume. Hence, the variation (\ref{variation-d9}) is
\beqa
\delta_\lambda S_{D9}=\lambda \int_{10d} N \delta_6(D3) Y^{(1)} =N \,\lambda\int_{4d} Y^{(1)}\, .
\eeqa 
This has the precise form to cancel the anomaly from the 4d fermions in the D3-D9 sector, including the multiplicity $N$ of fermions in the fundamental of $SU(n)$. The picture is that the fermion anomaly implies a violation of $SU(n)$ charge conservation at the location of the D3-brane, which is compensated by an inflow coming from the 10d bulk of the D9-brane worldvolume. 

\smallskip

{\bf The gravity dual (with no ETW boundaries)}

The above discussion in flat space turns out to be useful in the understanding of the gravity dual. We take the number $N$ of D3-branes to be large, and backreact the D3-branes replacing them by the near horizon AdS$_5\times\IS^5$. We ignore the backreaction of the D9-branes, because their number is small so we work in the quenched approximation, in analogy with \cite{Karch:2002sh}. In this approximation, the holographic dual theory is still a 4d CFT, in agreement with ignoring the backreaction on the AdS$_5\times \IS^5$ gravity dual.

We are interested in the theory after reduction on the $\IS^5$, which has a 5d $U(n)\times U(n)'$ gauge symmetry, corresponding via the holographic dictionary to the global symmetry of the 4d holographic dual field theory. Focusing on the $SU(n)$ for concreteness, there is a 5d Chern-Simons term, arising from the 10d term (\ref{d9-action}) upon integrating by parts, using the descent (\ref{descent}) and reducing on the $\IS^5$. The term and its gauge variation are given by
\beqa
S_5=N\int_{5d}  Y^{(0)} \quad , \quad \delta_\lambda S_5=N\,\lambda \int_{4d} Y^{(1)}\, ,
\label{5d-cs}
\eeqa
where the prefactor $N$ arises from the 5-form RR flux over the $\IS^5$. The variation reproduces precisely the anomaly of the 4d theory, in a 5d description of the same inflow explained above in flat space. Its role in the gravity dual is to encode the anomalies of the global symmetries of the 4d dual field theory, with the asymptotic value of the 5d gauge fields at the holographic boundary playing the role of background fields for those global symmetries. This is the realization of the anomaly theory in the holographic setup, analogous to that in the D4/D8-branes system in \cite{Sakai:2004cn} (see \cite{Rebhan:2009vc,Gynther:2010ed} for further clarifications on this anomaly). Equivalently, we may envision the 4d field theory as inducing a localized anomaly on the holographic boundary, which is canceled by the above inflow from the bulk. 

\smallskip

{\bf The gravity dual with ETW boundaries}

We now would like to study the modification of the above picture adding the ETW boundaries of section \ref{sec:d3-ns5-d5}. In the flat space configuration we have $N$ D3-branes along 012 and of semi-infinite extent in the coordinate 3, because they end on a system of NS5-branes (along 012456) and D5-branes (along 012789), now with all branes in the presence of $n$ spacetime filling D9-$\aD9$ pairs. We note that there is no incompatibility in the simultaneous presence of the branes. 

In the holographic boundary, we have the 4d $\NN=4$ $SU(N)$ theory, defined on half-space with 3d $\NN=4$ BCFT boundary conditions, plus the additional fermionic flavours in the $(N,{\ov n},1)+({\ov N},1,n)$ of the gauge $SU(N)$ and the global $U(n)\times U(n')$ groups. Hence, this system belongs to a class of non-trivial ETW boundary configurations for chiral theories (see \cite{Angius:2024pqk} for a recent discussion).

Let us consider the gravity side, which is given by the asymptotic AdS$_5\times\IS^5$, the region including the 5-brane sources which successively peel of the 5-form flux, and the ultimate boundary where the (fluxless) $\IS^5$ shrinks. We simply need to dress this up with the D9-$\aD9$ pairs\footnote{One may fear that the non-trivial $H_3$ flux (\ref{the-fluxes}) on the $\IS^3$'s around NS5-brane stacks may lead to a Freed-Witten anomaly (actually, its non-torsion version \cite{Maldacena:2001xj} extending the torsion case \cite{Freed:1999vc}) on the D9- (and $\aD9$) branes, but one should notice that the $\IS^3$ is non-trivial in the neighbourhood of a 5-brane stack, but it is trivial on the $\IS^5$, so no inconsistency arises.}. Since we are working with topological effects, we just need to focus on the topological structure of the gravitational background after reduction to 5d, which is the SymTFT Fan in \cite{GarciaEtxebarria:2024jfv}, c.f. figure \ref{fig:poincare}. In fact we are free to deform it continuously, so let us consider collapsing the whole SymTFT Fan, namely, we collapse the wedges between the 5-branes into a single ETW 4d boundary of the 5d bulk. Then, given that the 5d bulk theory with the D9-$\aD9$ pairs contains the topological Chern-Simons coupling in the first half of (\ref{5d-cs}), its gauge variation leads to a 4d boundary contribution in the second half of (\ref{5d-cs}), now not just on the holographic boundary, but also on the 4d ETW boundary. 

Because the 4d ETW boundary is at finite distance, this implies that there must be a dynamical set of $N$ fermions in the antifundamental of $U(n)$ and $N$ fermions in the fundamental of $U(n)'$ on the ETW boundary. This is something which is not obvious from the holographic dual of the BCFT$_3$, so it is a remarkable implication of the bulk gravitational dynamics. Postponing momentarily the detailed microscopic explanation of these fermions, let us just accept their presence, and continue the discussion about their implications regarding the relative quantum gravity viewpoint.

By choosing judiciously the 5-brane configuration as explained in section \ref{sec:d3-ns5-d5}, we can get an ETW configuration supporting localized 4d gravity. From the 4d ETW boundary perspective, we have a 4d gravity theory on AdS$_4$, with 4d fermions coupled to a $U(n)\times U(n)'$ global symmetry with anomaly proportional to $N$. As mentioned in the bottom-up discussion, one may argue that there is no problem because the anomalous symmetry is a global symmetry from the 4d perspective. On the other hand, the no global symmetry constraint requires the 4d gravity theory to be relative to a 5d bulk gravity theory, where the symmetry is gauge. This would pose a problem because of the localized anomaly, but the coupling to the 5d theory comes with the appearance of a 5d Chern-Simons coupling which cancels the anomaly. Note that both the coupling to the 5d theory and the anomaly are parametrically controlled by the same quantity, $N$.

\smallskip

{\bf The microscopic origin of the boundary fermions}

We now come back to the origin of the dynamical fermions in the 4d ETW boundary. Let us unfold the SymTFT Fan into the bunch of 5-brane stacks that compose it, as in figure \ref{fig:poincare}. Recall that we have stacks of $n_a$ NS5-branes on AdS$_4\times \IS^2$, and of $m_b$ D5-branes on AdS$_4\times \IS^2$, leading to a 4d gauge symmetry $\otimes_a U(n_a)\times \otimes_b U(m_b)$, dual to the global flavour symmetry of the Gaiotto-Witten BCFT$_3$. Recall also that the 5-branes carry worldvolume gauge fluxes (\ref{worldvolume-fluxes}), which satisfy the sum rule (\ref{the-sum}).
 
 Let us now consider the interplay of the D9-branes with these 5-branes (again, the $\aD9$-branes work similarly), focusing on the D5-branes for concreteness. In the D9-D5$_b$ open string sector, one gets a 6d hypermultiplet in the bifundamental $(m_b, {\ov n})$ of $U(m_b)\times U(n)$. This is compactified on an $\IS^2$ with ${\tilde L}_b$ units of worldvolume flux, so using the index theorem this leads to ${\tilde L}_b$ copies of a 4d fermion in the bifundamental $(m_b, {\ov n})$. The overall number of fermions in the fundamental of $SU(n)$ is thus given by $ m_b{\tilde L}_b$ (no sum), precisely matching the induced D3-brane charge. Morally, the 4d fermions are coming from a D3-D9 sector, corresponding to the induced D3-brane charge on the D5-branes. For the NS5-branes, the microscopic derivation cannot be carried out in terms of open strings, but we assume a similar result. Namely, that the NS5$_a$-branes produce $K_a$ copies of 4d fermions in the bifundamental $(n_a,{\ov n})$. This means that, using (\ref{the-sum}), the NS5- and D5-branes reproduce $N$ 4d fermions in the antifundamental of $U(n)$, as expected from the inflow argument.
 
This concludes our discussion of anomalous effective global symmetries in gravity theories, made consistent by coupling to a higher-dimensional gravity theory, where the symmetries are gauged and their anomaly is canceled by anomaly inflow. It is amusingly satisfactory that the same mechanism (coupling to higher-dimensional theory) which promotes the localized global symmetry into a gauged one, implements the inflow required for cancellation of its anomaly.

\subsection{The Species Scale and Emergence}
\label{sec:emergence}

In this section we explore the description of the Karch-Randall setup of AdS$_d$ ETW boundaries of AdS$_{d+1}$ in the intermediate picture in section \ref{sec:ads} (see also appendix \ref{sec:intermediate-action}), and argue that it produces explicit quantitative descriptions of the species scale (see \cite{Dvali:2007hz,Dvali:2007wp,Dvali:2008ec,Dvali:2009ks,Dvali:2010vm,Dvali:2012uq} for early references and \cite{Castellano:2022bvr,vandeHeisteeg:2022btw,Cribiori:2022nke,Cribiori:2023ffn,Cribiori:2023sch,Calderon-Infante:2023uhz,Basile:2024dqq,Herraez:2024kux} for recent explorations in the swampland context) due to CFT$_d$ degrees of freedom, and reproduce the different definitions of the species scale. We use this picture to quantitatively explain that the global symmetries turning into gauge ones in section \ref{sec:no-global-symmetries} corresponds, in the intermediate picture, to the emergence of gauge symmetries and their interactions (see \cite{Heidenreich:2017sim,Grimm:2018ohb,Heidenreich:2018kpg,Palti:2019pca,Marchesano:2022axe,Castellano:2022bvr,Castellano:2023qhp,Blumenhagen:2023yws,Blumenhagen:2023tev,Blumenhagen:2023xmk,Hattab:2023moj,Blumenhagen:2024ydy} for proposals of emergence in the swampland context).

The quantitative aspect is particularly worth emphasizing, because, in contrast with usual weak-coupling discussions of emergence, in this case the result arises from quantum effects of a strongly coupled theory, yet it is computable using a gravity dual. The results in this section are well established in the holographic literature (see \cite{Panella:2024sor} for the case $d=3$), but the interpretation in terms of emergence in the spirit of the swampland program is new.

\subsubsection{The Species Scale}
\label{sec:species-scale}

In this section we point out that the computation of the $d$-dimensional brane action using the holographic renormalization prescription sketched in section \ref{sec:actual-double} (see also appendix \ref{sec:intermediate-action}) can be interpreted as yielding the species scale as the cutoff of the AdS$_d$ theory. Our discussion follows the review \cite{Panella:2024sor} (see also \cite{Emparan:1999wa,Emparan:1999fd,Emparan:2002px,Emparan:2020znc,Emparan:2021hyr} for related ideas). 

Starting from the $d-$dimensional action (\ref{brane-action-general}) 
\begin{equation}
   S_d = \frac{1}{16 \pi G_d} \int_{\mathcal{B}} d^dx \sqrt{-h} \left[ R-2 \Lambda_d + \frac{L^2_{d+1}}{(d-2)(d-4)} \left( R_{\mu \nu} R^{\mu \nu} - \frac{d R^2}{4 (d-1)} \right) \right] 
   \label{brane-action-3d}
\end{equation}
with induced Newton's constant (\ref{eq:effGd}), and AdS$_d$ cosmological constant and curvature length scale (\ref{eq:curvscale})
\begin{equation}
    G_d = \frac{G_{d+1} (d-2)}{L_{d+1}} \; , \quad \Lambda_d = - \frac{(d-1)(d-2)}{2L_d^2} \; , \quad \frac{1}{L_d^2}= \frac{2}{L_{d+1}^2} \left( 1- \frac{8 \pi G_{d+1} L_{d+1}}{d-1} \tau \right),
    \label{parameters-3d}
\end{equation}

with the parameter $\ell$ related to the brane tension $\tau$ by
\begin{equation}
    \ell= \frac{d-1}{8 \pi G_{d+1} \tau} \;.
    \label{brane-tension}
\end{equation}

%
The $(d+1)-$dim length scale $L_{d+1}$ can be traded for a $d-$dim AdS$_d$ length scale $\ell_d$ which includes the CFT$_d$ backreaction by
\begin{equation}
    L_{d+1} = \left(\frac{1}{\ell^2}+\frac{1}{\ell_d^2}\right)^{-1/2}
    \label{eq:bulkAdS4length}
\end{equation}
We are interested in the regime $L_{d}\approx \ell_{d}$, when the ETW brane is near the holographic boundary and we have a parametrically light graviton. Hence, (\ref{eq:bulkAdS4length}) becomes
\begin{equation}
  \frac{1}{L_{d}^{2}}=\frac{1}{\ell_{d}^{2}}\left[1+\frac{\ell^{2}}{4\ell_{d}^{2}}+\mathcal{O}\left(\frac{\ell^{4}}{\ell_{d}^{4}}+...\right)\right]  
\end{equation}
with $\ell\sim L_{d+1}\ll \ell_{d}$. The $d-$dim action (\ref{brane-action-3d}) becomes
  \begin{equation}
  S_d = \frac{1}{16 \pi G_d} \int_{\mathcal{B}} d^dx \sqrt{-h} \left[ R+ \frac{2}{\ell^2_d}+ \frac{\ell^2}{(d-2)(d-4)} \left( R_{\mu \nu} R^{\mu \nu} - \frac{d R^2}{4 (d-1)} \right) \right] +I_{CFT} \;,
  \label{eq:inducactqbtzneu2}
  \end{equation}
  
The expansion in higher-curvature terms is controlled by the effective cutoff length scale $\ell$. Hence, according to \cite{vandeHeisteeg:2023ubh}, it is a natural candidate for the species scale in the present theory. Indeed it can be shown to be parametrically larger than the Planck length by a factor related to the number of degrees of freedom of the CFT$_d$, dovetailing the definition of species scale in \cite{Dvali:2007hz}. This effective number of species can be characterized by the CFT$_d$ central charge 
%
%
\begin{equation}
    c_{d}\sim\frac{L_{d+1}^{d-1}}{G_{d+1}}\approx (d-2)\,\frac{\ell^{d-2}}{G_d}\left[1-\frac{d-2}2\frac{\ell^{2}}{\ell_{d}^{2}}+\mathcal{O} \left(\frac{\ell^{4}}{\ell_{d}^{4}}\right)+... \right]\;,
\end{equation}
where the last expansion is for small $\ell$. Hence in this regime we have
\begin{equation}
    \ell\sim \left( c_{d} G_{d} \right)^{\frac 1{d-2}}\gg M_{p,d}^{-1}\;,
\end{equation}
Finally, we can use the quantum black hole solutions in appendix \ref{sec:bhs} to relate $\ell$ to the definition of species scale in terms of black holes \cite{Dvali:2007hz,Dvali:2007wp}. From the metric (\ref{eq:naiveBTZ}), we see that the $\mu\ell/r$ term is a semi-classical correction making the size of the quantum black hole much larger than the Planck length. 

We find it extremely satisfactory that the intermediate picture provides a quantitative characterization of the species scale, in particular in a strong coupling context, and allows to connect the different definitions in the literature (see \cite{Calderon-Infante:2023uhz} for related link between the three definitions of the species scale in a different context). This is even more remarkable because, as already emphasized, the computation is carried out in a strongly coupled theory with a large number of degrees of freedom, thanks to the use of the dual bulk picture.

\subsubsection{Emergence of Gauge Dynamics}
\label{sec:emergence-gauge}

In this section we explain that the trading of global symmetries for gauge ones in section \ref{sec:no-global-symmetries} arises, from the intermediate picture viewpoint, from the emergence of gauge symmetries and their interactions. Let us go back to the discussion of double holography and the brane action computation in section \ref{sec:actual-double} and generalize it to our present case in which some symmetry is present, which for simplicity we take to be a $U(1)$. The discussion can be carried out for general dimension $d$ (see appendix \ref{sec:intermediate-action} for the computations), with an AdS$_d$ KR ETW brane in an AdS$_{d+1}$ bulk (see the review \cite{Panella:2024sor} and also e.g. \cite{Taylor:2000xw,Feng:2024uia,Climent:2024nuj} for related references). In the bulk picture we have a $(d+1)-$dimensional  Einstein-Maxwell theory of gravity and $U(1)$ gauge interactions, with action
\begin{equation}
  S_d= \frac{1}{16 \pi G_{d+1}} \int_{\mathcal{M}} d^{d+1}x \sqrt{-G} \left[ R_{d+1} -2 \Lambda_{d+1} - \frac{\ell^2_{\ast}}{4} F^2\right] - \frac{1}{8 \pi G_{d+1}} \int_{\partial \mathcal{M}} d^dx \sqrt{-h}  K \; ,  
  \label{einstein-maxwell}
\end{equation}
where $\ell^{2}_{\ast}=\frac{16\pi G_{d+1}}{g_{\ast}^{2}}$, with $\ell_{\ast}$ the coupling constant with dimensions of length and $g_{\ast}$ is the dimensionless gauge coupling constant in $d$ dimensions. This action is complemented with a purely tensional KR brane as in section \ref{sec:double-holography}. Hence, we have the same junction conditions for gravity, and we refer to the literature for details on the junction conditions for the gauge field.

Recall that in the intermediate picture, we have the holographic CFT$_d$ with a $U(1)$ global symmetry defined on flat $d-$dimensional half-space with no dynamical gravity, coupled with transparent boundary conditions to gravity in AdS$_d$ coupled to  this same CFT$_d$, but with a UV cutoff. Hence we have the striking situation of having dynamical gravity and a global symmetry, in principle in a full quantum theory. Note that in this intermediate picture there is no coupling to higher-dimensional gravity, which only arises in the bulk picture. What saves the day in the intermediate picture is the coupling to the CFT$_d$. As we now argue, upon integrating out its fast modes above the cutoff,  it generates kinetic terms for the background fields of the global symmetry, which thus become dynamical gauge bosons. The global symmetry is promoted to a gauge symmetry in the full theory including quantum effects of the CFT$_d$. This is therefore a complete realization of emergence in which even the tree level gauge kinetic terms arise from proper inclusion of quantum effects. 

As already mentioned,  the quantum effects most remarkably arise from a strongly coupled theory with a large number of degrees of freedom. Hence, in analogy with the pure gravity case in sections \ref{sec:actual-double} and \ref{sec:species-scale}, the computation is performed using holographic renormalization. Namely, we allow for a general metric and backgrounds for the global symmetry on the KR brane, and derive an effective action for them by invoking the bulk picture and performing the semiclassical path integral of the $(d+1)-$dimensional Einstein-Maxwell theory over the wedge between the $d-$dimensional holographic boundary and the KR brane. Following \cite{Panella:2024sor}, the result, already expressed as an expansion in $\ell$, is 
\begin{equation}
\begin{split}
   S_d= &  \frac{1}{16 \pi G_{d}} \int d^dx \sqrt{-h} \left\lbrace R_d \left[ h \right] + \frac{2}{\ell_d^2} +16 \pi G_d A_{\mu} j^{\mu} - \frac{\ell^2_{\ast}}{4} \tilde{F}^{(0)2} +  \right. \\
   & \left. + \frac{\ell^2}{(d-2)(d-4)} \left[ R_{\mu \nu} \left[ h \right] R^{\mu \nu} \left[ h\right] - \frac{R^2_d \left[ h \right]}{4(d-1)} \right]+ ... \right\rbrace + I_{CFT}\, , \\
\end{split}
\label{emergent-gauge}
\end{equation}
where we have introduced an effective $d-$dimensional gauge coupling
\beqa 
\tilde{\ell}_{\ast}^{2}=\frac{16\pi G_{d}}{g_{d}^{2}}\;,\quad g_{d}^{2}=\frac{4(d-1)(d-4)}{d^2-d-16} \frac{g^2_{\ast}}{L_{d+1}}\approx \frac{4(d-1)(d-4)}{d^2-d-16} \frac{g^2_{\ast}}{\ell}\;,
\label{3d-couplings-gauge}
\eeqa
such that $\ell_{\ast}^{2}=4(d-1)(d-4) \tilde{\ell}^2_{\ast}/(d-2)(d^2-d-16)$. Note that although the expression \eqref{emergent-gauge} is ill-defined when $d=4$, the correct coefficients for this case are computed in the equation \eqref{d-action:gauge_4d} of the appendix \ref{sec:intermediate-action}.

The expression (\ref{emergent-gauge}) as an expansion in $\ell$ again makes it clear that this defines the species scale, as in the previous section. It is in fact straightforward to relate it to the various definitions of species scale, just like in the previous section. In the present context, we focus instead on its role in the gauge kinetic terms. It is clear that in the limit $\ell\to0$, the coupling $g_{d}\to\infty$ becomes non-dynamical. This corresponds to the fact that when the cutoff is taken to the UV, the gauge kinetic terms vanish and the symmetry is effectively global. When the cutoff is lowered, quantum effects turn on the gauge kinetic terms and the symmetry becomes gauged. Hence, the setup provides an explicit quantitative realization of emergence, where even the tree level gauge couplings arise from quantum backreaction effects.

The picture matches our interpretation in the bulk picture that the symmetry on the brane is effectively global. In fact, the intermediate picture can provide a description of the screening phenomenon by the CFT degrees of freedom in section \ref{sec:no-global-symmetries-bu}. In the case $d=3$, this is possible using the charged quantum black hole solutions in appendix \ref{sec:charged-bh}, not yet known at the time of the screening proposal in \cite{Kaloper:2000xa}. Specifically, the charge $q$ of the black hole solution of the bulk theory does {\em not} correspond to a charge of the 3d black hole solution (\ref{chargedbh-solution}), since in the limit $\ell\to0$ the $q$-dependent correction vanishes. This is a manifestation of the the result in section \ref{sec:no-global-symmetries} that charges are screened on the brane, and the symmetry is effectively global. The gauge charge of the 3d quantum black hole is a consequence of quantum backreaction of the fast CFT$_3$ modes.

\subsection{Cobordism Conjecture and Relative Defects}
\label{sec:cobordism}

In this section we discuss the cobordism conjecture in relative gravity theories. The Cobordism Conjecture \cite{McNamara:2019rup}  (see \cite{GarciaEtxebarria:2020xsr,Ooguri:2020sua,Montero:2020icj,Dierigl:2020lai,Hamada:2021bbz,Blumenhagen:2021nmi,Andriot:2022mri,Dierigl:2022reg,Debray:2023yrs} for topological applications, and \cite{Buratti:2021yia,Buratti:2021fiv,Angius:2022aeq,Blumenhagen:2022mqw,Blumenhagen:2023abk} for dynamical cobordisms\footnote{See also 
\cite{Dudas:2000ff,Blumenhagen:2000dc,Dudas:2002dg,Dudas:2004nd,Hellerman:2006nx,Hellerman:2006ff,Hellerman:2007fc} for related early work and \cite{Basile:2018irz,Antonelli:2019nar,GarciaEtxebarria:2020xsr,Mininno:2020sdb,Basile:2020xwi,Mourad:2021qwf,Mourad:2021roa,Basile:2021mkd,Mourad:2022loy,Angius:2022mgh,Basile:2022ypo,Angius:2023xtu,Huertas:2023syg,Mourad:2023ppi,Friedrich:2023tid,Angius:2023uqk,Delgado:2023uqk,Angius:2024zjv,Mourad:2024dur,Mourad:2024mpg,GarciaEtxebarria:2024jfv,Angius:2024pqk} for recent developments.}) posits that any two quantum gravity theories should admit some interpolating configuration relating them. At the purely geometrical level, this is implemented by cobordism theory, hence the name, but in the context of e.g. string theory it may involve other ingredients, such as spin (and other) structures, gauge fields, and charged objects. Hence,  when we discuss cobordism, unless we explicitly mention {\em geometric cobordism}, we mean the suitably generalized version of cobordism appropriate to the ingredients present in the configuration.

The Cobordism Conjecture was related in \cite{McNamara:2019rup} to the no global symmetry conjecture in quantum gravity theories. However, we have argued in section \ref{sec:no-global-symmetries} that the no global symmetry conjecture is not obeyed in the usual way in relative gravity theories. Namely, it is possible to have gravity theories with effectively global symmetries if the theory is completed by coupling it to a higher-dimensional gravity theory. This opens up the possibility that the cobordism conjecture has a similar behaviour in relative gravity theories.

In this section we indeed argue that there exist gravity theories which cannot be connected even by (even generalized) cobordisms, but which can be connected when regarded as relative gravity theories, i.e. when they are coupled to a higher-dimensional gravity theory. We provide some general arguments, and back them up with a class of examples, which admit an explicit string theory embedding.

\subsubsection{General picture and relative bordisms}

We are interested in exploring the possibility that relative gravity theories violate the usual formulation of the cobordism conjecture, in the following sense. We may have two $d$-dimensional gravity theories ${\cal X}_d$ and ${\cal X}_d'$ which are not related by an interpolating configuration in $d$-dimensions. Regarding them as relative theories, they must be completed by coupling them as ETW branes to some corresponding $(d+1)$-dimensional gravity theories ${\cal Y}_{d+1}$, ${\cal Y}_{d+1}'$. Conversely, the resulting $(d+1)$-dimensional configurations basically describe bordisms to nothing of the asymptotic $(d+1)$-dimensional theories. Now, since the $(d+1)$-dimensional theories are assumed to be UV complete, they should obey the usual cobordism conjecture, namely there should exist some interpolating domain wall configuration between ${\cal Y}_{d+1}$ and ${\cal Y}_{d+1}'$. This means that the two $d$-dimensional theories ${\cal X}_d$ and ${\cal X}_d'$ can be related by a domain wall in the $(d+1)$-dimensional theory to which they couple\footnote{Such configurations are closely related to the intersections of different ETW branes, which were considered in \cite{Angius:2023uqk,Angius:2024zjv,Ruiz:2024gzv} as a generalization of dynamical cobordisms \cite{Buratti:2021fiv,Angius:2022aeq,Blumenhagen:2022mqw,Blumenhagen:2023abk}, see also
  \cite{Dudas:2000ff,Blumenhagen:2000dc,Dudas:2002dg,Dudas:2004nd,Hellerman:2006nx,Hellerman:2006ff,Hellerman:2007fc}
  for early references, and
\cite{Basile:2018irz,Antonelli:2019nar,GarciaEtxebarria:2020xsr,Mininno:2020sdb,Basile:2020xwi,Mourad:2021qwf,Mourad:2021roa,Basile:2021mkd,Mourad:2022loy,Angius:2022mgh,Basile:2022ypo,Angius:2023xtu,Huertas:2023syg,Mourad:2023ppi,Friedrich:2023tid,Angius:2023uqk,Delgado:2023uqk,Angius:2024zjv,Mourad:2024dur,Mourad:2024mpg,GarciaEtxebarria:2024jfv,Angius:2024pqk}
  for recent works.}. In this way, they violate the usual form of the cobordism conjecture as $d$-dimensional theories, yet satisfying it when regarded as relative gravity theories, i.e. when the $(d+1)$-dimensional gravity completion is implemented.

One may consider that the domain wall between the configurations ${\cal Y}_{d+1}$ and ${\cal Y}_{d+1}'$ must necessarily intersect the ETW boundary, and this intersection defines a domain wall configuration, a bordism, between the theories ${\cal X}_d$ and ${\cal X}_d'$,with a direct interpretation in the $d$-dimensional sense, decoupled from the $(d+1)$-dimensional interpretation.
This is not necessarily the case, and indeed, in the next section we will see a class of examples, with an explicit string theory embedding, in which such a decoupled $d$-dimensional interpretation does not exist. The  bordism of the $d$-dimensional theories only exists if coupled as a boundary to the bordism of the $(d+1)$-dimensional theory, see figure \ref{fig:cobordism} for a later example. We dub this kind of configuration {\em relative bordism}, for obvious reasons.

A realization of the above ideas was put forward in \cite{Ruiz:2024gzv} (albeit not specifically in the context of localized gravity theories), for purely geometric cobordisms. The basic setup is to consider a quantum gravity theory arising from compactification on an $n$-dimensional geometry $X_n$. We now consider two different bordism to nothing of this geometry, namely two manifolds $B_{n+1}$, $B_{n+1}'$, with boundary $\partial B_{n+1}=X_n$, $\partial B_{n+1}'=X_n$. We now consider the question of whether these manifolds $B_{n+1}$, $B_{n+1}'$ are bordant to each other. As argued in \cite{Ruiz:2024gzv}, the answer is positive, and there always exist a bordism between the bordisms, namely an $(n+2)$-dimensional manifold ${\tilde B}_{n+2}$, with boundary $\partial {\tilde B}_{n+2}=B_{n+1}-B_{n+1}'$.  

In our present context, the original theory on $X_n$ leads to the bulk $(d+1)$-dimensional theories. Letting one of the direction of the bordisms $B_{n+1}$, $B_{n+1}'$ (dubbed the Morse-Bott coordinate in \cite{Ruiz:2024gzv}) correspond to one spacetime direction, the two different bordisms define two different ETW configurations. Finally, letting a further coordinate of the bordism of bordisms ${\tilde B}_{n+2}$ correspond to a further spacetime dimension, the configuration describes a domain wall of the $(d+1)$ dimensional theory, which effectively interpolates between the two different ETW configurations. This is much in the spirit of relative bordisms introduced above for general cobordism.

In the following section we describe a class of examples realizing the above ideas, with explicit embedding in string theory. 

\subsubsection{Mismatch of Global Symmetry Anomalies and Cobordism}
\label{sec:anomaly-jumps-bu}

As already emphasized, relative gravity theories can effectively coexist with global symmetries, and this opens up the possibility of violating the usual form of the cobordism conjecture in relative gravity theories. Namely, the cobordism conjecture in the $d$-dimensional theory is realized via relative bordisms, i.e. necessarily invoking the coupling to a $(d+1)$-dimensional gravity theory. In this section we describe a class of gravity theories which at face value cannot be connected to each other, but which can be connected when they are coupled to a higher-dimensional gravity theory. The mechanism preventing the connection among the initial theories is that they have the global symmetry, but with different anomalies for it. Morally, the lack of 't Hooft anomaly matching forbids theories with different anomalies (for the same global symmetry) to be connected (see \cite{Antinucci:2024izg} for related ideas in a non-gravitational setup).

Let us consider a simple example in a bottom-up perspective (a similar example will be subsequently embedded in a top-down string theory example in section \ref{sec:anomaly-jumps-td}). Consider a theory of 4d gravity coupled to a set of $k$ Weyl fermions with charge $+1$ under a $U(1)$ global symmetry group. There are cubic $U(1)^3$ and mixed $U(1)$ gravitational anomalies, proportional to $k$, as discussed in section \ref{sec:anomaly}. A key property in theories with global symmetries is that, in order to be related, the anomalies of their (unbroken) global symmetries must match. This 't Hooft anomaly matching has been extensively exploited in the study of non-perturbative gauge dynamics, to match UV and IR descriptions of the theory (see \cite{Intriligator:1995au} for a review in the 4d $\NN=1$ context). For the same reasons, it can be similarly applied to constraining possible domain walls separating phases of the theory.

In our case, theories of this kind with different numbers of fermions, $k=N$ and $k=N+m$, cannot be joined by a domain wall configuration, since it is forbidden by 't Hooft anomaly matching. Hence, the cobordism conjecture is not satisfied, as a direct consequence of the existence of (potentially anomalous) global symmetries. 

The theories in this class actually do not make sense as quantum gravity theories on their own, as precisely signaled by the presence of global symmetries, which are not admissible in quantum gravity. As explained in section \ref{sec:no-global-symmetries}, the theories can however be completed by regarding them as relative gravity theories, i.e. by coupling them to a 5d gravity theory. In doing so, as explained in section \ref{sec:anomaly}, the global symmetry becomes gauged in 5d and the anomaly is encoded in a Chern-Simons term (\ref{5d-cs}), with coefficient $k$. Applying this to both 4d relative gravity theories, we end up with two 5d quantum gravity theories which differ in the level of their Chern-Simons couplings.

We can now search for a domain wall interpolating between the two 5d gravity theories with different topological couplings. This is guaranteed to exist if the 5d theories are full-fledged quantum gravity theories, by applying the cobordism conjecture. Indeed, it is possible to describe the structure of such domain wall in EFT terms. Since the domain wall represents a codimension 1 boundary for each of the 5d theories it separates, the Chern-Simons couplings induce inflows into it, proportional to $-N$ and $N+m$, where the minus sign in the first is due to the opposite orientation of the two regions, so there is a net anomaly inflow proportional to $m$. This implies that the domain wall must support an anomalous dynamical field content to cancel this inflow. The simplest such field content is $m$ 4d fermions charged under the gauge $U(1)$.

\begin{figure}[htb]
\begin{center}
\includegraphics[scale=.3]{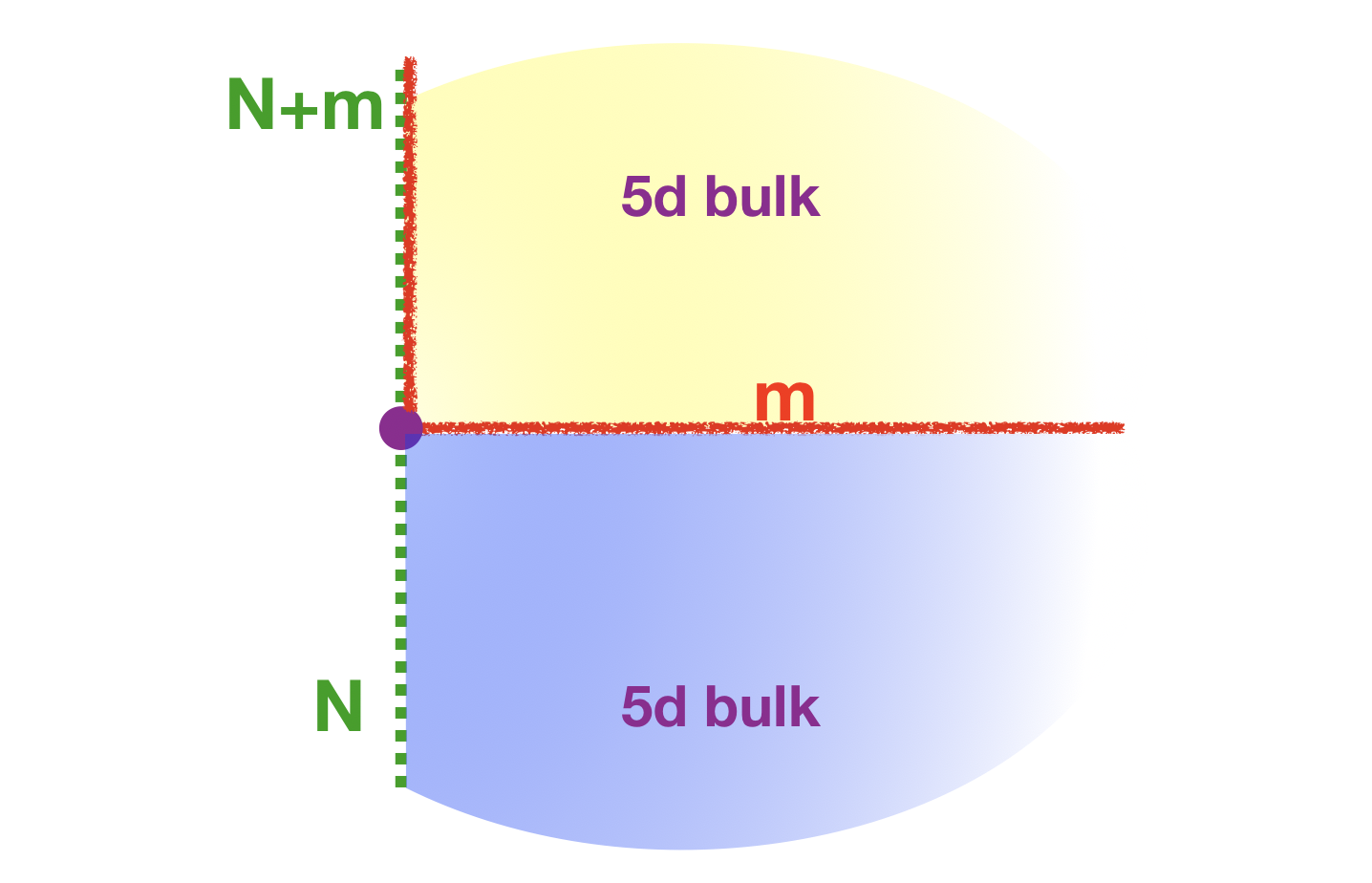}
\caption{\small Structure of 4d relative gravity theories and domain wall separating the two 5d bulk theories to which they couple. The green dashed line corresponds to the two ETW boundaries, with different coefficients for the anomaly of the global symmetries, while the red line describes a 4d domain wall defect in the 5d bulk, carrying 4d dynamical fermions, which reaches the ETW brane and produces the anomaly mismatch.}
\label{fig:cobordism}
\end{center}
\end{figure}

Such domain walls can be used to build the configuration shown in figure \ref{fig:cobordism}, in which the domain wall can be seen to reach the ETW boundary theories. Note that there is continuity of the anomalies across the junction of the two 4d ETW theories and the 4d domain wall in the 5d bulk, as depicted using the coloring of the lines in figure \ref{fig:cobordism}. This kind of configuration allows the two 4d theories to be connected when regarded as relative gravity theories, i.e. when they are completed by their couplings to the 5d bulk gravity (plus gauge) theory. It is interesting that the cobordism becomes possible when the global symmetry which prevented it becomes gauged. 

It is appropriate to emphasize a subtle point. Note that one might think that the 3d boundary of the 4d domain wall can be regarded as a 3d domain wall separating the two 4d gravity theories, hence allowing them to be connected and satisfy the cobordism conjecture in a 4d sense. However this is incorrect, as the corresponding 3d object does not make sense by itself as a domain wall in the 4d gravity theory. Rather, it should be regarded as a {\em relative} defect, which only makes sense as the boundary of a 4d object extending in the extra dimension of the 5d bulk. This is most easily understood in terms of the necessity of an extra direction to maintain the continuity of the anomaly flows. The mechanism is basically analogous to that in holographic realizations of the cobordism conjecture in \cite{Ooguri:2020sua}. 

The generalizations to other symmetries and field contents is clear. We skip them and move on to the realization of the above example in a top-down string theory model.

\subsubsection{String theory embedding of Global Anomaly Mismatch}
\label{sec:anomaly-jumps-td}

In this section we construct a string theory embedding of the example in the previous section using the top-down configuration of ETW branes in AdS$_5\times \IS^5$ in section \ref{sec:d3-ns5-d5}. In order to reproduce the global symmetry and its anomalies due to dynamical 4d chiral fermions, we resort to the introduction of additional branes, in particular $n$ D9-$\aD9$ brane pairs as discussed in section \ref{sec:anomaly-td}. Although this setup is non-supersymmetric and unstable, it is a simple toy model which suffices to illustrate the topological aspects in a complete top-down construction. It is easy to carry out a similar construction in the more involved but fully supersymmetric setup with D7-branes in appendix \ref{sec:anomaly-d7s}.

We thus consider the configuration in section \ref{sec:anomaly-td}. In the 10d brane construction we have a stack of $N$ D3-branes ending on a configuration of NS5- and D5-branes, with $n$ spacetime filling D9-$\aD9$ pairs introducing an $U(n)\times U(n)'$ global symmetry from the D3-brane worldvolume perspective. The gravity dual picture corresponds to an asymptotic AdS$_5\times\IS^5$ region ending on a ETW configuration which provides a string theory embedding of an AdS$_4$ KR ETW brane, and there are $n$ spacetime filling D9-$\aD9$ pairs introducing an $U(n)\times U(n)'$ gauge symmetry. The  AdS$_4$ ETW brane supports $N$ 4d dynamical fermions transforming in the $(\fund,1)+(1,\antifund')$ of $U(n)\times U(n)'$, whose anomaly is canceled by the inflow (\ref{5d-cs}) from the 5d bulk Chern-Simons coupling. For $n=1$ we recover the $U(1)$ symmetry picture in the previous section, for $k=N$. 

We would now like to introduce a domain wall in the bulk theory, separating two regions with different values of the Chern-Simons level, $k=N$, $k=N+m$. The level of the Chern-Simons coupling is given by the 5-form flux on the $\IS^5$, therefore the necessary domain wall is a stack of $m$ D3-branes. In the following we consider $m\ll N$, so that we may introduce them as probes. This also allows to check that the D3-D9 and D3-$\aD9$ open string spectrum produces a set of $m$ 4d chiral fermions in the  $(\fund,1)+(1,\antifund')$ of $U(n)\times U(n)'$, whose anomaly precisely matches the net inflow from the Chern-Simons terms on both sides of the domain wall. 

\begin{figure}[htb]
\begin{center}
\includegraphics[scale=.25]{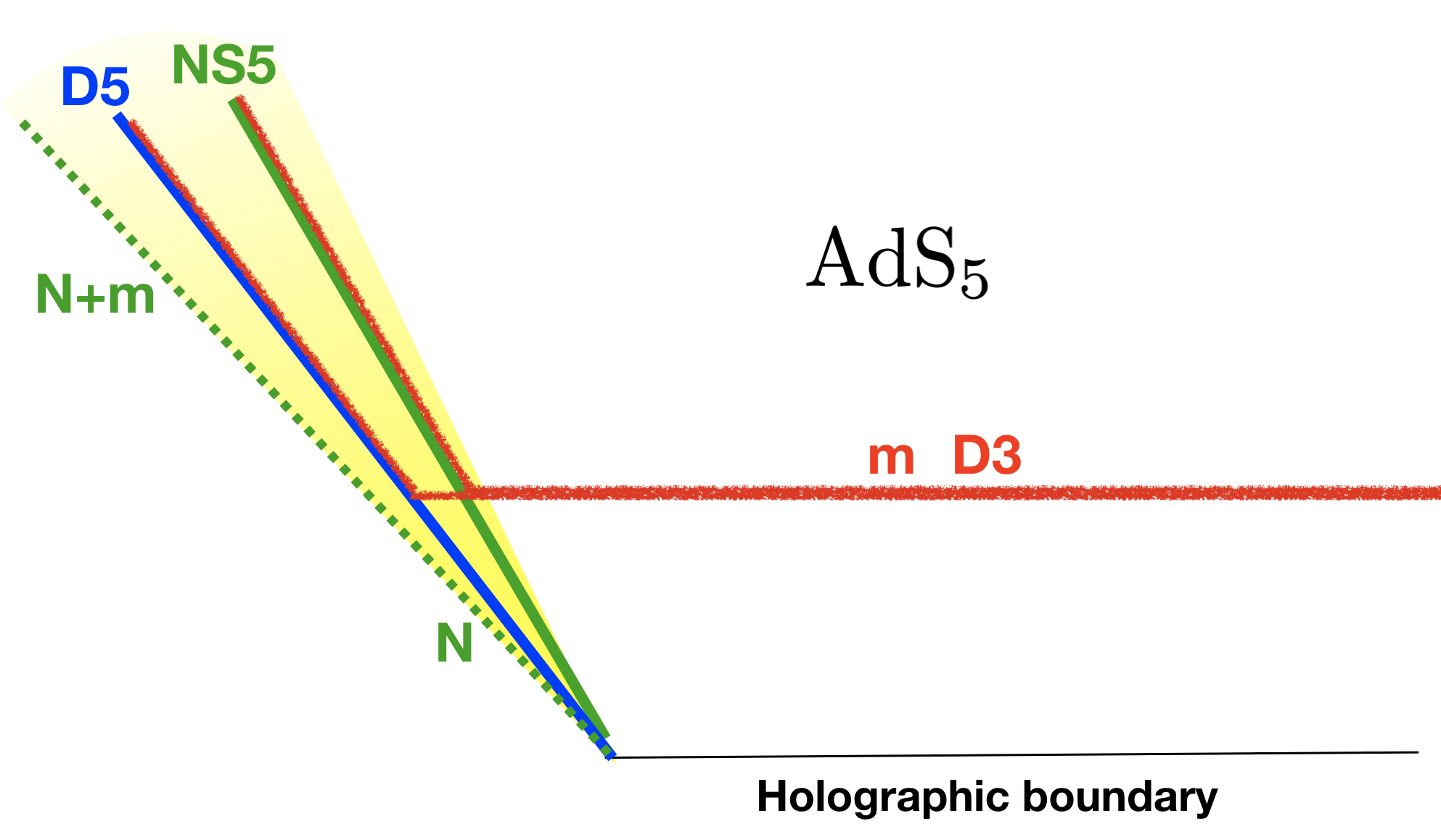}
\caption{\small The top-down string version of 4d relative gravity theories with a D3-brane domain wall (red line) separating the two 5d bulk theories to which they couple. The D3-branes reach the 5-branes (green and blue lines) in the ETW configuration, and turn into induced D3-brane charge in their worldvolumes. The 4d fermions on the D3-branes explain the mismatch in anomaly coefficients in the two ETW regions.}
\label{fig:cobordism-ads}
\end{center}
\end{figure}

We must now describe the configuration in which the corresponding domain wall reaches the ETW 4d boundary. This is obtained by a simple adjoint Higgsing from the holographic gauge theory perspective. Namely, in the flat space 10d brane configuration we consider $m$ D3-branes slightly separated from the original stack of $N$, corresponding to a Higgsing $SU(N+m)\to SU(N)\times SU(m)$. Note that, since the D3-branes are ending on the 5-brane configuration, this motion is partially restricted because the motion of the $m$ branes must be along some subset of the 5-branes, e.g. moving them along the NS5-branes in 456 or along the D5-branes in 789 (in general, we can allow for a partition of $m$ among the 5-branes and move the D3-branes at some common distance in the corresponding directions). In the gravity dual, adjoint Higgsing corresponds to locating the D3-branes at a fixed position away from the holographic boundary of AdS$_5$, in its foliation in 4d Minkowski slices (namely, as a Randall-Sundrum domain wall brane), see figure \ref{fig:cobordism-ads}. These D3-branes reach the AdS$_4$ ETW configuration and end on the 5-branes inside it according to the partition of $m$ mentioned above. This modifies the linking numbers of the 5-branes (i.e. the worldvolume monopole charges), hence there is a jump in the induced D3-brane charges on both sides of the domain wall, i.e. in the two regions of the AdS$_4$ ETW brane closer or further the holographic boundary. Since the induced D3-brane charges determine the number of dynamical fermions localized on the ETW brane, c.f. section \ref{sec:anomaly-td}, this precisely reproduces the cobordism between the two 4d ETW theories, made possible by their coupling to the 5d bulk gravity theory.

The top-down embedding sheds light on the subtle point mentioned in the previous section, i.e. the nature of the 3d defect separating the two 4d ETW theories as a relative defect. In the string theory realization, if the 4d ETW theories are taken as decoupled from the 5d AdS$_5$ bulk theory, it would be impossible to have the 3d defect we have considered. Indeed, the decoupling from the AdS$_5$ corresponds to considering the AdS$_4\times\IX_6$ theory with compact $\IX_6$, i.e. by removing the AdS$_5\times \IS^5$ spike. In such configuration it is not possible to introduce a 3d domain wall in AdS$_4$ corresponding to an isolated boundary of the stack of $m$ D3-branes. This is because such D3-branes must extend in some extra direction, necessarily inside $\IX_6$, which is a compact geometry, so there is no room for the D3-branes to escape away. The coupling to the 5d bulk theory precisely introduces a  non-compact spike in the geometry of $\IX_6$, through which the D3-branes can escape to infinity, hence allowing the  stack of $m$ D3-branes to have just one boundary, ending on the ETW brane (or rather, its 5-branes). This beautifully illustrates the correlation between the coupling to higher-dimensional gravity and the validity of the cobordism conjecture for relative gravity theories.

The fact that the non-compactness of $\IX_6$ allows to consider new objects, relative defects, which are otherwise not possible if $\IX_6$ would be compact, has interesting implications for the completeness conjecture, which we study next.

\section{Gauge symmetries and Swampland Constraints}
\label{sec:gauge-swampland}

In this section we discuss conjectures for theories of gravity coupled to gauge symmetries, in the context of relative gravity setups. As seen in section \ref{sec:no-global-symmetries}, bulk gauge symmetries are observed as effective global symmetries from the perspective of the lower-dimensional theory. Therefore we focus on setups having gauge symmetries localized on the $d$-dimensional gravity theory and not living in the bulk. This has important implications on how some of the most familiar swampland constraints apply to relative gravity theories, in particular the completeness conjecture and the weak gravity conjecture. We study them in turn.

\subsection{The Completeness Conjecture}
\label{sec:completeness}

The completeness conjecture \cite{Polchinski:2003bq} states that in a consistent quantum gravity theory for every gauge field there must exist all possible electric and magnetic sources compatible with Dirac quantization condition. Morally, even if the theory would not have fundamental degrees of freedom charged under the gauge fields, it should contain black holes carrying those charges.

Although the conjecture is expected to hold for general higher form symmetries, we will restrict our discussion to standard 1-form gauge fields, in particular to $U(1)$ symmetries. In this section we explore setups in which the completeness constraint can be violated on a $d$-dimensional gravity theory, but with the missing states arising when the theory is completed by coupling it to a $(d+1)$-dimensional bulk gravity theory.

We need to discuss the $d$-dimensional gravity theory coupled to $d$-dimensional $U(1)$ gauge symmetries, and ultimately embed this system as an ETW brane in a bulk $(d+1)$-dimensional gravity theory. From our discussion in section \ref{sec:no-global-symmetries}, the $d$-dimensional gauge symmetry cannot arise from a bulk $(d+1)$-dimensional gauge symmetry, because the latter behave as effectively global symmetries from the $d$-dimensional perspective. Therefore, we have to consider gauge symmetries which are localized on the $d$-dimensional theory and which do not propagate in the bulk.

One relevant observation in the context of the Karch-Randall or Randall-Sundrum bottom-up setups  reviewed in section \ref{sec:double-holography}, is that the black holes for the $d$-dimensional gravity theory are built from black hole solutions of the bulk $(d+1)$-dimensional theory, see appendix \ref{sec:bhs}. This implies that this class of black holes is automatically uncharged under the localized $d$-dimensional gauge symmetries. Therefore, it is not possible to use them to directly argue that the $d$-dimensional theory must necessarily satisfy the completeness constraint. In fact, we will show that 
for certain gauge symmetries electrically and magnetically charged states may not exist as genuine $d$-dimensional objects, but only as relative ones, namely, as endpoints of extended objects stretched in the extra dimension in the bulk $(d+1)$-dimensional theory. This generalizes the discussion of relative domain wall defects in section \ref{sec:cobordism}.

Let us make this discussion concrete using the top-down string construction of AdS$_5\times\IS^5$ with 5-brane ETW configurations in section \ref{sec:d3-ns5-d5}, but in a language adapted to the bottom-up interpretation. In the string theory setup, the stacks of NS5- and D5-branes in the ETW configuration carry worldvolume gauge symmetries, which are therefore localized and do not extend to the AdS$_5\times \IS^5$ bulk. For simplicity let us focus on the illustrative class of models in section \ref{sec:KSU}, with one 5-brane stack of each kind. Also, even though the gauge symmetry on each stack of $N_5$ 5-branes is $U(N_5)$, we just focus on the center of mass $U(1)$.

Let us focus on the gauge symmetry on the D5-branes, and on electric charges (magnetic charges are discussed later on). In the string theory construction, the only states charged under the worldvolume gauge symmetry of the D5-brane stack are fundamental strings with endpoints on the D5-branes, but these are in the adjoint representation, hence neutral under the center of mass $U(1)$. Hence there are no states charged under the $U(1)$ in the 4d theory\footnote{Actually, it is possible to have a charge $n$ state by considering a baryonic D5-brane\cite{Witten:1998xy} wrapped on an $\IS^5$ with $n$ units of flux in the ETW configuration, so it emits $n$ fundamental strings, which can end on the D5-brane stack. The flux jumps in crossing NS5- and D5-brane stacks, but in our configurations, the only non-trivial flux occurs for $r> k,l$, and is $n=N=2N_5P$. In the regime of interest to get a localized light graviton, $N_5$ is large, so the only charged states from this mechanism carry very large charge, and presumably do not suffice to satisfy the completeness constraint.\label{foot:baryon1}}. On the other hand, when it is completed by coupling to the 5d theory, it is possible to consider fundamental strings which escape in the extra dimension, so that they have only one endpoint on the D5-branes, and are charged under the $U(1)$. These states therefore allow to satisfy the completeness conjecture in the full theory, even though it seems violated in the 4d gravity theory before completing it by coupling to the 5d bulk.

A similar argument applies to gauge fields on the worldvolume of NS5-branes, where electric charges arise from the endpoint of D1-branes. Also, we note that a similar argument applies in the general case in which there are several stacks of $n_a$ NS5-branes and $m_b$ D5-branes. Focusing on the latter (a similar argument holds for NS5-branes), the AdS$_4$ ETW brane has a localized gauge group with several factors $U(m_b)$, and the open strings stretching among them give rise to states in bifundamental representations $(\fund_b,\antifund_{b'})$, carrying charges $+1$ and $-1$ under the corresponding $U(1)\subset U(m_b)$ and $U(1)'\subset U(m_{b'})$. These states are localized in the ETW configuration, but do not suffice to satisfy the completeness constraint, since all those states are uncharged under the diagonal ``center of mass'' linear combination of the $U(1)$'s. Obtaining electrically charged states for the latter requires fundamental strings with only one endpoint on one, at most one of the D5-brane stacks, so they must extend in the extra dimension of the 5d bulk. 

Notice that the existence of the extra states required to satisfy the completeness constraint in the localized theory, regarded as a relative gravity theory, is ultimately linked to the existence of extended objects stretched in the extra dimension. One interesting observation is that the existence of the latter of required by the completeness constraint of the bulk theory! In the above string theory setup, the existence of electric charges under the 4d $U(1)$ gauge field $A_1$ is linked to the existence of the strings in the 5d bulk theory, which are required to exist by the completeness conjecture for the NSNS 2-form field $B_2$, for which they provide electric charged sources. The relation between both is transparent from the familiar mixing of both fields in the gauge transformation $B_2\to B_2+d\Lambda_1$, $A_1\to A_1-\Lambda_1$. We expect that this link between the completeness conjecture in relative gravity theories and in their higher-dimensional completions will hold in other realizations of localized gravity theories with brane-localized gauge dynamics.

We would like emphasize an important point. We have seen that some sets of charges may not exist in the $d$-dimensional theory and only arise the completion by coupling to a $(d+1)$-dimensional theory. However, this does not imply that all sets of charges should behave in this way. This is illustrated by the discussion of magnetic charges under the $U(1)$'s on 5-branes at the ETW configuration of AdS$_5\times\IS^5$, which, as we now argue, satisfy the completeness constraint directly in the ETW configuration, without having to resort to the completion by coupling to the bulk.

Let us again focus on the D5-branes (NS5-branes work similarly) in an ETW configuration for AdS$_5\times\IS^5$, for simplicity with only one stack of each kind. For the worldvolume center of mass $U(1)$ gauge field on the D5-branes, the magnetic charges correspond to the boundary of D3-branes ending on the D5-branes. Since the D5-branes span an AdS$_4\times\IS^2$, the magnetic monopole of the 4d $U(1)$ after reduction on the $\IS^2$ is given by a D3-brane wrapped on the $\IS^2$ and ending on the D5-branes. One may be tempted to think that these D3-branes must extend infinitely in the extra dimension of AdS$_5$, so that the corresponding state exists only in the full theory after coupling to the AdS$_5$ bulk. However, it is clear that there is no non-trivial $\IS^2$ in the $\IS^5$ of the asymptotic AdS$_5\times\IS^5$ bulk for the D3-brane to wrap. In other words, although the $\IS^2$ is non-trivial on the D5-brane worldvolume, it is trivial in the $\IS^5$. So the actual monopole is given by a D3-brane wrapped on a 3-chain $D_3$ whose boundary is the $\IS^2$ on which the D5-brane wraps, $\partial D_3=\IS^2$. Now the crucial point is that the triviality of $\IS^2$ in $\IS^5$ already takes place in the ETW configuration, as is clear in Figure \ref{fig:quadrant}, without having to extend to the asymptotic AdS$_5\times\IS^5$ region. In fact, it is energetically preferred for the D3-brane to wrap a minimal volume 3-chain, which corresponds to the hemisphere filling the interior of the $\IS^2$ in the $\IS^5$ at the position $r$ of the D5-brane. Namely, fibering the $\IS^2$ over the coordinate $\varphi$, as given by only the second equations in (\ref{s5-split}) (or only the first, for the case of NS5-branes). Hence, magnetic charges for the D5- or NS5-brane worldvolume $U(1)$'s are given by D3-branes on 3-chains localized on the ETW configuration. The completeness constraint for magnetic charges is satisfied by the localized gravity theory on its own, without need to invoke its completion by coupling to the higher-dimensional theory. 

Let us finally add that similar discussions may apply to other generalized $p$-form gauge symmetries, and the corresponding extended charged objects. An example is provided by the bulk D3-brane defects in the string theory setup in section \ref{sec:anomaly-jumps-td}, used to define a domain wall defect between two different 4d gravity theories. The 3d boundary of the D3-branes in the AdS$_4$ spanned by e.g. a D5-brane stack is electrically charged under a D5-brane worldvolume 3-form (6d dual to the worldvolume 1-form gauge field), such that it acts as domain wall for the worldvolume flux 
\beqa
f_0=\int_{\IS^2} {\cal F}_2\quad ,\quad df_0= \delta_1(\partial {\rm D3})\, ,
\eeqa 
where $\delta_1(\partial {\rm D3})$ is a bump 1-form Poincar\'e dual (in 4d) to the domain wall. The above flux actually corresponds to the monopole charge on the D5-brane worldvolume (i.e. the linking number), or the induced D3-brane charge on the D5-brane, and its jump explains the difference in anomalies for global theories between the two AdS$_4$ gravity theories, c.f. section \ref{sec:anomaly-jumps-td}.

We hope this discussion suffices to illustrate the role of relative defects (and the situations in which they are not needed) in satisfying the completeness conjecture in relative gravity theories, and refrain from entering the discussion of further examples.

\subsection{Weak Gravity Conjecture}
\label{sec:weak-gravity}

In this section we consider the weak gravity conjecture (WGC) \cite{Arkani-Hamed:2006emk} (see also \cite{Cheung:2014vva,Rudelius:2015xta,Montero:2015ofa,Heidenreich:2015nta,Heidenreich:2016aqi,Lust:2017wrl,Heidenreich:2019zkl,Cordova:2022rer,Etheredge:2022opl} for related developments and \cite{Harlow:2022ich} for a review) in relative gravity theories. There are several formulations, but we focus on the most popular, which requires that for each $U(1)$ gauge symmetry with coupling $g$ there must exist a charge $q$ particle with mass $m$ obeying the bound
\beqa
m< g\,Q M_p\, .
\label{wgc-bound}
\eeqa
For multiple $U(1)$'s there are convex hull formulations\cite{Cheung:2014vva}, but for simplicity we will not consider these generalizations.

One first observation, building on the results of the previous section, is that, since the completeness conjecture can be effectively violated in the $d$-dimensional gravity theory, the same follows for the weak gravity conjecture. In particular, let us focus on the top-down string theory 5-brane ETW configurations for AdS$_5\times\IS^5$ in section \ref{sec:d3-ns5-d5-dual}. As discussed in section \ref{sec:completeness}, there is a center of mass $U(1)$ on the 5-branes for which there are no genuinely 4d charged states (see later, footnote \ref{foot:baryon2}, for the baryonic D5 state), and hence the weak gravity conjecture is violated. We also argued that including the coupling to the AdS$_5(\times\IS^5$) bulk, there are charged states from relative defects, namely open strings with one endpoint on the 5-branes and stretched in the extra dimension, which provide the missing states for the completeness conjecture. In the following we check that these states also satisfy the weak gravity bound, hence reconciling the 4d gravity theory (regarded as a relative one) with the weak gravity conjecture.

Since the object we want to study is extended in the extra dimension i.e. it lives in 5d, but the gauge symmetry is 4d, testing (\ref{wgc-bound}) requires some care. The simplest way to proceed is to consider we introduce a cutoff in the extra dimension\footnote{This can be taken to be a merely formal cutoff, or a physical one e.g. in the form of a second ETW brane, a setup related to wedge holography \cite{Akal:2020wfl,VanRaamsdonk:2021duo}.}, and to effectively reduce the discussion to 4d, hence using a 4d Planck scale and a 4d mass for the extended state. 
For our purposes it will suffice to study the scaling of the different quantities with the length scale $L_4$ of AdS$_4$ and of the internal space $\IX_6$ (and possibly with the length scale $L_5$ of AdS$_5$ and $\IS^5$). For simplicity we work in $M_{p,10}=1$ units, and assume an order 1 value for the string coupling, so we also have $M_s\sim 1$. Then the 4d Planck scale goes as $M_p\sim L_4^{\,3}$, from compactification over the 6d space $\IX_6$; and $g\sim L_4^{-1}$, from compactification of the 6d 5-brane gauge interaction over $\IS^2\subset \IX_6$. Finally, since we are working in the regime $L_4\gg L_5$, the length of the open string stretched in the extra dimension, corresponding to the $q=1$ charged state, is dominated by $L_4$, so $m\sim L_4$. This is parametrically smaller than the combination $gq \,M_p \sim L_4^2$, so, as anticipated, the state allows to satisfy the weak gravity conjecture in the full theory\footnote{Actually, as discussed in footnote \ref{foot:baryon1}, there is a charge $q=N$ particle already present in the ETW configuration, arising from the baryonic D5-brane wrapped on $\IS^5$. Barring that its enormous charge may not allow it to avoid the black hole evaporation arguments underlying the WGC, one can show that it violates the weak gravity bound (\ref{wgc-bound}). Indeed, using the scalings with $L_4$ just discussed, the gauge coupling and 4d Planck mass scale as $g\sim L_4^{-1}$, $M_p\sim L_4^{\, 3}$. Also, for the wrapped D5-brane state, we have a charge $q=N\sim N_5\sim L_4^2$, where the last estimations holds for the model in section \ref{sec:KSU}, see (\ref{size-x6}). We hence have $qgM_p\sim L_4^{\, 4}$.  Its mass scales as $m\sim L_4^{\,5}$ from the $\IS^5$ volume (as the mass contribution of the $N$ attached strings is subleading). Hence $m\gg qgM_p$, as claimed above.\label{foot:baryon2}}.

Actually, there are ongoing developments about the precise formulation of the weak gravity conjecture (see e.g. \cite{Heidenreich:2015nta,Heidenreich:2016aqi,Palti:2017elp,Lust:2017wrl,Heidenreich:2019zkl,Etheredge:2022opl}), in particular in non asymptotically flat spacetimes such as AdS, in which it is presumably more subtle than (\ref{wgc-bound}). In particular it has been proposed that, in the spirit of the repulsive force conditions \cite{Palti:2017elp,Heidenreich:2019zkl}, the WGC in AdS should be formulated as a positive binding conjecture \cite{Aharony:2021mpc}. As proposed there, the corresponding condition admits a simple translation into the dual CFT, in the form of the CFT charge convexity conjecture \cite{Aharony:2021mpc}. Schematically, in the abelian case, it posits that the conformal dimension of the lowest dimension CFT operator with charge $q$ under a $U(1)$ global symmetry is a convex function of $q$. Although there exists a counterexample for the validity for general CFTs \cite{Sharon:2023drx}, it remains a solid proposal at least for CFTs admitting a gravity dual, which suffices to imply the WGC in gravity theories.

We can try to apply this strategy to our setup, and explore the WGC in relative gravity theories in AdS setups from the perspective of the holographic dual CFT realization. In other words, we consider AdS$_d$ gravity theories and explore to what extent they can violate the WGC, but such that it can be satisfied by coupling it as an ETW boundary to an AdS$_{d+1}$ gravity theory. One may be tempted to claim that the above explained charge convexity CFT conjecture \cite{Aharony:2021mpc}, imply, when applied to the BCFT$_{d-1}$, that the WGC automatically holds for the AdS$_d$ theory. However, the result relies on the crucial assumptions that the CFT should be local (conserved energy-momentum tensor) and unitary, and this is not satisfied by the BCFT$_{d-1}$, because its coupling to the CFT$_d$. Only the combined CFT$_d$/BCFT$_{d-1}$ is local and unitary, so we can only expect the WGC to hold for this combined system.

This fits perfectly with the relative quantum gravity picture. The BCFT$_{d-1}$ has in principle the potential to violate the charge convexity conjecture, which means that the gravity theory in AdS$_d$ may violate the WGC. However, the combined BCFT$_{d-1}$/CFT$_d$ system should satisfy the charge convexity conjecture (assuming it holds similarly for CFTs with boundaries). This implies that the AdS$_d$ gravity theory can be completed by coupling it to the higher-dimensional bulk AdS$_{d+1}$ gravity theory, such that the combined system satisfies the WGC in the familiar way.

It would be interesting to find explicit realizations of these ideas, to improve the qualitative estimates provided at the beginning of this section. We leave this interesting question for future work.

\section{Distance Conjectures}
\label{distance-swampland}

In this section we explore the applicability of distance conjectures to relative gravity theories. These posit the appearance of infinite towers of states becoming light, and characterized their rate when some parameter (continuous or discrete) is taken to infinity. We study the AdS and the ordinary distance conjectures in turn.

\subsection{AdS Distance Conjecture and Scale Separation}
\label{sec:ads-distance}

In this section we explore the realization of the AdS distance conjecture in relative gravity theories. The AdS Distance Conjecture \cite{Lust:2019zwm}states that, in a quantum gravity in an AdS$_d$ spacetime with cosmological constant $\Lambda$, in the flat space limit $\Lambda\to 0$ there appears an infinite tower of states with mass scale $m$ becoming light as 
\beqa 
m\sim |\Lambda|^\alpha\, ,
\eeqa
(in Planck units), with $\alpha$ a positive order 1 coefficient. The strong version of the conjecture is that $\alpha=1/2$ for supersymmetric AdS vacua. 

A concept related to the strong AdS distance conjecture is the no scale separation property (see \cite{Coudarchet:2023mfs} for a review), which is the statement that in compactifications leading to AdS vacua for the non-compact dimensions,  it is not possible to make the internal and non-compact length scales parametrically separated. If correct, it would mean that AdS vacua would not correspond to actual compactifications, as the KK cutoff is parametrically related to the AdS cosmological constant. We however note that there are several explicit classes of supersymmetric AdS vacua \cite{DeWolfe:2005uu,Camara:2005dc} 
displaying scale separation (see also \cite{Buratti:2020kda,Farakos:2020phe,Cribiori:2021djm,Apers:2022zjx,Shiu:2022oti,Carrasco:2023hta} for discussions of scale separation in these and other models, and \cite{Coudarchet:2023mfs} for a review), whose existence has passed non-trivial tests, see e.g. \cite{Marchesano:2020qvg,Montero:2024qtz}.

In order to study the AdS distance conjecture in relative gravity theories, we consider setups of localized gravity in an AdS$_d$ brane in a higher-dimensional spacetime, specifically embedded as a KR ETW brane inside an AdS$_{d+1}$ bulk, as reviewed in section \ref{sec:ads}. We study the bottom-up and top-down perspectives in turn.

\subsubsection{Bottom-up perspective with AdS Karch-Randall branes}
\label{sec:ads-distance-bu}

Let us start with some considerations about the AdS distance conjecture limit from the perspective of the bottom-up Karch-Randall construction in section \ref{sec:double-holography}. The relevant observation is that in this setup it is possible to change the AdS$_d$ cosmological constant by changing the brane tension, which in the bottom-up description is a continuous parameter \footnote{As discussed in the next section, in top-down string theory constructions it is determined by discrete parameters in the construction, but is still tunable in a discrete way. The conclusions are therefore essentially unchanged.}, while one maintains the $(d+1)$-dimensional bulk cosmological constant fixed\footnote{It is possible to consider mixed limits, in which both the AdS$_d$ and the AdS$_{d+1}$ cosmological constants are scaled to zero. It is tantalizing to speculate that there may be a convex hull formulation of the AdS distance conjecture, controlling the rate at which combinations of individual towers become light as one takes the limit along different trajectories. We leave this interesting question for future work.}. In fact, the possibility to tune the cosmological constant of the localized gravity theory has been explored in the so-called extended thermodynamics of quantum (i.e. brane-localized) black holes \cite{Frassino:2022zaz} (see \cite{Panella:2024sor} for review).
We can thus consider the flat space limit of the AdS$_d$ gravity theory by the above tuning by simply considering the KR ETW brane to cutoff the AdS$_{d+1}$ spacetime at an AdS$_d$ slice closer and closer to the holographic boundary, as discussed in section \ref{sec:actual-double}. 

Hence the flat space limit of the AdS$_d$ theory corresponds to removing the infrared cutoff in the gravitational bulk AdS$_{d+1}$ theory. One may be confused by the fact that, equivalently, this corresponds to the removal of the UV cutoff of the CFT$_d$ on AdS$_d$, which would seems contrary to the AdS distance conjecture statement that there is an infinite tower of states becoming light in the limit and lowering its UV cutoff. Actually, there is no contradiction: the removal of the CFT$_d$ UV cutoff in the limit in the intermediate picture corresponds via the UV/IR holographic correspondence to the fact that the CFT$_d$ reconstructs more and more of the AdS$_{d+1}$ spacetime geometry as one removes the IR cutoff in the bulk picture. But this clearly has to do with the non-compact $(d+1)$-dimensional gravitational AdS$_{d+1}$ modes of the theory, and these need not be (and in general are not) those in the AdS$_d$ distance conjecture tower. Understanding the latter requires a UV completion of the bulk picture, which we discuss in string theory setups in the next section.

\subsubsection{Top-down view from explicit string theory ETW configurations}
\label{sec:ads-distance-td}

In this section we consider a string theory top-down analysis of the AdS distance conjecture tower of states arises in the flat space limit, using the 5-brane ETW configurations of AdS$_5\times\IS^5$ studied in section \ref{sec:d3-ns5-d5-dual}. In particular, we consider the class of examples  presented in section \ref{sec:KSU}, where the ETW configuration includes one 5-brane stack of each kind, as explained there.

The geometry is determined by the two harmonic functions (\ref{harm.func}), which we repeat for convenience
\beqa
h_1&=&4r\sin \varphi +2{\rm d}\log\left(\frac{r^2+\delta^2+2r\delta\sin\varphi}{r^2+\delta^2-2r\delta\sin\varphi}\right)\, ,\\
h_2&=&-4r\cos \varphi -2{\rm d}\log\left(\frac{r^2+\delta^2+2r\delta\cos\varphi}{r^2+\delta^2-2r\delta\cos\varphi}\right)\, ,
\label{harm.func1}
\eeqa
where recall that ${\rm d}=N_5/32\pi^2$, with $N_5$ the number of 5-branes at each stack, and $\delta=P/32\pi$ is the  position of these two stacks in the coordinate $r$ in the quadrant c.f. figure \ref{fig:quadrant}. The number of D3-branes is $N=2N_5P$. Using these harmonic functions we can find the 10d metric, see appendix \ref{sec:string-embedding} for details. 

We are now interested in using this family of solutions to explore the AdS distance conjecture, which arises in the  flat space limit of the AdS$_4$ ETW configuration. In order to analyze it in detail, it is useful to distinguish two interesting regions of the spacetime, the asymptotic AdS$_5(\times\IS^5)$ throat and the ETW configuration. The former arises in the $r\to\infty$ limit, where the metric becomes:
\beq
ds^2=\frac{2\sqrt{2}\,r^2}{\sqrt{\delta {\rm d}}}\, ds_{AdS_4}^2+16\sqrt{2}\sqrt{\delta {\rm d}}\, \frac{dr^2}{r^2}+16\sqrt{2}\sqrt{\delta {\rm d}}\, ds_{S^5}^2\, .
\label{metric1}
\eeq
Making the change $r=\sqrt{\delta {\rm d}}\, e^x$, we get
\beq
ds^2=16\sqrt{2}\sqrt{\delta {\rm d}}\,\left[ \frac{e^{2x}}{4}\, ds_{AdS_4}^2+ dx^2+ ds_{S^5}^2\right]\, ,
\eeq
which is the metric of AdS$_5\times \IS^5$ in Poincar\'e coordinates in the $x\to\infty$ limit, with length scale $L_5\sim (\delta {\rm d})^{1/4}$, so we indeed see that $N\sim L_5^4\sim \delta {\rm d}\sim N_5P$.

To study the dynamics of the ETW configuration, we drop the first term in both harmonic functions \eqref{harm.func1}, and keep only the logarithm (see \cite{Huertas:2023syg} for a discussion of this kind of limit). This can be implemented  physically by sending ${\rm d}\to\infty$ or $\delta\to0$, or as discussed in section \ref{sec:KSU}, sending $N_5/P\to\infty$. As explained there, this limit corresponds to the flat space limit of the AdS$_4$. 

In this limit, the neck to the asymptotic AdS$_5\times\IS^5$ pinches off and the  coordinate $r$ becomes effectively compact. The limit effectively isolates the ETW region, which is given by AdS$_4\times \IS^2\times\IS^2$ fibered over ${\ov \Sigma}$, a compact version of the Riemann surface $\Sigma$, obtained by closing off the asymptotic region of the quadrant. The $\IS^2\times\IS^2$ fibration over ${\ov \Sigma}$ defines a compact version of $\IX_6$. We can  explicitly compute the metric from the logarithmic terms in \eqref{harm.func1}, but the resulting expressions are fairly involved. We may get the gist of the result by making the change $r\to\delta\,  r$ in \eqref{ansatz}, so the metric reads:
\beq
ds^2={\rm d}\left[f_4^2(r,\varphi)\, ds_{AdS_4}^2+ \rho^2(r,\varphi)\, ds_{\IX_6}^2\right]\, ,
\eeq
where now the warp factors $f_4$ and $\rho$, which are constructed using the harmonic functions (see \eqref{dilaton} and below), are independent of $\delta$ and ${\rm d}$.

As anticipated above, the limit ${\rm d}\to \infty$ corresponds to the flat space limit of the AdS$_4$. In our setup it is easy to identify the nature of the tower of states becoming light in this limit. Indeed, the above metric shows that in the ETW configuration the characteristic lengths of AdS$_4$ and the compact space $\IX_6$ are parametrically the same:
\beqa 
L_{\IX_6}\sim L_{{\rm AdS}_4}\sim {\rm d}^{1/2}\sim N_5^{1/2}\, .
\label{size-x6}
\eeqa
Namely, there is no scale separation in the ETW configuration.
In the flat space limit, the size of $\IX_6$ grows, and there is a tower of light Kaluza-Klein states with a mass scale $m_{\IX_6}\propto L_{6}^{-1}\sim N_5^{-1/2}$. Hence, the AdS distance conjecture is satisfied, with $m\sim |\Lambda_{{\rm AdS}_4}|^{1/2}$, so in fact in its strong version, as befits a supersymmetric setup. 

Let us conclude the discussion by mentioning that in the above limit, we should be careful to keep the 5-form flux in the asymptotic AdS$_5\times \IS^5$ fixed. In particular, this is important since this flux fixes the value of the AdS$_5$ length scale, and also the $\IS^5$ size, due to absence of scale separation in AdS$_5\times\IS^5$. This implies that if the AdS$_5$ is taken to its flat space limit, there is a further tower of light KK modes with mass scale $m_{\IS^5}\sim  (N_5P)^{-1/4}$. In fact the mass scales of both towers are related by
\beqa 
m_{\IS^5}\sim  m_{\IX_6} \left(\frac{N_5}{P}\right)^{1/4}\, ,
\eeqa 
so in the limit $N_5/P\to \infty$ the tower of KK modes in $\IX_6$ is parametrically lighter than that of the $\IS^5$.
It would perhaps be interesting to explore combined flat limits of both AdS spaces, and possibly develop a convex hull version of the AdS distance conjecture. We leave these questions for future work.

\subsection{Distance Conjecture and CFT Distance Conjecture}
\label{sec:distance-conjecture-cft}

In this section we discuss the Distance Conjecture \cite{Ooguri:2006in} (see also \cite{Klaewer:2016kiy,Grimm:2018ohb,Ooguri:2018wrx,Corvilain:2018lgw,Grimm:2018cpv,Buratti:2018xjt,Marchesano:2019ifh,Lee:2019xtm,Lee:2019wij,Baume:2019sry,Gendler:2020dfp,Calderon-Infante:2020dhm, Buratti:2021fiv,Angius:2022aeq,Etheredge:2023odp,Castellano:2023jjt,Castellano:2023stg} for related developments) in relative gravity theories. The distance conjecture states that when some modulus moves to infinite distance in moduli space, there is an infinite tower of states becoming exponentially light. As is familiar by now, we focus on AdS$_d$ KR ETW boundaries of AdS$_{d+1}$.

In the context of gravity theories in AdS, the conjecture can be recast in terms of the holographic dual CFT \cite{Baume:2020dqd,Perlmutter:2020buo}, see \cite{Baume:2023msm,Calderon-Infante:2024oed} for related recent developments. In short, moduli in AdS correspond to marginal couplings of the CFT, the moduli space in AdS corresponds to the space of marginal couplings, i.e. the conformal manifold of the CFT, the moduli space metric corresponds to the CFT Zamolodchikov metric, and the masses of states in the tower correspond to the conformal dimensions of operators in a tower (rather, the difference of their conformal dimensions with respect to the value given by the unitarity bound). In this language, one can formulate the distance conjecture in terms of the CFT by says that, when one moves a marginal coupling towards a point at infinite distance in the conformal manifold, there is an infinite tower of operators of the CFT whose conformal dimensions approach the unitarity bound exponentially fast. This is part of the content of the CFT distance conjecture as expressed in \cite{Perlmutter:2020buo}, which also includes the statement of a one to one correspondence between infinite distance points in the conformal manifolds and Higher Spin points (theories with an enhanced higher spin symmetry, with infinite towers of operators of increasing spin becoming light, which are associated to emergent string points in the moduli space of the AdS gravity side). The CFT distance conjecture has been extensively studied and tested and partially proven, see e.g. \cite{Baume:2020dqd,Perlmutter:2020buo,Baume:2023msm,Calderon-Infante:2024oed}, also \cite{Kusuki:2024gss} for the 2d CFT case.

We can try to apply this strategy to our setup, and explore the distance conjecture in relative gravity theories in AdS setups from the perspective of the holographic dual CFT realization. In other words, we consider $d$-dimensional gravity theories in AdS$_d$ and explore to what extent they can violate the distance conjecture, albeit in a consistent way thanks to their coupling to a $(d+1)$-dimensional gravity theory in AdS$_{d+1}$. 
Let us explore the distance conjecture in the gravity theory on AdS$_d$, in terms of its dual (B)CFT$_{d-1}$. One may be tempted to claim that the arguments in \cite{Baume:2020dqd,Perlmutter:2020buo,Baume:2023msm,Calderon-Infante:2024oed} supporting the CFT distance conjecture, imply, when applied to the BCFT$_{d-1}$, that the distance conjecture automatically holds for the AdS$_d$ theory. However, the crucial assumptions of the CFT distance conjecture is that the CFT should be local (conserved energy-momentum tensor) and unitary, and this is not satisfied by the BCFT$_{d-1}$, because its coupling to the CFT$_d$. Only the combined CFT$_d$/BCFT$_{d-1}$ is local and unitary, so we can only expect the CFT distance conjecture to hold for this combined system.

This fits perfectly with the relative quantum gravity picture. The BCFT$_{d-1}$ has in principle the potential to violate the CFT distance conjecture, which means that the gravity theory in AdS$_d$ may violate the distance conjecture. However, the combined BCFT$_{d+1}$/CFT$_d$ system is local and unitary, so we expect it to satisfy the CFT distance conjecture (assuming it holds similarly for CFTs with boundaries). This implies that the AdS$_d$ gravity theory can be completed by coupling it to the higher-dimensional bulk AdS$_{d+1}$ gravity theory, such that the combined system satisfies the distance conjecture in the familiar way.

It would be interesting to provide a string theory top-down realization of this proposal, in particular with CFT/BCFT systems with a gravitational dual bulk picture, so as to directly realize it in terms of the usual distance conjecture for infinite distance is moduli spaces of gravity theories. We leave this for future work.

An amusing observation is that the prominent role played by unitarity in the above argument is tantalizingly reminiscent to the role of unitarity in the Page curve in black hole evaporation, as studied using ETW branes in the quantum island literature (see \cite{Almheiri:2020cfm} for a review). It would be interesting to explore these connections further.

\section{Constraints on Vacua}
\label{sec:vacua}

Several swampland constraints deal with the structure of possible vacua in quantum gravity theories, such as the absence of non-supersymmetric stable AdS vacua \cite{Ooguri:2016pdq}, or the de Sitter conjecture \cite{Obied:2018sgi,Garg:2018reu,Ooguri:2018wrx}. In this section we discuss aspects of the applicability of such conjectures to relative gravity theories.

\subsection{No Stable Non-susy AdS Vacua Conjecture}
\label{sec:no-nonsusy-ads}

An implication of the generalization of the weak gravity conjecture to higher-form symmetries is the proposal in \cite{Ooguri:2016pdq} that non-supersymmetric AdS vacua should be unstable in quantum gravity (see \cite{Buratti:2018onj} for a generalization). In string theory AdS vacua supported by fluxes, this often occurs via bubbles of nothing \cite{Ooguri:2017njy}, or nucleation of charged domain wall shells which discharge the fluxes (see e.g. \cite{Maldacena:1998uz,Bena:2020xxb,Marchesano:2021ycx,Casas:2022mnz,Marchesano:2022rpr}). 

In this section we would like to explore some aspects of this proposal in relative gravity theories. In particular we would like to explore if it may be possible for a $d$-dimensional gravity theory to admit non-supersymmetric stable AdS vacua, but such that they become unstable upon coupling to a higher $(d+1)$-dimensional gravity theory, so that the conjecture is obeyed by the $d$-dimensional theory as a relative gravity theory. 

As usual, we will consider the higher-dimensional gravity theory to have an AdS$_{d+1}$ vacua. One may be tempted to think that, given a putative stable non-susy AdS$_d$ vacuum, the natural way to render it unstable upon coupling it to an AdS$_{d+1}$ gravity vacuum is to require that the latter is non-supersymmetric. If so, one could invoke the no stable non-supersymmetric AdS vacuum conjecture for the $(d+1)$-dimensional bulk theory to render it unstable, and with it its embedded AdS$_d$. Our purpose in this section is to show that the situation is far more interesting, and that the whole configuration can be unstable even if the AdS$_{d+1}$ bulk is supersymmetric.

Since the formulation of the conjecture and its analysis deal with supersymmetry, and this is not a property which plays an important role in the bottom-up Karch-Randall constructions in section \ref{sec:double-holography}, we will work in terms of string theory top-down constructions. Actually, we need to consider some variant of this construction, in which the AdS$_4$ ETW configurations are non-supersymmetric while the bulk AdS$_5$ configuration is supersymmetric. In particular, we will consider the configurations in appendix \ref{sec:nonsusy-ads4}, which are based on NS5- and D5-brane ETW configurations of an asymptotic AdS$_5\times\IS^5/\IZ_k$, in the presence of D7-branes. The bulk AdS$_5$ ($\times \IS^5/\IZ_k$) is supersymmetric even after including the D7-branes, whereas the ETW NS5- and D5-branes do not preserve those supersymmetries, see appendix \ref{sec:nonsusy-ads4} for details. Note that in these setups the breaking of supersymmetry is only due to the D7-branes, i.e. it is broken by a small amount, subleading in the numbers of D3- or of 5-branes.

Actually, the orbifold $\IZ_k$ is not essential to obtain the features required for our purposes in this section. It is possible to consider a simpler analogous model in terms of the 5-brane ETW configurations in AdS$_5\times\IS^5$ in section \ref{sec:d3-ns5-d5-dual} with the addition of D7-branes along 01236789 (D7$_1$-branes, in the conventions of appendix \ref{sec:nonsusy-ads4}). The bulk AdS$_5$ ($\times\IS^5$) is supersymmetric even after including the D7-branes, whereas the ETW NS5- and D5-branes do not preserve those supersymmetries.

Let us thus focus on the AdS$_4$ ETW configuration, which is non-supersymmetric, and let us assume for the time being that it is stable (we will revisit this point later on). What we would like to show is that, when it is coupled to the bulk AdS$_5$ configuration, even if the latter is supersymmetric, the combined system develops an instability. 
This is easy to argue, for instance by checking a particular instability channel related to the D5-branes (there may be other decay modes involving the NS5-branes, but they are harder to characterize), as follows. The D5-D7$_1$ open string sector is non-supersymmetric, and has 2 DN+ND boundary conditions. This kind of mixed sector produces a complex tachyon in the bifundamental representation (with string scale negative mass squared, well below the BF bound), which signals an instability against dissolving the D5-branes as a worldvolume gauge flux on the D7-branes  (see e.g. \cite{Gava:1997jt,Loaiza-Brito:2001yer}). In this process, the induced D3-brane charge on the D5-branes also ends up dissolved on the D7-branes. Since the D7-branes are non-compact in the extra direction in AdS$_5$, this decay channel thickens the ETW configuration until the D3-brane charge is fully delocalized in the transverse direction. This effectively discharges the 5-form flux of the AdS$_5\times\IS^5$, and destroys the whole configuration, as in \cite{Ooguri:2016pdq}.

With this improved understanding of the nature of the instability, we may now revisit the original assumption that the initial non-supersymmetric ETW AdS$_4$ configuration was stable, and at least test its stability against this D5-D7 instability. This requires being more precise about the nature of the ETW AdS$_4$ configuration. Recall from section \ref{sec:d3-ns5-d5} that it corresponds to the gravity dual of the BCFT$_3$ with no asymptotic D3-branes, which in this case corresponds to AdS$_4\times\IX_6$, with the compact version of $\IX_6$ with the asymptotic spike producing the asymptotic AdS$_5\times\IS^5$ throat closed off, as in section \ref{sec:ads-distance-td}. Hence, in this case, even if the D5-branes has a local instability against dissolving into the D7-branes, the compactness of the internal space suggests that one may reach some non-trivial stable minimum with a continuous flux distribution. From this perspective, the coupling to the AdS$_5$ bulk theory opens up the spike in the internal space, allowing the dissolution of the flux to proceed indefinitely.

We think that the resulting picture, although qualitative, fits well with the possibility that the no stable non-supersymmetric AdS vacua works for relative gravity theories in the manner described, even when the bulk theory could have been expected to enjoy extra protection due to bulk supersymmetry.

\subsection{De Sitter Conjecture and the Festina Lente Bound}
\label{sec:desitter}

The de Sitter conjecture \cite{Obied:2018sgi,Garg:2018reu,Ooguri:2018wrx} posits that no de Sitter minima exist in quantum gravity theories (more in general, it proposes bounds on the derivatives of the potential, which in particular prevent positive cosmological constant vacua, and constrain rolling solutions). In this section we explore its possible realization in relative gravity theories.

\subsubsection{De Sitter Conjecture}
\label{sec:desitter-etw}

An immediate observation from the bottom-up perspective is that, as reviewed in section \ref{sec:kr}, it is possible to construct dS$_d$ Karch-Randall ETW branes of AdS$_{d+1}$ \cite{Karch:2000ct}, c.f. figure \ref{fig:ETWs}c. To recap, one considers a bulk AdS$_{d+1}$ with a tensionful ETW brane and solves the equations of motion and junction conditions. The resulting bulk solution is simply a region of AdS$_{d+1}$ bounded by an ETW brane (or the simple combination of two such regions glued along a common brane, which then defines a domain wall configuration). The ETW brane geometry depends on the relation of the brane tension to the bulk AdS$_{d+1}$ curvature length scale $L_{d+1}$. For a critical value, one recovers the Randall-Sundrum setup \cite{Randall:1999vf} of $d$-dimensional Minkowski slices. For subcritical values, one recovers the by now familiar AdS KR branes. Finally, for overcritical tension, the AdS$_{d+1}$ is foliated into dS$_d$ slices, and cut off at a dS$_d$ ETW brane, whose location is determined by the deviation with respect to the critical brane tension.

As reviewed in section \ref{sec:kr}, in the dS Karch-Randall case (and in the Randall-Sundrum setup) there is localized graviton on the ETW brane, which is exactly massless (in contrast with the massive but potentially parametrically light graviton in the AdS case). Hence, this bottom-up construction suggests a natural setup to avoid the de Sitter conjecture in gravity theories obtained as dS KR branes. This scenario has in fact appeared in the cosmological model building context in \cite{Muntz:2024joq} (see \cite{Banerjee:2018qey,Banerjee:2019fzz,Banerjee:2020wix} for related setups), as well as from the quantum black hole applications (see \cite{Panella:2024sor} for a review).

The KR construction thus provides  an in principle direct way to build relative gravity theories in de Sitter spacetime. The $d$-dimensional gravity theory admits a dS vacuum, thanks to the fact that it is ultimately embedded into a higher-dimensional AdS spacetime. We however emphasize that this construction is purely bottom-up, but signals a potential direction to avoid the de Sitter conjecture. Hence, a most important question is how to embed this scenario in a microscopically well defined setup, such as string theory. 

Interestingly, although we have fairly precise string theory embeddings of the Randall-Sundrum scenario \cite{Verlinde:1999fy}, as reviewed in section \ref{sec:rs}, and of AdS Karch-Randall branes, reviewed in section \ref{sec:d3-ns5-d5}, there is no known embedding of dS Karch-Randall branes in string theory so far. This may be due to a mere limitation of our ability to build these constructions (possibly because any such solution is necessarily non-supersymmetric). On the other hand, one may regard it as adding up to the difficulties to construct explicit de Sitter vacua in string compactifications in controllable regimes (see \cite{Danielsson:2018ztv} for review and discussions), which form much of the empirical support for the de Sitter conjecture. We leave this important question for future research, but we would like to make an interesting observation regarding a possible tension between the existence of a string theory realization of dS ETW branes and the Festina Lente bound.

\subsubsection{The Festina Lente Bound}
\label{sec:fl}

The study of evaporation of black holes charged under a $U(1)$ gauge symmetry in dS spacetime led in \cite{Montero:2019ekk} to the so-called Festina Lente bound for the mass of any charged particle, of the form 
\beqa
m^2\gtrsim \, q g M_p H\, ,
\eeqa
where $q$ is the charge, $g$ the gauge coupling and $H$ the Hubble scale of the dS. In particular, it implies that there cannot exist exactly massless charged particles in de Sitter spacetime (see e.g. \cite{Montero:2021otb,Guidetti:2022xct} for other discussions of Festina Lente).

In the following we argue that this condition may be in tension with a hypothetical string theory embedding of dS$_4$ (possibly dressed with some internal space) ETW boundaries in AdS$_5(\times\IS^5)$. The argument is essentially topological, and hence may admit generalization to other AdS$_5$ host spaces, or even to some other spacetime dimensions. However, because it is topological, there may be ways out of it which involves non-trivial dynamics, as we also point out.

Let us sketch the argument. Imagine there is a hypothetical string theory embedding of such a dS$_4$ ETW configuration in AdS$_5 \times \IS^5$. This would be some analogue of the AdS$_4$ ETW configurations in section \ref{sec:d3-ns5-d5-dual}, which will serve as inspiration, although we keep a general perspective and do not assume much specific details on the microscopic structure of the dS$_4$ ETW configuration. One fairly robust feature, however, is that, since the asympotic $\IS^5$ has $N$ units of RR 5-form flux, the dS$_4$ ETW region must necessarily contain some $N$ units of D3-brane charge, so that the flux in the $\IS^5$ can discharge. This D3-brane charge may not be explicit, i.e. in general it may be dissolved in other higher-dimensional branes. 

We now consider adding a (small) number $n$ of D9- and anti-D9 brane pairs, just like in section \ref{sec:anomaly-td}. The fact that these pairs may be added is a basic property of the K-theory classification of D-branes in string theory \cite{Witten:1998cd,Srednicki:1998mq}. Then, because of the presence of the $N$ units of RR 5-form flux on the  $\IS^5$ in the 5d bulk, we can repeat the inflow argument in section \ref{sec:anomaly-td}, and derive that on the dS$_4$ ETW boundary there must exist $N$ chiral fermions  in the $(\fund,1)+(1,\antifund)$ of the $U(n)\times U(n)'$ group on the D9-$\aD9$ pairs.

Since these chiral (hence necessarily massless) fermions are charged under the gauge symmetry of the D9-branes, one could think that they lead to a violation of the Festina Lente bound. However, we should recall from section \ref{sec:no-global-symmetries} that bulk gauge symmetries look like global on the ETW branes. Hence there is no direct contradiction. We may however consider that the (explicit or induced) D3-branes in the ETW region presumably do carry some gauge symmetry (as for the 5-branes in the AdS$_4$ ETW case). In such case, this is a gauge symmetry on the ETW boundary, under which the above chiral fermions are charged, now leading to an actual violation of the Festina Lente bound.

This argument is essentially topological and hence fairly robust and may admit interesting generalization. It however has loopholes, which allow to circumvent it, albeit in interesting ways. One possibility is that the initial hypothetical solution describing dS$_4$ ETW branes in AdS$_5\times\IS^5$ is de-stabilized by the addition of the D9-$\aD9$ pairs. This is certainly possible, given that both the initial configuration and the extra ingredients are necessarily non-supersymmetric. On the other hand, the smallness of the number $n$ of D9-$\aD9$ pairs (compared e.g. with $N$, which can be taken as large as necessary) provides parametric control, so it may be possible that the addition may be carried out without a major disaster on the initial configuration. Still in this case, there may be additional dynamical ingredients which could kill the above argument. Indeed, the masslessness of the fermions is protected by chirality only as long as the $U(n)\times U(n)'$ symmetry is unbroken. Recalling that in the D9-$\aD9$ open string sector there is a tachyon in the bifundamental, it is possible that its profile in the actual solution upon including the D9-$\aD9$ pairs includes a non-trivial condensate in the ETW region. The breaking of the symmetry to the diagonal $U(n)$ (or its eventual disappearance due to brane-antibrane annihilation) would make the set of fermions non-chiral, hence allowing them to acquire a mass, which could be above the Festina Lente bound.

Hence, although the above argument is compelling, it is not possible to assess its validity without further progress in understanding better the key ingredients in the microscopic realization of dS ETW configurations in string theory, if at all possible. We hope our considerations provide useful input in this quest.

\section{Conclusions}
\label{sec:conclusions}

In this work we have initiated the first systematic study of the applicability of swampland constraints to theories of localized gravity. We have found that localized gravity theories can violate swampland constraints, and require being coupled to a higher-dimensional gravity theory to satisfy them. We have dubbed {\em relative quantum gravity theories} those gravity theories which, to become consistent at the quantum level (namely, to satisfy the swampland constraints), must be defined as relative to a host higher-dimensional gravity theory.

We have carried out the discussion in the codimension 1 case, both from the bottom-up and the string theory top-down perspectives. In the bottom-up approach, we have mostly focused on AdS$_d$ Karch-Randall ETW branes in an AdS$_{d+1}$ host spacetime. This has allowed a rich discussion of the swampland constraints using the different pictures provided by double holography. In particular we have uncovered remarkable quantitative realizations of the species scale and emergence of gauge dynamics in the brane effective action obtained using holographic renormalization. In the top-down string theory realizations, we have used 10d supergravity solutions describing ETW configurations for AdS$_5\times\IS^5$ using the holographic duals of semi-infinite D3-branes ending on NS5- and D5-branes. These provide an explicit string theory realization of the Karch-Randall AdS ETW branes, providing important additional microscopic ingredients which are crucial to test some of the swampland constraints.

We have discussed aspects of no global symmetry conjecture, the species scale, emergence of gauge interactions, the cobordism conjecture, completeness of the spectrum, the weak gravity conjecture, the AdS and CFT distance conjectures, the no stable non-susy AdS and the de Sitter conjectures. Our results have clarified many aspects of these swampland constraints in the context of localized gravity theories, and explained the precise sense in which they are relative to higher-dimensional ones. Moreover, our work has opened many interesting new questions, some of which are:

$\bullet$ One of the most pressing questions in the string theory realization of these ideas is the construction of embeddings of AdS Karch-Randall branes in setups with scale separation. As we have already suggested, a promising approach could be provided by the introduction of the ETW branes as cobordism defects in scale separated setups such as those in \cite{DeWolfe:2005uu,Camara:2005dc}.

$\bullet$ In a similar top-down spirit, it would be desirable to construct string theory embeddings of dS ETW branes (or domain walls) in AdS bulks, or to derive no-go results of such embeddings. This would serve to either strengthen the scope of the de Sitter conjecture, or to find ways around it for applications to real-world cosmology.

$\bullet$ The use of precision holography can be a powerful tool in the study of swampland constraints in localized gravity theories. In this respect, we expect that precise statements can be made regarding the extension of the CFT distance conjecture to the inclusion of boundaries coupled to BCFTs, and correspondingly in the discussion of the distance conjecture in their dual bulk picture.

$\bullet$ The notion of relative quantum gravity is fairly general, very much in the spirit of the swampland program. In this work we have restricted to its realization in the Karch-Randall (and Randall-Sundrum) approaches (i.e. coupling a bulk gravity theory to a purely tensional brane), and their string theory realizations. It would be interesting to explore the implementation of the idea of relative gravity theories in other setups of gravity localization.

$\bullet$ We have focused on theories of localized gravity in real codimension 1, a context which in fact encompasses most approaches to gravity localization. However, from a broader perspective it would be interesting to explore both bottom-up and top-down realizations of localized gravity and application of swampland constraints in higher codimension setups, as further test of the generality of the ideas pioneered in this work.

We hope to come back to these and other exciting questions in the near future.


\section*{Acknowledgments}

We are pleased to thank Jos\'e Calder\'on-Infante, Matilda Delgado, Karl Landsteiner, Miguel Montero, Juan Pedraza, Ignacio Ruiz and Irene Valenzuela for useful discussions. This work is supported through the grants CEX2020-001007-S, PID2021-123017NB-I00 and  ATR2023-145703 funded by MCIN/AEI/10.13039/501100011033 and by ERDF A way of making Europe. The work by E. A. is supported by the fellowship LCF/BQ/DI24/12070005 from ``La Caixa'' Foundation (ID 100010434). R.A. is supported by the ERC Starting Grant QGuide-101042568 - StG 2021. J. H. is supported by the FPU grant FPU20/01495 from the Spanish Ministry of Education and Universities. C. W. is supported by the grant ATR2023-145703 funded by MCIN/AEI/10.13039/501100011033 and by ERDF A way of making Europe.

\newpage

\appendix

\section{The Brane Action in the Intermediate Picture}
\label{sec:intermediate-action}

In this section we describe the computation of the AdS$_d$ brane action in the intermediate picture of double holography introduced in section \ref{sec:double-holography}. We follow and adapt the discussion from \cite{Panella:2024sor} and references therein. 

In the bulk picture we work with classical dynamical Einstein gravity in AdS$_{d+1}$ minimally coupled with a Maxwell field:
\begin{equation}
    I_{bulk}= \frac{1}{16 \pi G_{d+1}} \int_{\mathcal{M}} d^{d+1}x \sqrt{-G} \left[ R_{d+1} -2 \Lambda_{d+1} - \frac{l^2_{\ast}}{4} F^2\right] - \frac{1}{8 \pi G_{d+1}} \int_{\partial \mathcal{M}} d^dx \sqrt{-h}  K . 
    \label{action:d_dim_bulk}
\end{equation}
to which we add an ETW Karch-Randall brane with tension $\tau$ at a fixed small distance from the AdS$_{d+1}$ boundary:
\begin{equation}
    I_{\tau}= - \tau \int_{\mathcal{B}_{l}} d^dx \sqrt{-h}.
\end{equation}
In this expression $\mathcal{B}_l$ is the AdS$_d$ slice filled by the brane and located at distance $l$ from the AdS$_{d+1}$ boundary (in the foliation in AdS$_d$ slices). $G_{d+1}$ is the $(d+1)$-dimensional Newton's constant, $G_{MN}$ is the bulk metric and $\Lambda_{d+1}= - \frac{d(d-1)}{2 L^2_{d+1}}$ is the cosmological constant, while $h_{\mu \nu}$ is the induced metric on the brane, $K$ is the trace of the extrinsic curvature providing a Gibbons-Hawking York (GHY) boundary term. The parameter $l^2_{\ast}$ encodes the information on the $U(1)$ coupling constant via $l^2_{\ast}=\frac{16 \pi G_{d+1}}{g_{d+1}^2}$. 

In braneworld holography it is well established that the presence of the ETW brane plays the same role of an IR cutoff in the bulk theory necessary to remove divergences of the gravity partition function. This means that we can exploit the computations done in the gravity side (bulk) to regularize and renormalize the action \eqref{action:d_dim_bulk} in order to get information about the theory living in the brane (intermediate picture). The result will be a $d$-dimensional theory with dynamical gravity coupled to the holographic CFT$_d$ endowed with a UV cutoff fixed by the bulk IR cutoff $\epsilon$. The dynamics of the gravity and the gauge field in the brane is induced by integrating out the CFT degrees of freedom above the UV cutoff, we can interpret the process as the emergence of the kinetic terms in the effective action and identify the UV cutoff with the \textit{emergence scale} from the perspective of $d$-dimensional theory, as explained in section \ref{sec:emergence}. 

In the following we report the explicit computations of the regularization of the bulk action \eqref{action:d_dim_bulk} from which we will be able to extract the $d$-dimensional brane action. \\
According to \cite{Fefferman:1985feg}, the bulk AdS$_{d+1}$ spacetime metric can asymptotically be expressed in the Fefferman-Graham gauge as:
\begin{equation}
    ds^2_{d+1} = G_{MN}dx^{M} dx^N = L^2_{d+1} \left( \frac{d \rho^2}{4 \rho^2} + \frac{1}{\rho} g_{\mu \nu} (x, \rho) dx^{\mu} dx^{\nu} \right)\, ,
\end{equation}
where the conformal boundary is located at the position $\rho=0$. The length scale $L_{d+1}$ is related to the cosmological constant as $\Lambda_{d+1}=- \frac{d(d-1)}{2L^2_{d+1}}$. For $d$ odd, the $d$-dimensional metric $g_{\mu \nu}$ admits the following expansion in the radial coordinate:
\begin{equation}
     g_{\mu \nu} = g_{\mu \nu}^{(0)} (x) + \rho g_{\mu \nu}^{(2)} (x) + \rho^2 g_{\mu \nu}^{(4)} (x) + ... \, ,
     \label{g:perturbative_exp_odd}
\end{equation}
where all the $g^{(k)}$ are covariant tensors with respect to $g^{(0)}$. For $d$ even, a logarithmic term emerges at order $d/2$ \cite{Henningson:1998hsk}: 
\begin{equation}
  g_{\mu \nu} = g_{\mu \nu}^{(0)} (x) + \rho g_{\mu \nu}^{(2)} (x) + \rho^2 g_{\mu \nu}^{(4)} (x) + ... + \rho^{d/2} g_{\mu \nu}^{(d)} (x) + \rho^{d/2} \log \rho h_{\mu \nu}^{(d)}(x) + ... \, , 
  \label{g:perturbative_exp_even}
\end{equation}
breaking the covariance of the full tensor.

The equations of motion associated with the bulk action are:
\begin{equation}
\begin{split}
    & R_{MN} - \frac{1}{2} G_{MN} R_{d+1} = - \Lambda_{d+1} G_{MN} - \frac{l^2_{\ast}}{8} F^2 G_{MN} + \frac{l^2_{\ast}}{2} G^{AB}F_{MA}F_{NB}, \\
    & \partial_M \left( \sqrt{-G} G^{MN} G^{AB} F_{NA} \right) =0. \\
\end{split}
\end{equation}
For the subsequent calculations it is useful to rewrite the Einstein equations in the following equivalent form:
\begin{equation}
    \begin{split}
        & R_{MN} = - \frac{d}{L_{d+1}^2} G_{MN} - \frac{l^2_{\ast}}{4(d-1)} F^2 G_{MN} + \frac{l^2_{\ast}}{2} G^{AB}F_{MA} F_{NA} \\
        & R_{d+1} = - \frac{d (d+1)}{L_{d+1}^2} + \frac{(d-3)}{4(d-1)} l^2_{\ast} F^2.
\end{split}
\end{equation}
Let us choose the following gauge for the $U(1)$ gauge field:
\begin{equation}
    A_{\rho} =0\, ,
\end{equation}
and expand the remaining components as:
\begin{equation}
    A_{\mu} (\rho,x) = A_{\mu}^{(0)} (x) + \rho A_{\mu}^{(1)} (x) + \rho^2 A_{\mu}^{(2)} (x) +... \, .
    \label{A_M:gauge_field_expansion}
\end{equation}
The field strength associated to the gauge field is by definition:
\begin{equation}
    F_{AB}= \partial_{A} A_B - \partial_B A_A.
\end{equation}
Using \eqref{A_M:gauge_field_expansion}, we can express its various components as an expansion in the radial coordinate:
\begin{equation}
    \begin{split}
         F_{\mu \nu} & = \left[ \partial_{\mu} A_{\nu}^{(0)} - \partial_{\nu} A_{\mu}^{(0)} \right] + \rho \left[ \partial_{\mu} A_{\nu}^{(1)} - \partial_{\nu} A_{\mu}^{(1)} \right] + \rho^2 \left[ \partial_{\mu} A_{\nu}^{(2)} - \partial_{\nu} A_{\mu}^{(2)} \right] + \rho^3 \left[ \partial_{\mu} A_{\nu}^{(3)} - \partial_{\nu} A_{\mu}^{(3)} \right] + ... \, , \\
        & = F_{\mu \nu}^{(0)} + \rho F_{\mu \nu}^{(1)} + \rho^2 F_{\mu \nu}^{(2)} + \rho^3 F_{\mu \nu}^{(3)} + o(\rho^4) \, ,\\
        F_{\rho \mu} & = - F_{\mu \rho}= A_{\mu}^{(1)} +2 \rho A_{\mu}^{(2)} + 3 \rho^2 A_{\mu}^{(3)} +4 \rho^3 A_{\mu}^{(4)} + o(\rho^4)  \, ,\\
        F_{\rho \rho} & = 0 \, ,
    \end{split}
\end{equation}
from which:
\begin{equation}
    F^2 = \frac{\rho^2}{L_{d+1}^4} \left[ F^{(0)2} + \rho \left( 8 A^{(1)2} -2 g^{(0) \alpha \mu} g^{(0) \beta \nu} g^{(2)}_{\mu \nu} g^{(0) \gamma \delta} F^{(0)}_{\alpha \gamma} F^{(0)}_{\beta \delta} + 2 F^{(0)} F^{(1)} \right) + o(\rho^2)\right]\, ,
\end{equation}
where all contractions use $g^{(0)}$.
Solving the Einstein equations order by order we obtain the following results for the first coefficients of the metric expansion:
\begin{equation}
    g^{(2)}_{\mu \nu} = \frac{1}{d-2} \left[ \frac{\hat{R}^{(0)}_d}{2(d-1)} g^{(0)}_{\mu \nu} - \hat{R}^{(0)}_{\mu \nu} \right]\, ,
\end{equation}
\begin{equation}
\begin{split}
    & g^{(4)}_{\mu \nu}  = - \frac{g^{(4) \lambda}{}_{\lambda}}{d-4} g^{(0)}_{\mu \nu} - \frac{l^2_{\ast}}{8L_{d+1}^2(d-4)(d-1)} F^{(0)2} + \frac{l^2_{\ast}}{4 L_{d+1}^2(d-4)} g^{(0) \alpha \beta} F^{(0)}_{\mu \alpha} F^{(0)}_{\nu \beta} \\
    & + \frac{g^{(0)}_{\mu \nu}}{2(d-4)(d-2)^2} \left[ \frac{(2-3d) \hat{R}^{(0)2}_d}{4(d-1)^2}+ \hat{R}^{(0)}_{\alpha \beta} \hat{R}^{(0) \alpha \beta} \right] +  \frac{1}{(d-4)(d-2)^2} \left[  \frac{\hat{R}^{(0)}_d}{(d-1)} \hat{R}^{(0)}_{\mu \nu} -  g^{(0) \alpha \beta}\hat{R}^{(0)}_{\alpha \mu} \hat{R}^{0}_{\beta \nu} \right]\, ,\\
\end{split}
\end{equation}
with
\begin{equation}
    g^{(2) \lambda}{}_{\lambda} = - \frac{\hat{R}_d}{2(d-1)}\, ,
\end{equation}
and
\begin{equation}
    g^{(4) \lambda}{}_{\lambda}= \frac{l^2_{\ast}}{16(d-1)L_{d+1}^2}F^{(0)2} + \frac{1}{4(d-2)^2} \left[ \frac{4-3d}{4(d-1)^2} \hat{R}^2_d + \hat{R}^{(0)\mu \nu} R^{(0)}_{\mu \nu}\right]\, .
\end{equation}
In order to isolate the IR divergences of the theory let us introduce a cutoff surface at $\rho= \epsilon$ near the asymptotic boundary at $\rho =0$. The regulated action becomes:
\begin{equation}
    I_{bulk}^{reg}= \frac{1}{16 \pi G_{d+1}} \int_{\mathcal{M}_{\epsilon}} d^{d+1}x \sqrt{-G} \left[ R_{d+1} -2 \Lambda_{d+1} - \frac{l^2_{\ast}}{4} F^2\right] - \frac{1}{8 \pi G_{d+1}} \int_{\partial \mathcal{M}} d^dx \sqrt{-h}  K \big\vert_{\rho=\epsilon}. 
    \label{action:regulated}
\end{equation}
where $\mathcal{M}_{\epsilon}$ indicates the bulk volume up to the IR cutoff surface.\\
We now want to write explicitly the various divergent contributions in the action. To achieve this, we evaluate the action on-shell and express the determinant of the bulk metric in terms of the determinant of the $d$-dimensional metric $g_{\mu \nu}$:
\begin{equation}
\begin{split}
     I_{bulk}^{reg}  = & \frac{L_{d+1}^{d-1}}{16 \pi G_{d+1}} \int d^{d}x \left[ \int_{\rho \geq \epsilon} d \rho \frac{\sqrt{-g(\rho,x)}}{\rho^{d/2+1}} \left( -d + \frac{L_{d+1}^2 l^2_{\ast}}{4(d-1)} F^2\right) + \right. \\
     & \left. +\frac{1}{\rho^{d/2}}\left( -d \sqrt{-g(\rho,x)}+2 \rho \partial_{\rho} \sqrt{-g(\rho,x)} \right)_{\rho=\epsilon} \right].  \\
\end{split}
\label{action:bulk_regulated_g}
\end{equation}
Using the expansion of the metric, we can expand the determinant\footnote{$det(g+h)=det(g)det(1+b^{-1}h)$.} in powers of $\rho$:
\begin{equation}
    \sqrt{-g (\rho,x)} = \sqrt{-g^{(0)}} \left[ 1 + \frac{1}{2} \rho {\rm Tr}\,(g^{(2)}) + \rho^2 \left( \frac{1}{2} {\rm Tr}\,(g^{(4)}) + \frac{1}{8} \left( {\rm Tr}\,(g^{(2)}) \right)^2 - \frac{1}{4} {\rm Tr}\,(g^{(2)})^2 \right) +... \right].
\end{equation}
For $d$ even we have the same power series with covariant coefficients up to and including the $\rho^{d/2}$ terms
\begin{equation}
    \rho^{d/2} \left( {\rm Tr}\,(g^{(d)}) + \log \rho {\rm Tr}\,(h^{(d)}) \right).
\end{equation}
The higher-order non-covariant corrections will not affect the subsequent analysis.
Integrating over the radial coordinate $\rho$ in \eqref{action:bulk_regulated_g} we can now separate the divergent contributions from the finite terms:
\begin{equation}
    I^{reg}_{bulk}= I_{fin} + I_{div}\, ,
\end{equation}
with
\begin{equation}
\begin{split}
    &I_{fin}=  \frac{L_{d+1}^{d-1}}{16 \pi G_{d+1}} \int d^dx \sqrt{-g^{(0)}(x)} \left\lbrace 2 \rho^{-d/2} + \frac{d {\rm Tr}\,(g^{(2)})}{d-2} \rho^{-d/2+1} + \frac{2\rho^{-d/2+2}}{d-4} \left[ d \left( \frac{1}{2} {\rm Tr}\,(g^{(4)}) + \right. \right. \right. \\
    & \left. \left. \left. + \frac{1}{8} \left( {\rm Tr}\,(g^{(2)}) \right)^2 - \frac{1}{4} {\rm Tr}\,(g^{(2)})^2 \right) + \frac{l^2_{\ast}}{4L_{d+1}^2(d-1)} F^{(0)2} \right] -\frac{2 \rho^{-d/2+3}}{d-6} \left[ \frac{l^2_{\ast}}{4L^2_{d+1}(d-1)} \left( 8 A^{(1)2} + \right. \right. \right. \\
    & \left. \left. \left. -2 g^{(0) \alpha \mu} g^{(0) \beta \nu} g^{(2)}_{\mu \nu} g^{(0) \gamma \delta} F^{(0)}_{\alpha \gamma} F^{(0)}_{\beta \delta} + 2 F^{(0)} F^{(1)} \right) + \frac{l^2_{\ast}}{8L_{d+1}^2(d-1)} F^{(0)2} \right] + ... \right\rbrace\, , \\
    \end{split}
    \label{action:finite_contribution}
\end{equation}
and
\begin{equation}
     I_{div}= -\frac{L^{d-1}_{d+1}}{16 \pi G_{d+1}} \int d^dx \sqrt{-g^{(0)}} \left\lbrace a^{(0)} \epsilon^{-d/2} +a^{(1)} \epsilon^{-d/2+1} + a^{(2)} \epsilon^{-d/2+2}+ o( \epsilon^{-d/2+3})  \right\rbrace\,
     \label{I_divergent:coefs_a}
\end{equation}
with
\begin{equation}
    \begin{split}
        & a^{(0)}=-2(d-1)\, ,\\
        & a^{(1)}=- \frac{(d-1)(d-4)}{d-2} {\rm Tr}\,(g^{(2)})\, , \\
        & a^{(2)}= -2 \frac{d^2-9d+16}{d-4} \left( \frac{1}{2} {\rm Tr}\,(g^{(4)}) + \frac{1}{8} \left( {\rm Tr}\,(g^{(2)}) \right)^2 - \frac{1}{4} {\rm Tr}\,(g^{(2)})^2 \right) + \frac{l^2_{\ast}}{2L_{d+1}^2 (d-1)(d-4)} F^{(0)2} \, ,\\
    \end{split}
\end{equation}
for $d$ odd. For $d$ even, there are additional contributions to both the divergent and finite parts of the action given by:
\begin{equation}
    \frac{L_{d+1}^{d-1}}{16 \pi G_{d+1}} \int d^dx \int_{\epsilon} d \rho \frac{d}{\rho^{d/2+1}} \left( \frac{1}{2} \rho^{d/2} {\rm Tr}\,(g^{(d)}) \right) = \frac{L_{d+1}^{d-1}}{16 \pi G_{d+1}} \int d^dx a^{(d)} \left( \log \rho - \log \epsilon \right).
\end{equation}
The term proportional to $h^{(d)}$ does not give contribution because ${\rm Tr}\,(h^{(d)})=0$.

Re-expressing everything in \eqref{I_divergent:coefs_a} in terms of the induced metric $h_{\mu \nu}$ we have:
\begin{equation}
\begin{split}
   I_{div}= &  \frac{L_{d+1}}{16 \pi G_{d+1}(d-2)} \int d^dx \sqrt{-h} \left\lbrace 2 \frac{(d-1)(d-2)}{L_{d+1}^2}+R_d \left[ h \right] - \frac{l^2_{\ast}(d-2)(d^2-d-16)}{16(d-1)(d-4)} \tilde{F}^{(0)2} +  \right. \\
   & \left. + \frac{L_{d+1}^2}{(d-2)(d-4)} \left[ R_{\mu \nu} \left[ h \right] R^{\mu \nu} \left[ h\right] - \frac{R^2_d \left[ h \right]}{4(d-1)} \right]+ ... \right\rbrace\, , \\
\end{split}
\label{master-eq}
\end{equation}
where the ellipsis correspond to higher derivative terms entering in the action with higher powers of $L_{d+1}^2$ and $\tilde{F}^{(0)2}$ indicates that the contraction is performed with the metric $h$.\\
The induced theory on the brane is given by:
\begin{equation}
    I_{brane} = I_{div}+ I_{\tau} +I_{fin}\, ,
\end{equation}
where $I_{fin}$ is given by \eqref{action:finite_contribution}. It describes the CFT living on the brane with an UV cutoff induced by the bulk IR cutoff $\epsilon$. 

\subsection{Case $d=4$}
In the case of even $d$, the procedure for solving the equations of motion order by order in $d+1$ dimension using the perturbative expansion \eqref{g:perturbative_exp_odd} breaks down at order $d/2$, where a logarithmic term appears \eqref{g:perturbative_exp_even}. The tensors $g^{(k)}$, for $k=0,2,...,d-2$ are covariant combinations of the metric $g^{(0)}$, its covariant derivatives and its Riemann tensor. The same is not true for the coefficient $g^{(d)}$ and the higher order corrections. Truncating the expansion at the order where the logarithmic term appears and applying the same regularization procedure as in the previous section, we obtain:
\begin{equation}
    I^{reg}_{bulk} = I_{fin} - \frac{L^{d-1}_{d+1}}{16 \pi G_{d+1}} \int d^dx \sqrt{-g^{(0)}} \left\lbrace a^{(0)} \epsilon^{-d/2} +a^{(1)} \epsilon^{-d/2+1} + ... + a^{(d)} \log \epsilon  \right\rbrace.
    \label{I_reg:d_even}
\end{equation}
Note that a term of the form $\int d^dx a^{(d)} \log \rho$ is also present in $I_{fin}$. In holographic renormalization, after subtracting local divergent counterterms, including the logarithmic term, it was proven in \cite{Henningson:1998hsk} that this term induces an anomaly of the form:
\begin{equation}
    \delta I_{fin} = \int d^dx \sqrt{-g^{(0)}} \mathcal{A} \delta \lambda = \frac{1}{16 \pi G_{d+1}} \int d^dx \sqrt{-g^{(0)}} a^{(d)} \delta \lambda
\end{equation}
under a conformal transformation $\delta g^{(0)} = 2 \delta \lambda g^{(0)}$ of the metric. From the perspective of the intermediate picture, this is the anomalous breaking of conformal invariance when coupling the CFT$_d$ to $d$-dimensional gravity with a cutoff $\epsilon\neq 0$. In the bulk picture this comes from an IR effect, as familiar in the UV/IR mixing in holography.

Specializing the computations to the case $d=4$ for illustration, the action \eqref{I_reg:d_even} becomes:
\begin{equation}
    I^{reg}_{bulk}=I_{fin} - \frac{L_5^3}{16 \pi G_5} \int d^4x \sqrt{-g^{(0)}} \left\lbrace - \frac{6}{\epsilon^2} - \log \epsilon \left[ \frac{l^2_{\ast}}{8L^2_5} F^{(0)2} + \frac{1}{8} \left( \frac{1}{3} R_d^{(0)2} - R^{(0)}_{\mu \nu} R^{(0) \mu \nu}\right)\right] \right\rbrace.
\end{equation}
Re-writing everything in terms of the induced metric $h_{\mu \nu}$ we have:
\begin{equation}
    \begin{split}
         I^{reg}_{bulk} = & I_{fin} + \frac{L^{-1}_5}{32 \pi G_5} \int d^4x \sqrt{-h} \left\lbrace 12+R[h] - \frac{l^2_{\ast} L_5^2}{4} \tilde{F}^{(0)2} - \frac{1}{8} \left( R_{\mu \nu} [h] R^{\mu \nu} [h] - \frac{1}{3}R^2[h] \right) + \right.\\
        & \left. + \log \epsilon \left[  \frac{l^2_{\ast} L^2_5}{4} \tilde{F}^{(0)2} + \frac{L^4_5}{4} \left( \frac{R^2 [h]}{3} - R_{\mu \nu} [h] R^{\mu \nu} [h]\right)\right]\right\rbrace. \\
    \end{split}
    \label{d-action:gauge_4d}
\end{equation}

\section{Overview of Quantum Black Holes in AdS$_3$}
\label{sec:bhs}

In this appendix we review several quantum black hole solutions in AdS$_3$ gravity theories coupled to a CFT$_3$ (with a cutoff) with an AdS$_4$ gravity dual. The quantum black hole includes the backreaction of the fast CFT$_3$ modes above the UV cutoff, and it is effectively computed using the 4d bulk picture and translating to the 3d intermediate picture. In practice one constructs a black hole solution of the 4d AdS$_4$ gravity theory satisfying the junction conditions of a Karch-Randall ETW brane and extracts the induced metric on the latter. We simply quote several results collected in  the review \cite{Panella:2024sor}, to which we refer the reader for details.

\subsection{Neutral Quantum BTZ Black Hole}
\label{sec:neutral-bh}

The starting point is Einstein-Maxwell theory in the $\text{AdS}_{4}$ bulk. We are interested in a specific sub-class of solutions, known as the AdS$_4$ C-metrics, which describe accelerating black holes in an AdS$_4$ background. The neutral non-rotating solution of this kind is
\beq
ds^{2}=\frac{\ell^{2}}{(\ell+xr)^{2}}\left[-H(r)dt^{2}+\frac{dr^{2}}{H(r)}+r^{2}\left(\frac{dx^{2}}{G(x)}+G(x)d\phi^{2}\right)\right]\;,
\label{eq:AdS4cmetBLstat}\eeq
with metric functions
\beq
H(r)= \frac{r^{2}}{\ell_{3}^{2}}+\kappa-\frac{\mu\ell}{r} \;,\qquad G(x)=1-\kappa x^{2}-\mu x^{3}\;.
\label{eq:metfuncsstatCmet}
\eeq
Here $\mu$, $\kappa$, $\ell$ and $\ell_3$, are real parameters characterizing the solution. For instance the non-negative parameter $\mu$ is related to the mass of the bulk black hole. When interpreted as an accelerating black hole, the parameter $\ell\geq0$ equals the inverse acceleration, and is related to the bulk $\text{AdS}_{4}$ length scale $L_{4}$ via (\ref{eq:bulkAdS4length}), with $\ell_{3}$ the $\text{AdS}_{3}$ brane curvature scale. The parameter values $\kappa=+1,-1$ or $0$ respectively correspond to foliating AdS$_4$ intro global AdS$_3$, BTZ or Poincar\'e AdS$_3$ slices.

For $\ell=0$, the above solution describes a static, neutral AdS$_4$ black hole. For non-zero $\ell$, the black hole is accelerated and the horizon is distorted, in fact developing a conical singularity, describing a string-like spike, which pulls the black hole and supports its acceleration.

A crucial feature of the C-metric (\ref{eq:AdS4cmetBLstat}) is that the $x=0$ hypersurface is \emph{umbilic}, meaning that the extrinsic curvature $K_{ij}$ of the hypersurface is proportional to its induced metric $h_{ij}$. This implies that it obeys the junction conditions for an ETW AdS$_3$ brane to be located at $x=0$, cutting off the spacetime behind it. Remarkably, this excised part of spacetime is that containing the string-like spike, so in the cut-off configuration the black hole is kept accelerated by its attachment to the tensionful ETW brane. Indeed the tension of a brane at $x=0$ is (\ref{brane-tension}), hence proportional to the black hole acceleration. In the tensionless limit, $\ell\to\infty$ the AdS$_3$ ETW brane cuts the bulk AdS$_4$ in half. We are rather interested in the limit $\ell \to 0$, where the AdS$_3$ ETW brane is pushed to the holographic boundary of AdS$_4$.

The quantum black hole is obtained from the induced metric at the ETW brane, obtained from (\ref{eq:AdS4cmetBLstat}) by restriction at $x=0$, namely:
\beq
\label{eq:naiveBTZ}
ds^2|_{x=0} = -\left(\frac{r^{2}}{\ell_{3}^{2}}+\kappa-\frac{\mu\ell}{r} \right)dt^2 +\left(\frac{r^{2}}{\ell_{3}^{2}}+\kappa-\frac{\mu\ell}{r} \right)^{-1}  dr^2 + r^2 d\phi^2 \;,
\eeq
To express this as a more manifest black hole solution we perform a redefinition
\beq
t = \eta \bar{t}\ ,  \qquad r = \frac{\bar{r}}{\eta} \ , \qquad \phi= \eta \bar{\phi} \quad ,\quad {\rm with}\;\;
\eta \equiv \frac{\Delta \phi}{2\pi} = \frac{2 x_1}{3- \kappa x_1^2} \ .
\label{eq:etacanon}
\eeq
where $x_1$ is a root of $G(x)$ in (\ref{eq:metfuncsstatCmet}). The ETW brane metric is now 
\beqa 
\hspace{-2mm}{ds^{2}_{\text{qBTZ}} }= -H(\bar{r})d\bar{t}^2 +H(\bar{r})^{-1}  d\bar{r}^2 + \bar{r}^2 d\bar{\phi}^2 \quad\quad  {\rm with}\;\;\; H(\bar{r})=\frac{\bar{r}^{2}}{\ell_{3}^{2}}-8 \mathcal{G}_3 M-\frac{\ell F(M)}{\bar{r}} \, ,\nonumber
\label{eq:qBTZ}
\eeqa 
where we introduced the 3d mass $M$, a renormalized 3d Newton's constant $\mathcal{G}_3$
\beq
M \equiv -\frac{\kappa}{8G_3}\frac{\ell}{L_4}\eta^2 =-\frac{1}{2\mathcal{G}_3}\frac{\kappa x_1^2}{(3-\kappa x_{1}^{2})^2} \ , \qquad \mathcal{G}_3\equiv G_3 \frac{L_4}{\ell} =\frac{G_{3}}{\sqrt{1-\nu^{2}}}\ ,
\label{eq:massqBTZ}\eeq
and the function $F(M)$
\beq 
F(M)\equiv \mu\eta^{3}=8\frac{(1-\kappa x_{1}^{2})}{(3-\kappa x_{1}^{2})^{3}}\;.
\label{eq:formfuncFM}
\eeq
This metric is a solution of the induced brane action (\ref{brane-action-3d}), i.e. the action of the AdS$_3$ theory in the intermediate picture. In the limit $\ell\to 0$, it corresponds to a classical neutral  BTZ black hole in AdS$_3$. The full solution for $\ell\neq 0$ includes the quantum backreaction of the CFT$_4$ fast modes, and is known as a quantum black hole.

\subsection{The Charged Quantum Black Hole}
\label{sec:charged-bh}

We now similarly review charged black holes in the bulk theory. The charged AdS$_{4}$ C-metric is of the form  (\ref{eq:AdS4cmetBLstat}), but with an additional term in the functions (\ref{eq:metfuncsstatCmet}) :
\beq H(r)=\frac{r^{2}}{\ell_{3}^{2}}+\kappa-\frac{\mu\ell}{r}+\frac{q^{2}\ell^{2}}{r^{2}}\quad,\quad G(x)=1-\kappa x^{2}-\mu x^{3}-q^{2}x^{4}\;.
\eeq
This is a solution to 4d Einstein-Maxwell gravity theory (\ref{einstein-maxwell}), for a $U(1)$ gauge field background 
\beq 
A=A_{a}dx^{a}=\frac{2\ell}{\ell_{\ast}}\left[e\left(\frac{1}{r_{+}}-\frac{1}{r}\right)dt+g(x-x_{1})d\phi\right]\;,\label{eq:gaugpot}
\eeq
where $e$ and $g$ are the electric and magnetic charges, with $q^{2}=e^{2}+g^{2}$.

The $x=0$ slice is umbilic, and junction conditions fix the brane tension as in the neutral case. There are also junction conditions for the gauge field, which we skip. The induced metric and gauge field at $x=0$ define a quantum black hole of the brane theory in the intermediate picture, namely a solution of the equations of motion for the 3d action (\ref{emergent-gauge}). 

The metric is recast in a more standard form by redefining $(t,r,\phi)=(\eta\bar{t},\eta^{-1}\bar{r},\eta\bar{\phi})$:
\beq
\begin{split}
ds^2 = & -H(\bar{r})d\bar{t}^{2}+H^{-1}(\bar{r})d\bar{r}^{2}+\bar{r}^{2}d\bar{\phi}^{2}\;,\\
&H(\bar{r})=-8M\mathcal{G}_{3}+\frac{\bar{r}^{2}}{\ell_{3}^{2}}-\frac{\ell F(M,q)}{\bar{r}}+\frac{\ell^{2}Z(M,q)}{\bar{r}^{2}}\;.
\end{split}
\label{chargedbh-solution}\eeq
Here the mass is
\beq M\equiv-\frac{\kappa}{8\mathcal{G}_{3}}\eta^{2}=-\frac{\kappa}{8\mathcal{G}_{3}}\frac{4x_{1}^{2}}{(3-\kappa x_{1}^{2}+q^{2}x_{1}^{4})^{2}}\;,
\label{eq:masscqbtz}
\eeq
and the functions
\beq 
F(M,q)\equiv \mu\eta^{3}=8\frac{1-\kappa x_{1}^{2}-q^{2}x_{1}^{4}}{(3-\kappa x_{1}^{2}+q^{2}x_{1}^{4})^{3}}\quad ,\quad  Z(M,q)\equiv q^{2}\eta^{4}=\frac{16 q^{2}x_{1}^{4}}{(3-\kappa x_{1}^{2}-q^{2}x_{1}^{4})^{4}}\;. \nonumber
\eeq
Even though the solution arises from a charged black hole in the bulk picture, in the limit $\ell\to0$ the $q$-dependent correction vanishes. This is a manifestation of the the result in section \ref{sec:no-global-symmetries} that charges are screened on the brane, and the symmetry is effectively global. The gauge charge of the braneworld black hole is a consequence of backreaction, i.e. arises via emergence, as explained in section \ref{sec:emergence}.

\section{String theory embedding of ETW branes in AdS$_5\times\IS^5$}
\label{sec:string-embedding}

In this appendix we gather the explicit description of the supergravity solution describing the near horizon limit of D3-branes ending on NS5- and D5-branes \cite{DHoker:2007zhm,DHoker:2007hhe,Aharony:2011yc,Assel:2011xz,Bachas:2017rch,Bachas:2018zmb} (see also \cite{Raamsdonk:2020tin,VanRaamsdonk:2021duo,Demulder:2022aij,Karch:2022rvr,DeLuca:2023kjj,Huertas:2023syg,Chaney:2024bgx} for recent applications).

The ansatz for the 10d metric is
\beqa
ds^2= f_4^2 ds^2_{AdS_4}+f_1^2 ds^2_{\IS_1^2}+f_2^2 ds^2_{\IS_2^2}+4\rho^2 |dw|^2\, .
\label{ansatz}
\eeqa
The metric describes a fibration of AdS$_4\times\IS^2\times\IS^2$ over a Riemann surface $\Sigma$. Here $f_1$, $f_2$, $f_3$, $\rho$ are functions of a complex coordinate $w$ of $\Sigma$. There are also non-trivial backgrounds for the NSNS and RR 2-forms and the RR 4-form, for which we refer the reader to the references. 

There are closed expressions for the different functions in the above metric. As an intermediate step, we define the real functions
\beqa
W= \partial_w h_1\partial_{\bar w}h_2+\partial_w h_2\partial_{\bar w}h_1 
\; \;,\;\; N_1= 2h_1h_2|\partial_w h_1|^2-h_1^2 W \;\; ,\;\;N_2= 2h_1h_2|\partial_w h_2|^2-h_2^2 W\, ,\nonumber 
\label{wnn}
\eeqa
in terms of two functions $h_1,h_2$ to be specified below. The dilaton is given by
\beqa
e^{2\Phi}=\frac{N_2}{N_1}\, ,
\label{dilaton}
\eeqa
and the functions are given by
\beqa
\rho^2=e^{-\frac 12 \Phi}\frac{\sqrt{N_2|W|}}{h_1h_2}\; ,\;\; f_1^2=2e^{\frac 12\Phi} h_1^2\sqrt{\frac{|W|}{N_1}}\; ,\;\; f_2^2=2e^{-\frac 12\Phi} h_2^2\sqrt{\frac{|W|}{N_2}}\; ,\; \;f_4^2=2e^{-\frac 12\Phi} \sqrt{\frac{N_2}{|W|}}\, .\nonumber
\label{the-fs}
\eeqa
The solutions describing one asymptotic AdS$_5\times\IS^5$ region ending on an ETW configuration are described with a  Riemann surface given by the quadrant $w=re^{i\varphi}$, with $r\in (0,\infty)$ and $\varphi\in \left[\frac{\pi}{2},\pi\right]$. The $\IS_1^2\times \IS_2^2$ is fibered such that $\IS^2_1$ shrinks to zero size over $\varphi=\pi$ (negative real axis) and $\IS^2_2$ shrinks to zero size over $\varphi=\pi/2$ (positive imaginary axis). There are also NS5- and D5-brane sources, described as punctures on $\Sigma$, and $r\to\infty$ region corresponds to the asymptotic region AdS$_5\times\IS^5$ dual the 4d $\CN=4$ $SU(N)$ SYM  on the semi-infinite D3-branes. 

These features follow from the quantitative expression for the functions $h_1$, $h_2$, which for this class of solutions have the structure 
\begin{subequations}
\begin{align}
h_1&= 4{\rm Im}(w)+2\sum_{b=1}^m{\tilde d}_b \log\left(\frac{|w+il_b|^2}{|w-il_b|^2}\right)\, ,\\
h_2&=-4{\rm Re}(w)-2\sum_{a=1}^n d_a \log\left(\frac{|w+k_a|^2}{|w-k_a|^2}\right)\, .
\end{align}
\label{the-hs}
\end{subequations}
In particular, the 5-brane sources are as follows: The NS5-branes are along the $\varphi=\pi$ axis, with stacks of multiplicity $n_a$ at positions $w=-k_a$, and the D5-branes are along the $\varphi=\pi/2$ axis, with stacks of multiplicity $m_b$ at positions $w=il_b$. The multiplicities of 5-branes are related to the parameters $d_a$, ${\tilde d}_b$ via
\beqa
n_a= 32\pi^2 d_a\in\IZ \quad ,\quad m_b= 32 \pi^2 {\tilde d}_b\in\IZ\, .
\label{quant1}
\eeqa
As described in the main text, each NS5-brane spans an AdS$_4\times \IS^2_2$ (respectively a D5-brane spans AdS$_4\times\IS^2_1$), with $n_a$ units of NSNS 3-form flux (respectively $m_b$ units of RR 3-form flux) on the $\IS^3$ surrounding the 5-brane source, with the structures of cycles depicted in Figure \ref{fig:quadrant}. There are also non-trivial NSNS and RR 2-form integrals over those $\IS^2$'s. In short, in the $\IS^2\times\IS^3$ geometry around the stacks of $n_a$ NS5-branes and of $m_b$ D5-branes, there are non-trivial integrals  \cite{Aharony:2011yc}
\beqa
\int_{\IS^2_2} C_2=K_a\quad,\quad \int_{\IS^3_1} H_3=n_a\quad ;\quad \int_{\IS_1^2} B_2={\tilde L}_b\quad,\quad \int_{\IS^3_2} F_3=m_b\, .
\label{the-fluxes}
\eeqa
These values imply that the 5-form flux over the $\IS^5$ jumps when each of the 5-brane punctures is crossed, describing that the flux can escape through the 5-branes. This is the gravitational manifestation that some number of D3-branes ends on the 5-brane, so the change in the 5-form flux encodes the linking number of the 5-branes in the corresponding stack.
The change in the flux 
\begin{equation}
{\tilde F_5}=F_5+B_2F_3-C_2H_3\, ,
\label{the-f5}
\end{equation}
with $F_5=dC_4$, upon crossing an NS5-brane (resp. D5-brane) is given by $n_aK_a$ (resp. $m_b{\tilde L}_b$), with no sum over indices. The 5-form flux decreases in such crossings until we reach the region $r<|k_a|,|l_b|$, in which there is no leftover flux, so that the $\IS^5$ shrinks and spacetime ends in an smooth way at $r=0$, see Figure \ref{fig:quadrant}. Hence, the asymptotic 5-form flux $N$ and the 5-brane parameters are related as in (\ref{sum-linkings})
\begin{equation}
    N=\sum_a n_a K_a+\sum_b m_b{\tilde L}_b\, .
\end{equation}
Finally, the positions of the 5-brane punctures $k_a$, $l_b$ are determined by the linking numbers $K_a$, ${\tilde L}_b$ via
\begin{equation}
    K_a=32\pi \left(k_a +2\sum_b{\tilde d}_b \arctan\left(\frac {k_a}{l_b}\right)\right)\;\; , \;\;
    {\tilde L_b}=32\pi \left(l_b -2\sum_a d_a \arctan\left(\frac {k_a}{l_b}\right)\right)\, .\quad \quad
    \label{positions-linking}
\end{equation}
We refer the reader to the references for a derivation of this result and other related details.

\section{Supersymmetric String Models with ETW Anomaly Inflow}
\label{sec:anomaly-d7s}

In this section we present an explicit fully supersymmetric string theory model realizing the ideas discussed in section \ref{sec:anomaly}. The model is based on NS5- and D5-brane ETW configuration for (an orbifold of) AdS$_5\times\IS^5$, with additional D7-branes to produce the gauge symmetry in the AdS$_5$ bulk, a localized 4d anomaly on the ETW brane, and the required 5d topological Chern-Simons couplings to cancel it. At the conceptual level the example works as the D9-brane model in section \ref{sec:anomaly-td}, and the new ingredients are simply technical aspects of the more involved construction. We also construct a class of supersymmetric bulk models with ETW configurations with controlled supersymmetry breaking, to be used in section \ref{sec:no-nonsusy-ads}.

\subsection{$\NN=1$ flavours for 4d $\NN=2$ orbifold quivers, and their gravity duals}
\label{sec:flavouring}

Let us start by considering the theories before the introduction of the 5-brane ETW configuration, namely we consider first the basic holographic AdS/CFT dual pair. The models are obtained from a supersymmetric system of D3- and D7-branes (of different kinds) at a $\IC^2/\IZ_k$ orbifold singularity, along the lines of section 2 in \cite{Park:1998zh,Park:1999eb}, to which we refer the reader for further details. We give a quick review of the relevant results.

\smallskip

{\bf D3-branes at $\IC^2/\IZ_k$ and the gravity dual}

We first consider D3-branes on a $\IC^2/\IZ_k$ singularity, where the generator $\theta\in\IZ_k$ acts on the $\IC^3$ transverse to the D3-branes as
\beqa
\theta:(z_1,z_2,z_3)\to (e^{\frac{2\pi i}k} z_1,e^{-\frac{2\pi i}k}z_2,z_3)\, .
\label{zk-action}
\eeqa
The relation between these complex coordinates and the real ones used in the main text will be specified later on, when we discuss the introduction of 5-branes.

The action of $\theta$ on the D3-branes is defined via the Chan-Paton matrix
\beqa
\gamma_{\theta,3}=\diag (1_{n_0},e^{2\pi i\frac{1}{k}}1_{n_1}, \ldots, 
e^{2\pi i\frac{k-1}{k}}1_{n_{k-1}})\, .
\label{cpthree}
\eeqa
The resulting theory on the D3-branes has 4d $\NN=2$ supersymmetry, but it is convenient to carry out the orbifold projection in $\NN=1$ multiplets. The projections on the 4d $\NN=1$ vector multiplet $V$ and chiral multiplets $Z_i$, $i=1,2,3$, are given by
\beqa
&V=\gamma_{\theta,3} V \gamma_{\theta,3}^{-1} \;  \rightarrow  \; \otimes_a U(n_a) \, ,\quad
& \quad Z_1=e^{2\pi i/k}\gamma_{\theta,3} Z_1 \gamma_{\theta,3}^{-1} \; \rightarrow  \;\; (n_a,{\ov n}_{a+1})\, ,\nonumber \\
& Z_3=\gamma_{\theta,3} Z_3 \gamma_{\theta,3}^{-1} \; \rightarrow \; \; {\rm Adj}_{\; a}  \, ,
& \quad Z_2=e^{-2\pi i/k}\gamma_{\theta,3} Z_2 \gamma_{\theta,3}^{-1}  \; \rightarrow  \;\;({\ov n}_a,n_{a+1}) \, ,
\label{33-projection} 
\eeqa
for $a=0,\ldots, k-1$. The $U(1)_a$ symmetries are massive due to St\"uckelberg couplings \cite{Ibanez:1998qp}, and are ignored from now on. There is also a 4d $\NN=2$ superpotential, which will not be needed for our purposes. 

The theory is non-chiral, so the $n_a$ are unconstrained by anomaly cancellation. Equivalently, there are no twisted RR tadpole cancellation constraints because the singular locus has the non-compact direction $z_3$ transverse to the D3-branes, so the fluxlines of twisted D3-brane charges can escape to infinity. We nevertheless focus on the case of regular D3-branes $n_a=N$, which corresponds to the 4d CFT case. In this case, the gravity dual is simply given by AdS$_5\times \IS^5/\IZ_k$. The action of $\IZ_k$ on $\IS^5$ is inherited from the restriction of (\ref{zk-action}) to the unit sphere
\beqa
|z_1|^2+|z_2|^2+|z_3|^2=1\, .
\label{unit-ball}
\eeqa
The action has fixed points at $z_1=z_2=0$, so that the quotient $\IS^5/\IZ_k$ contains an $\IS^1$ of $\IC^2/\IZ_k$ singularities, located at
\beqa
|z_3|^2=1\, .
\label{s1-singular}
\eeqa
This singularity supports a set of twisted sector fields \cite{Hanany:1998it,Gukov:1998kk}, including some RR forms relevant for anomaly cancellation later on.

\smallskip
{\bf Adding D7-brane flavours}

We now add D7-branes to introduce flavours in the 4d gauge theory on the D3-branes, and discuss their realization in the gravity dual. As in section \ref{sec:anomaly-td} we consider the number of additional branes to be small, so that the flavours are treated in the quenched approximation (and so do not modify the conformality of the 4d theory), and the D7-branes can be treated as probes in the gravity dual.

We consider 3 different kinds of D7-branes, denoting by D7$_i$-branes those located at $z_i=0$, $i=1,2,3$. The most familiar case are D7$_3$-branes, located at $z_3=0$ and spanning $z_1$ and $z_2$ (the whole $\IC^2/\IZ_k$). These D7$_3$-branes preserve 4d $\NN=2$ and have been considered e.g. in \cite{Park:1998zh}. In order to get chiral theories, we also introduce D7$_1$- and D7$_2$-branes, which only preserve 4d $\NN=1$, and which were considered in \cite{Park:1999eb}. 

The orbifold action on the D7$_i$-branes is defined via the Chan-Paton matrices:
 \beqa
\gamma_{\theta,7_1} & = & e^{\pi i \frac{1}{k}} \diag ( \id_{v_1},e^{2\pi i 
\frac{1}{k}} \id_{v_2},\ldots, e^{2\pi i\frac{k-1}{k}}\id_{v_k})\, ,\nonumber \\
\gamma_{\theta,7_2} & = &e^{\pi i \frac{1}{k}} \diag ( \id_{w_1},e^{2\pi i 
\frac{1}{k}} \id_{w_2},\ldots, e^{-2\pi i\frac{k-1}{k}} \id_{w_k})\, ,\nonumber \\
\gamma_{\theta,7_3} & = & \diag ( \id_{m_1},e^{2\pi i 
\frac{1}{k}} \id_{m_2},\ldots, e^{2\pi i\frac{k-1}{k}} \id_{m_k})\, .
\label{cp7}
\eeqa
The overall phase factors for D7$_1$- and D7$_2$-branes are determined by consistency of the projection in D3-D7$_i$ open sectors. The projections under the combined Chan-Paton and geometric orbifold actions have the structure (correcting some typos in \cite{Park:1999eb})
\beqa
37_1: & \;\; \lambda  = e^{-\frac{\pi i}{k}} \; \gamma_{\theta,3}\lambda 
\gamma_{\theta,7_1}^{-1}  \;\; \rightarrow\;\;(n_a,{\ov v}_{a-1})\, ,\quad\quad 
&7_1 3:  \;\; \lambda  = e^{-\frac{\pi i}{k}}\; \gamma_{\theta,7_1}\lambda 
\gamma_{\theta,3}^{-1} \;\; \rightarrow \;\;({\ov n}_a,v_a)\, ,
\nonumber \\
37_2: & \; \lambda  = e^{\frac{\pi i}{k}} \; \gamma_{\theta,3}\lambda 
\gamma_{\theta,7_2}^{-1} \; \;\rightarrow \;\; (n_a,{\ov w}_a)\, ,\quad\quad
& 7_2 3:  \;\;\lambda  = e^{\frac{\pi i}{k}}\; \gamma_{\theta,7_2}\lambda 
\gamma_{\theta,3}^{-1} \; \;\;\rightarrow \;\;\; ({\ov n}_a,w_{a-1})\, ,\nonumber \\
37_3: & \;\; \lambda  =   \gamma_{\theta,3}\lambda 
\gamma_{\theta,7_3}^{-1} \quad\quad\; \rightarrow  \;\;(n_a,{\ov m}_a)\, ,\quad\quad
&7_3 3:  \;\; \lambda  =  \gamma_{\theta,7_3}\lambda 
\gamma_{\theta,3}^{-1} \quad \quad\;\rightarrow \;\; \;\;({\ov n}_a,m_a)\, .\nonumber 
\label{proj37}
\eeqa
Here for clarity we have maintained the labels for representations of the D3-brane gauge factors, even though we will eventually set $n_a=N$. Notice that each D7$_3$-branes gives a full $\NN=2$ hypermultiplet to one D3-brane gauge factor, while each D7$_1$-, or D7$_2$-brane provides one $\NN=1$ chiral fundamental or antifundamental to two different gauge groups of the 4d theory. The latter statement corresponds to the flavour doubling effect in a T-dual Hanany-Witten IIA configuration \cite{Hanany:1997sa,Park:1999eb}. There are superpotential couplings (33)$_i$(37$_i$)(7$_i$3) among 33 states associated to $Z_i$ and states in the 37$_i$ and 7$_i$3 sectors, whose details will not be necessary for our purposes.

Since the extra flavours are chiral, there are constraints on the number of D7$_1$- and D7$_2$-branes from cancellation of non-abelian gauge anomalies for the D3-brane gauge factors\footnote{Cancellation of $U(1)$ anomalies is achieved via a 4d Green-Schwarz mechanism \cite{Ibanez:1998qp}, which involves the St\"uckelberg couplings mentioned above.}. Equivalently there are no directions in the singular locus which are transverse to the relevant D7-branes, so we have to cancel the RR twisted tadpoles, namely 
\beqa
\tr \gamma_{\theta^p,7_1}-\tr \gamma_{\theta^p,7_2}=0,\quad {\rm for}\; p=0,\ldots,k-1\, . 
\eeqa
Either way, the conditions read $v_{a-1}-v_a+w_a-w_{a-1}=0$, which are solved  by taking
\beqa
v_a-w_a=K\, ,\quad\forall a\, ,
\label{anomorbi}
\eeqa
with $K$ is a constant independent of $a$.

In our case of interest, $n_a=N$, the above D3-D7$_i$ spectrum leads to no localized 4d anomalies on the D7$_i$-branes. However, there are localized anomalies in the configuration arising in the  D7$_i$-D7$_j$ open string sectors, which we study next.

Let us start with the D7$_i$-D7$_i$ sector, which produces 8d fields propagating on the 4-cycle $z_i=0$. However, those surviving the orbifold projection are visible at the fixed locus (while non-invariant modes must vanish there). In 4d $\NN=1$ language, the reduction of the 8d vector multiplet produces one vector multiplet and 3 chiral multiplets, on which the projections have a structure analogous to (\ref{33-projection})
\beqa
&V=\gamma_{\theta,7_3} V \gamma_{\theta,7_3}^{-1} \;  \rightarrow  \;  \otimes_a U(m_a) \, ,\quad  & 
\quad Z_1=e^{2\pi i/k}\gamma_{\theta,7_3} Z_1 \gamma_{\theta,7_3}^{-1}  \; \rightarrow  \; \;  (m_a,{\ov m}_{a+1}) \, ,\quad \quad \nonumber \\
& Z_3=\gamma_{\theta,7_3} Z_3 \gamma_{\theta,7_3}^{-1}  \; \rightarrow \; \;  {\rm Adj}_{\; a} \, , 
& \quad Z_2=e^{-2\pi i/k}\gamma_{\theta,7_3} Z_2 \gamma_{\theta,7_3}^{-1}  \; \rightarrow  \; \; ({\ov m}_a,m_{a+1}) \, .
\label{77-projection-same}
\eeqa
For D7$_1$- and D7$_2$-branes we have a similar structure, but with gauge groups $\otimes_a U(v_a)$ and $\otimes_a U(w_a)$, respectively. These sectors thus produce localized 4d chiral multiplets in non-trivial representations of the 8d gauge groups.

We now consider the D7$_i$-D7$_j$ sectors for $i\neq j$, which are localized on the 2-cycle $z_i=z_j=0$ at the intersection of the D7-branes. Before the orbifold projection, there is one 6d hypermultiplet in the bifundamental representation, namely two 4d $\NN=1$ chiral multiplets (in the 7$_i$7$_j$ and 7$_j$7$_i$ sectors, respectively) in conjugate representations, whose orbifold invariant modes can lead to localized 4d anomalies. The projections under the combined Chan-Paton and geometric orbifold actions have the structure
\beqa
& 7_17_2:  \;\; \lambda  = \; \gamma_{\theta,7_1}\lambda \gamma_{\theta,7_2}^{-1}   \rightarrow (v_i,{\ov w}_{i})\, , \quad\quad
& \quad 7_27_1:  \;\; \lambda  = \; \gamma_{\theta,7_2}\lambda \gamma_{\theta,7_1}^{-1}   \rightarrow ({\ov v}_i, w_{i})\, ,
\nonumber \\
& 7_17_3: \;\; \lambda  = e^{\frac{\pi i}{k}}\; \gamma_{\theta,7_1}\lambda \gamma_{\theta,7_3}^{-1}  \rightarrow (v_i,{\ov m}_{i+1})\, ,
& \quad 7_37_1: \;\; \lambda  = e^{\frac{\pi i}{k}} \; \gamma_{\theta,7_3}\lambda \gamma_{\theta,7_1}^{-1}  \rightarrow ({\ov v}_i,m_i)\, ,
\quad\quad\quad\label{77-projection-different} \\
& 7_27_3: \;\; \lambda  = e^{-\frac{\pi i}{k}}\; \gamma_{\theta,7_2}\lambda \gamma_{\theta,7_3}^{-1}  \rightarrow (w_i,{\ov m}_i)\, ,
& \quad 7_37_2: \;\; \lambda  = e^{-\frac{\pi i}{k}}\;  \gamma_{\theta,7_3}\lambda \gamma_{\theta,7_2}^{-1}  \rightarrow ({\ov w}_{i-1},m_i)\, .
\nonumber 
\eeqa 
There are superpotential couplings in the $(7_i7_j)-(7_j3)-(37_i)$ sectors, which are not necessary for our purposes.

Note that not all of the D7-branes above must be simultaneously present. But for any non-trivial choice satisfying the RR tadpole cancellation condition (\ref{anomorbi}), there are localized 4d anomalies for the 8d gauge factors of the D7-branes present, which must be canceled by some inflow from the D7-brane bulk, as we show next.

\smallskip

{\bf D7-brane localized anomalies and cancellation by inflow}

Let us extract the 4d localized cubic non-abelian anomalies for the D7-brane gauge factors\footnote{There can also be mixed anomalies, which work similarly but with extra ingredients from 4d Green-Schwarz couplings \cite{Ibanez:1998qp}.}. As already mentioned, in the case of regular D3-branes, $n_a=N$, there are no contributions to the D7-brane anomalies from the D3-D7 open string spectrum (\ref{proj37}). Also, the D7$_i$-D7$_i$ spectrum (\ref{77-projection-same}) is non-chiral, so it does not produce localized anomalies. Hence, the 4d localized anomalies arise from the spectra (\ref{77-projection-different}), and read
\beqa
A_{SU(v_a)^3}=m_{a+1}-m_a=-A_{SU(w_a)^3}\; ,\; A_{SU(m_a)^3}= v_a - v_{a-1} -w_a + w_{a-1} \equiv 0\, ,\quad\quad 
\label{77-anomalies}
\eeqa
where in the last line we have used (\ref{anomorbi}). The automatic cancelation of the D7$_3$-brane anomalies $A_{SU(m_a)^3}$ is because the contributions of D7$_1$- and D7$_2$-branes to the spectrum in all sectors are conjugated. Morally, the D7$_1$- and D7$_2$-branes carry opposite twisted charges.
We will focus on configurations with no D7$_2$-branes ($w_a=0$), but including D7$_3$- and D7$_1$-branes (with general $m_a$ and, using (\ref{anomorbi}), $v_a=K_0$ for all $a$). This suffices to have non-trivial 4d localized anomalies, which arise on the D7$_1$-branes (the D7$_2$/D7$_3$ case is similar, and the case with all three kinds of branes is a mere superposition). 

The anomaly inflow arises from topological couplings of the D7-branes to twisted RR fields \cite{Douglas:1996sw}, specifically the RR 0-forms $c_0^p$ in the $\theta^p$-twisted sector, which propagates on the $z_3$ complex plane (times 4d spacetime), and its 6d dual 4-forms $c_4^p$. The D7$_1$-branes stretch along the $z_3$ complex plane, so they have a 6d topological coupling for their $SU(v_a)$ gauge factor with the structure
\beqa
S_{D7_{1,a}}=\int_{4d} \int_{z_3}\,\sum_{p=0}^k \tr(\gamma_{\theta^p_{7_1}}\lambda_a)\, c_0^p \,\tr F_a^3 =\int_{4d} \int_{z_3}\,\sum_{p=0}^k e^{\pi i \frac{(2a+1)p}{k}}\, c_0^p \,\tr F_a^3 \, ,
\label{71-coupling}
\eeqa
where in the first expression $\lambda_a$ is the Chan-Paton matrix for the $U(v_a)$ block. In the spirit of the inflow mechanism, we regard $\tr F_a^3$ as the anomaly polynomial of the 4d localized theory, so we denote it by $Y_a$ and use the descent relations (\ref{descent}), $Y_a=dY_a^{(0)}$, $\delta_\lambda Y_a^{0}=\lambda dY_a^{(1)}$. We now introduce $f_1^p=dc_0^p$ and integrate (\ref{71-coupling}) by parts, to write the action and its gauge variation as
\beqa
S_{D7_{1,i}}=-\int_{4d} \int_{z_3}\,\sum_{p=0}^k e^{\pi i \frac{(2a+1)p}{k}}\, f_1^p Y_a^{(0)} \;
\Rightarrow\;
\delta_\lambda S_{D7_{1,i}}=\lambda\int_{4d} \int_{z_3}\,\sum_{p=0}^k e^{\pi i \frac{(2a+1)p}{k}}\, df_1^p Y_a^{(1)}\, . \quad\quad\nonumber\\
\label{variation-twisted}\label{71-coupling-bis}
\eeqa
On the other hand the $b^{th}$ D7$_3$-branes, which do not span $z_3$, lead to a 4d coupling with the dual $\theta^p$-twisted RR 4-forms $c_4^p$, with the structure
\beqa
S_{D7_{3,b}}=\int_{4d} c_4^p \,\tr(\gamma_{\theta^p_{7_3}}^{-1}\lambda_b)= \int_{4d} c_4^p \,e^{-2\pi \frac {bp}k}\, ,
\eeqa
where the inverse of the $\gamma$ arises from an orientation flip as in \cite{Ibanez:1998qp}. This leads to the equation of motion
\beqa
df_1^p&=& \sum_{b=1}^k \, m_b\,  \sin \frac{\pi p}k\, e^{-2\pi \frac {bp}k}\, \delta_2(z_3)\, ,
\label{bianchi-twisted}
\eeqa
where $\delta_2(z_3)$ is a bump 3-form supported on the D7$_3$-brane location $z_3=0$, and the $\sin$ factor arises from the duality between the $c_0^p$ and the $c_4^p$, namely $dc_0^p= \sin \frac{\pi p}k *_{6d}  dc_4^p$. This can be checked from the closed string propagator in a direction twisted by $\theta^p$ \cite{Ibanez:1998qp}. 

\begin{figure}[htb]
\begin{center}
\includegraphics[scale=.3]{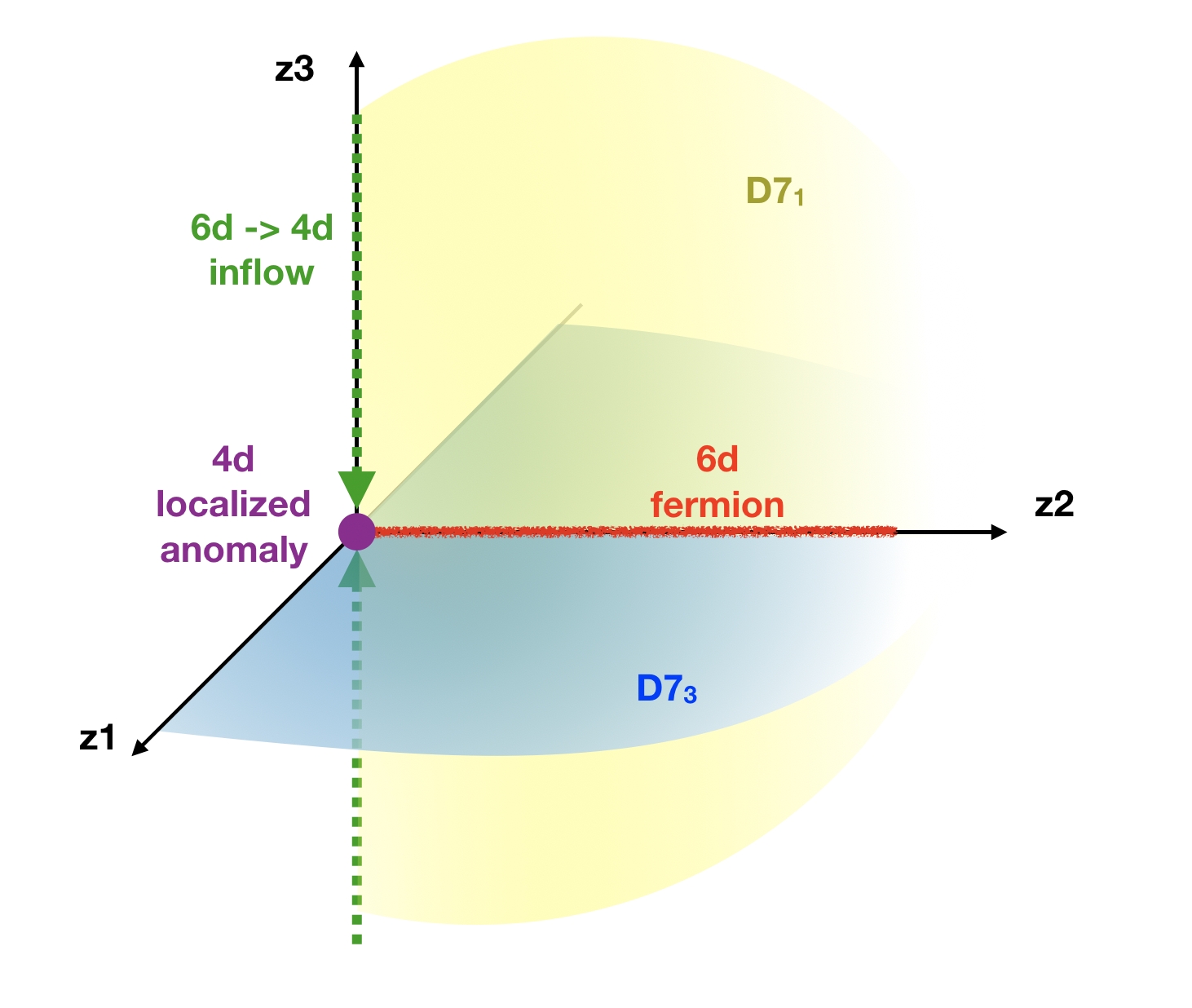}
\caption{\small The inflow for the brane configuration in flat space.}
\label{fig:inflow}
\end{center}
\end{figure}

Replacing now  (\ref{bianchi-twisted}) in (\ref{variation-twisted}) the variation is
\beqa
\delta S_{D7_{1,a}}&\sim&\int_{4d}\int_{z_3}\,\sum_{p=0}^k e^{\pi i \frac{(2a+1)p}{k}}\, \sum_{b=1}^k \, m_b\,  \sin \frac{\pi p}k\, e^{-2\pi \frac {bp}k}\,\, \delta_2(z_3) Y_a^{(1)} 
\sim (m_{a+1}-m_a)\int_{4d}Y_a^{(1)} \, ,\nonumber
\label{massage}
\eeqa
which has the structure necessary to cancel the localized 4d anomaly. A more precise picture of the anomaly inflow (see figure \ref{fig:inflow}) is as follows. Before the orbifold projection, the D7$_1$-D7$_3$ sector gives  a 6d bifundamental fermion propagating in $z_1=z_3=0$. The orbifold projection leads to a set of 4d zero modes localized at $z_1=z_2=z_3=0$, transforming as the bifundamentals in middle line in (\ref{77-projection-different}), and producing the 4d localized anomaly in the first entry in (\ref{77-anomalies}). This 4d anomaly is canceled by an inflow from the D7$_1$ brane worldvolume along the $z_3$ plane. Note that there is no inflow from the D7$_3$-branes because they do not span the complex plane $z_3$ on which the twisted RR forms propagate; this also matches the fact that the D7$_3$-brane gauge factors have identically zero localized anomaly, c.f. (\ref{77-anomalies}).

It is interesting to interpret the above story from the perspective of the 4d theory on the D3-branes, as follows. The $SU(v_a)$, $SU(m_b)$ are flavour global symmetries for the 4d quarks in the D3-D7 sectors. The $SU(v_a)$ global symmetries have an anomaly, whose anomaly theory is the 5d version of the above inflow mechanism, which is physically realized in the 5d gravity dual, as we now discuss.

\smallskip

{\bf The holographic picture of flavours}

We can now take the limit of large number $N$ of D3-branes and replace the flat space orbifold geometry by AdS$_5\times\IS^5/\IZ_k$, with the D7$_1$- and D7$_3$-branes as new ingredient. They are located at $z_1=0$ and $z_3=0$, respectively, so each spans AdS$_5$ times an $\IS^3$ given by the restriction of (\ref{unit-ball}), namely
\beqa
{\rm D7}_1:\, |z_2|^2+|z_3|^2=1\quad ,\quad {\rm D7}_3:\, |z_1|^2+|z_2|^2=1\, .
\label{s3-d7s}
\eeqa
Note that the D7$_1$-brane $\IS^3/\IZ_k$ includes the $\IS^1$  of singular points (\ref{s1-singular}), while the D7$_3$-brane $\IS^3/\IZ_k$ does not, it is a smooth Lens space. In either case, the 5d light modes from the $7_i7_i$ sector are obtained from the same orbifold projection as in flat space, c.f. (\ref{77-projection-same}), producing 5d gauge groups $U(v_a)$, $U(m_b)$ \footnote{For other fields, there are some extra subtleties which we skip. For instance the scalars are not really massless in AdS$_5$, rather they get tachyonic masses, but above the BF bound, so the system is stable.}.

Let us now discuss the $7_17_3+7_37_1$ sector. The two $\IS^3$'s (\ref{s3-d7s}) in the covering $\IS^5$ intersect over the $\IS^1$ given by  $|z_2|^2=1$. This intersection supports a 6d hypermultiplet in AdS$_5\times \IS^1$ in the bifundamental of the D7-brane gauge groups before the orbifold. This $\IS^1$ does not intersect the $\IS^1$ (\ref{s1-singular}) of singular points, so the $\IZ_k$ is freely acting an introduces a $\IZ_k$ Wilson line, hence the massless 5d states are those invariant under the projection (\ref{77-projection-different}). In particular, there are 5d fermions in AdS$_5$ in the bifundamentals $(v_a,{\ov m}_{a+1})$, $({\ov v}_a,m_a)$. These are 5d and hence non-chiral, so there are no anomalies.

On the other hand, the 5d theory should include 5d topological Chern-Simons couplings to account for the anomalies of global symmetries in the 4d holographic dual CFT discussed above. They arise from reducing on the internal space the D7-brane topological couplings to the RR twisted fields $c_0^p$, $c_4^p$. For this purpose, we introduce
\beqa
f_0^p=\int_{\IS^1/\IZ_k} f_1^p\, .
\eeqa
The $f_0^p$ are the (twisted sector axion) fluxes sourced by the D7$_3$ branes. These can be obtained from (\ref{bianchi-twisted}) by integrating over a 2d disk $D_2$ in the $z_3$ plane bounded by the $\IS^1/\IZ_k$ visible in the gravity dual. We have
\beqa
f_0^p=\int_{D_2}df_1^p &=& \sum_{b=1}^k \, m_b\,  \sin \frac{\pi p}k\, e^{-2\pi \frac {bp}k}\, ,
\label{twisted-fluxes}
\eeqa
where we used $\int_{D_2}\delta_2(z_3)=1$.
Reducing the D7$_1$ couplings (\ref{71-coupling-bis}) over $\IS^1/\IZ_k$, we get
\beqa
S_{D7_{1,a}}=\int_{5d} \,\sum_{p=0}^k e^{\pi i \frac{(2a+1)p}{k}}\, f_0^p I_a^{(0)} \, .
\label{action-5d}
\eeqa
The $f_0^p$ are just given by (\ref{twisted-fluxes}), so replacing them in (\ref{action-5d}), and operating as in (\ref{massage}), we get a 5d topological coupling
\beqa
S_5=\int_{5d} \sum_a (m_{a+1}-m_a)I_a^{(0)}\, .
\label{5d-inflow-couplings}
\eeqa
This is the anomaly theory of the 4d global symmetry anomalies, which are realized in the 4d holographic boundary. In other words, its gauge variation reproduces the anomaly of the holographic boundary theory, in a 5d version of the anomaly inflow mechanism of the flat space configuration.

The fact that the 5d theory has this topological coupling means that {\em any}  boundary that we put on this bulk theory must develop a localized 4d anomaly. We can then test this using physical 4d ETW boundaries of the Gaiotto-Witten kind

\subsection{Adding the 5-brane ETW configurations}
\label{sec:d3-singu-etw}

We now consider introducing the NS5- and D5-brane configuration to provide boundary conditions for the D3-brane gauge theory, and describe its gravity dual. We consider two possible ways to add such boundaries, distinguished by the embedding of the 5-branes in the $\IC^2/\IZ_k$ geometry. One class preserves 3d $\NN=1$ supersymmetry, while the second breaks all supersymmetries on the boundary. 

\subsubsection{Supersymmetric boundary conditions for the $\IC^2/\IZ_2$ theory}

Let us consider the flat space orbifold configuration of D3-, D7$_a$-branes at $\IC^2/\IZ_k$, and consider a way to add NS5- and D5-branes preserving supersymmetry. Let us regard the $\IC^2/\IZ_k \times \IC$ transverse to the D3-branes as a CY threefold, on which the D3- and D7-branes are wrapped on holomorphic (0- and 4-)cycles, so that the system preserves 4d $\NN=1$. We can add 5-branes preserving 3d $\NN=1$ supersymmetry by taking NS5- and D5-branes along 012 and spanning special lagrangian 3-cycles in the CY3 with different calibrating phases, namely
\beqa
&& J|_{\Pi_{\rm NS5}}=0 \quad ,\quad {\rm Im}\; \Omega|_{\Pi_{\rm NS5}}=0\, , \nonumber\\
&& J|_{\Pi_{\rm D5}}=0 \quad ,\quad {\rm Im}\; (e^{\pi/2}\Omega)|_{\Pi_{\rm D5}}=0 \, .
\label{slag}
\eeqa
Taking the complex and real coordinates be related by $z_1=x^4+i7$, $z_2=x^5+ix^8$, $z_3=x^6+ix^9$, this is precisely what is realized by taking the NS5-branes along 012456 and the D5-branes along 012789. These orientations are precisely those considered to define boundary conditions in section \ref{sec:d3-ns5-d5}.

\begin{figure}[htb]
\begin{center}
\includegraphics[scale=.3]{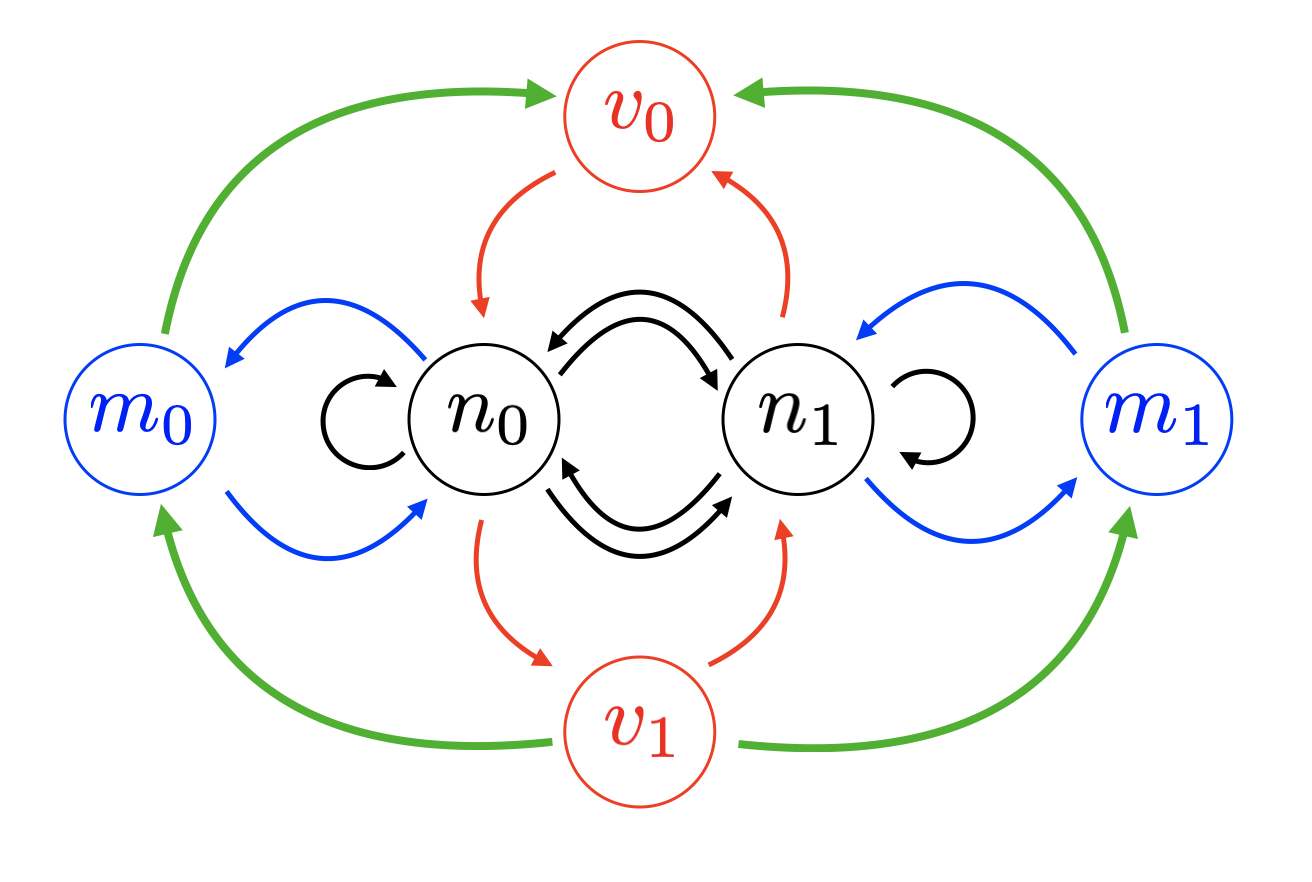}
\caption{\small Quiver diagram for the 4d theory on the D3/D7-brane system. Black circles correspond to the D3-brane gauge groups, with subindex labels for clarity (but recall we focus on $n_0=n_1=N$) and black arrows to 33 states. Blue and red circles correspond to D7-brane gauge groups (recall that consistency requires $v_0=v_1$), and arrows correspond to 37+73 states. Finally, green arrows correspond to $7_17_3+7_37_1$ states ($7_17_1$ and $7_37_3$ states are not shown to avoid clutter).}
\label{fig:d3d7z2}
\end{center}
\end{figure}

For general $\IC^2/\IZ_k$, the orbifold does not map the above 5-branes to themselves, so one is forced to introduce orbifold images to obtain an invariant configuration (see \cite{Feng:2001rh} for closely related models, also \cite{Blumenhagen:1999md,Blumenhagen:1999ev} for early references). These provide an interesting class of boundary conditions, but which do not have a known gravitational dual, due to the rotated orientations of the orbifold image 5-branes. However there is a special case\footnote{It is possible to build other similar examples, for instance using D3/D7-branes on a $\IC^3/(\IZ_2\times\IZ_2)$ singularity. We refrain from an exhaustive discussion and restrict to the simplest example.} where these images are not necessary. For $\IC^2/\IZ_2$, the generator acts as $\theta:(z_1,z_2,z_3)\to (-z_1,-z_2,z_3)$ and maps to themselves the NS5- and D5-branes along the original directions 012456 and 012789. Despite the relatively low order of the orbifold, it is easy to check that the  7$_1$-7$_3$ spectrum produces 4d chiral fermion zero modes and a localized 4d anomaly in the flat space configuration, as discussed above. Hence in this example we obtain 5-brane boundary conditions for the flavoured theory, with an explicit gravitational dual given by an orbifold of those in section \ref{sec:d3-ns5-d5-dual}. For convenience we describe the quiver of the gauge theory in figure \ref{fig:d3d7z2}.

Let us now turn to the discussion of the gravity dual for this particular $\IZ_2$ orbifold, first starting with no D7-brane flavour branes. The discussion is analogous to the NS5- and D5-brane ETW configurations for orbifolds AdS$_5\times\IS^5/\IZ_k$ discussed in \cite{Huertas:2023syg}. In particular, the solution is an orbifold of the AdS$_5\times\IS^5$ ending on the 5-brane ETW configuration. Let us focus on the geometry of the orbifold fixed locus $z_1=z_2=0$, namely $x^4=x^5=x^7=x^8=0$. Recalling the expression (\ref{s5-split}) of the $\IS^5$ as a $\IS^2\times\IS^2$ fibration over $\varphi$ at fixed $r$, we see that over a generic value of $\varphi$ there are 4 fixed points $x^6=\pm \cos \varphi$, $x^9=\pm \sin\varphi$. The four points collapse to only two by joining pairwise, in different ways at the endpoint values $\varphi=\pi/2$ and $\varphi=\pi$. The complete fibration over $\varphi$ thus corresponds to an $\IS^1$, which asymptotically matches the $\IS^1$ of singularities in the asymptotic AdS$_5\times\IS^5$ orbifold region. Including the radial coordinate $r$, the fixed locus is a complex plane, with the $\IS^1$ shrinking to zero size at $r=0$ (since so does the whole $\IS^5$). As explained, the action of the $\IZ_2$ maps the 5-branes to themselves.

The crucial new feature arising because of the introduction of the ETW configuration is that there is a new region, near $r=0$, where the geometry is locally AdS$_4\times \IR^6$, and the local physics is as in the original flat space configuration. This will be useful to understand the geometry and effects of the flavour D7-branes, which we study next.

We now introduce the flavour D7-branes and describe their geometry in the ETW configuration. From the description of the $\IS^5$ in (\ref{s5-split}) as a fibration over $\IS^2_1\times\IS^2_2$ over $\varphi$, the D7$_3$-branes, which are located at $x^6=x^9=0$, span an $\IS^3$ constructed as an $\IS^1\times\IS^1$ over $\varphi$, where the two $\IS^1$'s are the angular coordinate in the 45 and 78 planes. Near $r=0$, the $\IS^3$ shrinks and the D7$_3$-branes span a 4-plane. This implies that, even though they do not intersect the orbifold fixed locus in the asymptotic AdS$_5\times \IS^5/\IZ_2$ region, they do intersect it at $r=0$. Regarding the D7$_1$-branes, located at $x^4=x^7=0$, they span an $\IS^3$ constructed as an $\IS^1\times\IS^1$ over $\varphi$, where the two $\IS^1$'s are the polar angles in the 56 and 89 planes. In this case the $\IS^3$ includes the $\IS^1$ of the orbifold singularities even in the asymptotic region. Near $r=0$, they span a 4-plane, which includes the 2-plane of orbifold fixed points, exactly as in the original flat space geometry.

The intersection of the D7$_1$- and D7$_3$-brane $\IS^3$'s is an $\IS^1$, which arises as the fibration of 4 points over $\varphi$ given by $x^5=\pm\cos\varphi$, $x^7=\pm\sin\varphi$, degenerating pairwise in different ways at the two endpoint values $\varphi=\pi/2,\pi$. Near $r=0$, the intersection fills out a 2-plane. Therefore, thanks to the presence of the ETW configuration, there is a physical intersection of the 6d space on which the $7_17_3+7_37_1$ matter propagates and the orbifold fixed locus. This produces a set of physical fermions generating a localized 4d anomaly from the matter in the $7_17_3+7_37_1$ sector, exactly as in the flat space case. This 4d localized anomaly is canceled by the inflow mechanism from the 5d bulk topological couplings (\ref{5d-inflow-couplings}). The discussion is similar to the inflow towards the 4d holographic boundary in section \ref{sec:flavouring}, but with a physical 4d boundary and physical 4d fermion degrees of freedom whose anomaly is canceled by the inflow.

This completes our discussion of a fully supersymmetric model exhibiting anomaly inflow analogous to that in the non-supersymmetric model in section \ref{sec:anomaly-td}. As a final remark, we would like to note an interesting difference in the origin of the 4d dynamical fermions on the ETW boundary. In the example in section \ref{sec:anomaly-td}, there are no bulk fermions, they arise directly on the 4d boundary, as localized directly on the 5-branes. In the example described in this appendix there are 5d bulk fermions, and the 4d boundary fermions arise as localized zero modes of these. It is interesting that we can realize these two mechanisms using simple systems of D-branes, and that they both lead to very similar anomaly physics.

\subsubsection{Non-supersymmetric AdS$_4$ ETW configurations for AdS$_5\times\IS^5/\IZ_k$}
\label{sec:nonsusy-ads4}

We now quickly discuss a generalization of the above inflow mechanism for anomalous symmetries on the ETW brane to a class of non-supersymmetric ETW boundary configurations for AdS$_5\times\IS^5/\IZ_k$. The configuration without ETW branes is the gravity dual of the  configurations of D3-branes with D7$_1$- and D7$_3$-branes in section \ref{sec:flavouring}. We subsequently introduce NS5- and D5-branes which preserve 8 common supersymmetries with the D3-branes, as in section \ref{sec:d3-ns5-d5}, but which preserve no common supersymmetry with the D7-brane system. The system has a supersymmetric bulk corresponding to the gravity dual of a configuration of D3/D7-branes at $\IC^2/\IZ_k$ preserving 4 supersymmetries, and a 5-brane ETW configuration which preserves common supersymmetries with the dual of the D3-branes at $\IC^2/\IZ_k$, but breaking the supersymmetries preserved by the probe D7-branes. The boundary configuration is therefore non-supersymmetric, but with a the supersymmetry breaking controlled by the number of D7-branes, small compared with both the number $N$ of D3-branes and the number $N_5$ of 5-branes (which is taken much larger than $N$ in order to produce localized gravity).

The configuration of 5-branes is easily described as follows. Let us consider that the complex coordinates and the real coordinates relate by $z_1=x^4+ix^5$, $z_2=x^7+ix^8$, $z_3=x^6+ix^9$. Using the standard NS5-branes along 012456 and D5-branes along 012789, we see that they are oriented differently in the $\IC^2/\IZ_k\times \IC$, as compared with the previous section. In particular, each stack of 5-branes is mapped to itself under the orbifold action (\ref{zk-action}), even for general $\IZ_k$. From the general arguments in the previous section, the 5d bulk Chern-Simons couplings in the simultaneous presence of D7$_1$- and D7$_3$-branes implies that there must be 4d dynamical chiral fermion modes localized on the ETW boundary. These arise, as in the previous section, from localized modes of the higher-dimensional fermion at the D7$_1$-D7$_3$ intersection.

Let us expand a bit on the interplay of the ETW geometry and the $\IZ_k$ orbifold, following \cite{Huertas:2023syg}. The solution is an orbifold of the AdS$_5\times\IS^5$ ending on the 5-brane ETW configuration. Let us focus on the geometry of the orbifold fixed locus $z_1=z_2=0$, namely $x^4=x^5=x^6=x^7=0$. Using (\ref{s5-split}) to describe the $\IS^5$ as an $\IS^2\times\IS^2$ fibration over $\varphi$ at fixed $r$, we see that over a generic value of $\varphi$ there are 4 fixed points $x^8=\pm \cos \varphi$, $x^9=\pm \sin\varphi$. As in the previous section (with a mere exchange of the roles of $x^6$ and $x^8$), the pairwise collapse of the points at the endpoint values $\varphi=\pi/2$ and $\varphi=\pi$, leads to a whole $\IS^1$ of orbifold fixed point on $\IS^5$. Including the radial direction $r$, we fet a complex curve of fixed points in the full geometry. As emphasized above, the 5-branes are mapped to themselves under the $\IZ_k$ action, for general $k$.

Hence we have a class of 5-brane configurations for the gravity dual of D3/D7-branes at $\IC^2/\IZ_k$ for general $k$. The price to pay is that, although the 5-branes preserve common supersymmetries with the D3-branes, defining supersymmetric boundary configurations for AdS$_5\times \IS^5/\IZ_k$ \cite{Huertas:2023syg}, they break all the supersymmetries of the D7-brane probes in this geometry. For instance, the D5-branes span 012789 while the D7$_1$-branes span 01236789, i.e. only 2 DN+ND directions, and the D7$_3$-branes span 01234567, so have 6 DN+ND directions with the D5-branes. The breaking of D7-brane supersymmetry by the 5-brane ETW configuration will play a prominent role in section \ref{sec:no-nonsusy-ads}.

Let us explain in more detail the interplay of the D7-branes and the ETW geometry. From the description of the $\IS^5$ in (\ref{s5-split}) as a fibration over $\IS^2_1\times\IS^2_2$ over $\varphi$, the D7$_3$-branes, located at $x^8=x^9=0$, span an $\IS^3$ constructed as a fibration over $\varphi$ of $\IS^2$ times 2 points (i.e. two disjoint $\IS^2$'s) $x^7=\pm\sin\varphi$. The $\IS^2$'s collapse to zero size at $\varphi=\pi/2$, and they both get identified at $\varphi=\pi$, so the fibration corresponds to an $\IS^3$ (with each $\IS^2$ fibered over the interval corresponding to each hemisphere). Near $r=0$ the $\IS^3$ shrinks and the D7$_3$-branes span a 4-plane, which intersects the orbifold fixed locus (even though it doesn't in the asymptotic AdS$_5\times \IS^5/\IZ_k$ region). Regarding the the D7$_1$-branes, located at $x^4=x^5=0$, they span an $\IS^3$ constructed as a fibration of 2 points $x^6=\pm\cos\varphi$ times an $\IS^2$ (i.e. two disjoint $\IS^2$'s). In analogy with the D7$_1$-branes, this describes an $\IS^3$. Near $r=0$, they span a local 4-plane, which includes the orbifold singular 2-plane. As explained above, the $r=0$ region works locally exactly as in the original flat space geometry.

The intersection of the D7$_1$- and D7$_3$-brane $\IS^3$'s is an $\IS^1$ $x^4=x^5=x^7=x^8$, which arises as the fibration of 4 points over $\varphi$ given by $x^6=\pm\cos\varphi$, $x^9=\pm\sin\varphi$, which reconstructs an $\IS^1$ in a by now familiar way. Near $r=0$, the intersection fills out a 2-plane, which supports the 6d $7_17_3+7_37_1$ matter, and whose intersection with the orbifold fixed set provides the localized 4d chiral mode responsible for the anomaly to be cancelled by the inflow (\ref{5d-inflow-couplings}).

\bibliographystyle{JHEP}
\bibliography{mybib}

\end{document}